\documentclass[preprint,5p,times,twocolumn,authoryear]{elsarticle}

\usepackage{epsfig}
\usepackage{subfigure}
\usepackage{subcaption}
\usepackage{graphicx}
\usepackage{tikz}
\usepackage{multirow}
\usepackage{tabularx}
\usepackage{pdflscape}
\usepackage{physics}
\usepackage{float}
\usepackage{setspace}
\usepackage{amssymb}
\usepackage{amsmath}

\journal{International Journal of Heat and Fluid Flow}

\begin{document}
\captionsetup[figure]{labelfont={bf},labelformat={default},labelsep=period,name={Fig.}}

\begin{frontmatter}

\title{Spatial Features of Reynolds-Stress Carrying Structures in Turbulent Boundary Layers with Pressure Gradient}

 \author[label1]{Mehmet Ali Yesildag}
 \author[label1,label2]{Taygun Recep Gungor}
 
  \affiliation[label1]{organization={Faculty of Aeronautics and Astronautics, Istanbul Technical University},
 	city={Istanbul},
 	postcode={34469},
 	country={Turkey}}
 
 \affiliation[label2]{organization={Department of Mechanical Engineering, Universite Laval},
         	 city={Quebec City},
             postcode={G1V 0A6}, 
             state={QC},
             country={Canada}}

\author[label1]{Ayse Gul Gungor}
 
\author[label2]{Yvan Maciel}

\begin{abstract}

We investigate the Reynolds-shear-stress carrying structures in the outer layer of non-equilibrium pressure-gradient turbulent boundary layers using four direct numerical simulation databases, two cases of non-equilibrium pressure-gradient boundary layers and two of homogeneous shear turbulence. We examine and compare the spatial organization and shapes of the Reynolds-shear-stress structures, specifically sweeps and ejections, across all cases. The analysis includes five streamwise locations in the boundary layers, varying in pressure-gradient sign, intensity, and upstream history. For the boundary layers, two types of three-dimensional velocity fields are considered: fully spatial fields and spatio-temporal fields using Taylor's frozen turbulence hypothesis. Comparisons of the results indicate that the statistics of sweep and ejection shapes are sensitive to the choice of convection velocity in Taylor's hypothesis. The sweep and ejection shapes are consistent across all flows when their sizes range from 1 to 10 Corrsin length scales, suggesting that mean shear plays a similar role in all cases, driving the formation of Reynolds-shear-stress carrying structures and contributing to turbulence production. Sweeps and ejections of different types form side-by-side pairs, while structures of the same type align in an upstream-downstream configuration. This behavior persists regardless of pressure gradient variations or upstream history, emphasizing the dominant influence of local mean shear.

\end{abstract}

\begin{highlights}
	
\item Mean shear governs sweep and ejection spatial features in all flows.

\item Sweeps and ejections share consistent spatial features across flows.

\item Their dimensions scale with Corrsin length.

\item Their spatial arrangement also remains consistent across flows.
\end{highlights}


\end{frontmatter}

\section{Introduction}
\renewcommand{\arraystretch}{1.5}
\begin{table*}[h!]
	\centering
\caption{Properties of the DNS databases. DNS22 corresponds to the database of \citet{gungor2022}, while DNS23 refers to the database of \citet{gungor2024}. The domain dimensions are denoted by $L_x$, $L_y$, and $L_z$, representing the streamwise, wall-normal, and spanwise lengths, respectively. The boundary layer thickness at the domain exit is indicated by $\delta_e$. Grid sizes in the respective directions are given as $N_x$, $N_y$, and $N_z$. The minimum and maximum grid resolutions in the streamwise ($\Delta x^+$), wall-normal ($\Delta y^+$), and spanwise ($\Delta z^+$) directions are normalized by friction viscous scales, $\nu/u_{\tau}$, across the boundary layer. The Reynolds number based on momentum thickness and inlet velocity is represented by $Re_{\theta}$.}

	\label{tab::domain}
	\begin{tabularx}{\textwidth}{@{\extracolsep{\fill}}lrrrrrr@{}}
		\hline \hline
		& $(L_x, L_y, L_z) / \delta_{e}$ & $N_x \times N_y \times N_z$  & $\Delta x^+$ & $\Delta y^+$ & $\Delta z^+$ & $Re_{\theta}$\\
		\hline
		DNS22  & $14.6$, $2.8$, $4.1$ & $4609 \times 736 \times 1920$   &  $0.7$ - $11$&  $0.3$ - $17$ & $0.5$ - $7$ & $1569$ - $8648$\\ 
		DNS23  & $26.1$, $3.0$, $4.1$ & $12801 \times 770 \times 2700$ & $1.6$ - $9.8$ &  $0.5$ - $11$ & $1.2$ - $7.4$ & $1941$ - $12970$ \\
		\hline \hline
	\end{tabularx}
\end{table*}

Turbulent boundary layers developing under a pressure gradient are one of the most common flows encountered in engineering applications. The pressure force acting on the flow significantly affects the mean flow and the turbulent activity. When a turbulent boundary layer (TBL) is exposed to an adverse pressure gradient (APG), the mean velocity defect becomes significant, resulting in higher mean shear in the outer layer and reduced shear near the wall. These changes lead to significant alterations in turbulence characteristics. In canonical wall-bounded flows like channel flows or zero pressure gradient (ZPG) TBLs, turbulence is dominant in the inner region. However, in APG TBLs with large mean velocity defects, Reynolds stresses reach their highest values in the outer layer \citep{skarekrogstad1994, harun2013, gungor2016, lee2017, maciel2018, tanarro2020}. The increased turbulence production in the outer layer of APG TBLs is a crucial distinction from ZPG TBLs, where turbulence activity is primarily confined to the inner layer. Moreover, turbulence production in the outer layer of APG TBLs surpasses that in the inner layer when the defect is very large~\citep{gungor2016,gungor2022}.

In contrast, favorable pressure gradient (FPG) TBLs exhibit opposite effects, where the flow acceleration leads to a fuller mean velocity profile and diminished turbulence intensity in the outer region. The Reynolds stresses in FPG TBLs decay significantly in the outer layer, due to suppressed turbulence production in that region \citep{harun2013,volino2020,devenport2022}. This decay is attributed to the stabilizing effect of the FPG, which reduces mean shear and decreases the energy and activity of large-scale motions in the outer region \citep{piomelli2013}. \citet{joshi2014} demonstrated that an FPG suppresses the outward migration of near-wall turbulence which confines the turbulent activity closer to the wall.

While coherent structures and turbulence regeneration mechanisms have been extensively studied in canonical wall-bounded flows \citep{jimenezkawahara2010,jimenez_nearwall,Marusic_Adrian_2012}, a comprehensive theory of turbulence regeneration remains elusive \citep{jimenez2018}, particularly in pressure gradient TBLs (PG TBLs). The turbulence structures in pressure gradient turbulent boundary layers are influenced not only by the local pressure gradient but also by the upstream flow history. The cumulative effect of the upstream history further complicates the understanding of these flows, as it significantly shapes their turbulence structure and dynamics. Studies examining these cumulative effects include \citet{bobke2017,vinuesa2017,tanarro2020,volino2020,parthasarathy2023,gungor2024}. Among their findings, these studies reveal that the inner layer responds more rapidly to changes, while the outer layer exhibits a delayed response to upstream conditions, with differing behaviors between the mean flow and turbulence.

Furthermore, large-defect APG TBLs have been shown to exhibit characteristics similar to free-shear flows due to changes in the mean velocity profiles, suggesting that the wall's influence on turbulence in such flows is minimal \citep{gungor2016,kitsios2017}. In these flows, large-scale sweeps (Q4, $u > 0$ and $v < 0$) and ejections (Q2, $u < 0$ and $v > 0$), where $u$ and $v$ denote streamwise and wall-normal velocity fluctuations, respectively, and which carry most of the Reynolds shear stress, resemble those observed in homogeneous shear turbulence (HST) \citep{gungor2020}. \citet{dong} also demonstrated that large-scale sweeps and ejections in the overlap layer of channel flows exhibit similar characteristics to those in HST. These studies indicate that outer sweeps and ejections in wall-bounded flows are predominantly driven by the local mean shear, with minimal influence from the presence of a wall. \citet{deshpande2024} also studied sweeps and ejections in APG TBLs. Their findings highlight the critical role these structures play in distributing energy between the inner and outer regions.

Moreover, sweeps and ejections have been widely studied using one-point statistics and quadrant analysis.
\citet{maciel2017a} studied Q2s and Q4s in non-equilibrium APG TBLs using two methods: single-point quadrant statistics and quadrant analysis of strong $uv$ events. Their findings indicate that attached structures are predominant near the wall, as observed in ZPG TBLs. In contrast, in the outer layer of APG TBLs, detached structures contribute equally, differing from ZPG TBLs where attached structures dominate. \citet{maciel2017b} showed that in a large-defect APG TBL, near-wall $Q^-$ events are significantly less numerous than in a ZPG TBL, representing only 10\% of all $Q^{-} \mathrm{s}$ compared to 43\% in the ZPG case. Furthermore, the $Q^{-} \mathrm{s}$ in the APG TBL are primarily wall-detached and occupy larger volumes, while in the ZPG TBL, wall-attached $Q^{-} \mathrm{s}$ dominate the flow. \citet{abe2020} analyzed sweeps and ejections in APG TBLs, including near-separation and reattachment cases. Their findings indicate that, contributions of sweeps and ejections to Reynolds shear stress increase in APG compared to ZPG, even though their probability of occurrence decreases. Ejections and sweeps in APG regions become more intermittent yet more intense.

The spatial organization of sweeps and ejections in channel flows has been studied by \citet{lozano2012}, who found that same-kind Qs (e.g., Q2 with Q2) form an upstream-downstream configuration, while opposite kinds (e.g., Q2 with Q4) tend to arrange in side-by-side pairs. Later, \citet{dong} investigated sweeps and ejections in HSTs and outer-layer channel flows, and found similar results. \citet{gungor2020} studied the spatial organization of sweeps and ejections in the outer layer of APG TBLs and revealed that it is similar across all flows, consistent with the findings of \citet{lozano2012} and \citet{dong}.

In this work, we extend the study of \citet{gungor2020} by considering a wider variety of pressure gradient situations and flow histories of non-equilibrium TBLs. The primary objectives of this study are to analyze the spatial characteristics of Reynolds-shear-stress carrying structures in non-equilibrium PG TBLs, quantify their sizes, shapes, aspect ratios, and contributions to the Reynolds-shear-stress, and reveal how their spatial organization is influenced by the interplay of mean shear, pressure gradients, and upstream flow history. To achieve this, we employ two direct numerical simulation (DNS) databases: the non-equilibrium PG TBLs of \citet{gungor2022} (evolving from mild to strong adverse pressure gradients) and \citet{gungor2024} (undergoing adverse-to-favorable pressure gradient transitions). By employing both spatio-temporal and fully-spatial data, this study investigates how mean shear, pressure gradients, and upstream flow histories collectively influence the organization of Reynolds-shear-stress carrying structures.

\begin{table*}[ht!]
	\centering
	\caption{The main properties of the flow cases. ZPG indicates the database of \citet{sillero2013}. HST1 and HST2 are the homogoneous shear turbulence database of \citet{dong}. The skin friction coefficient, $C_f$ is defined as $C_f = \tau_w / (0.5\rho U_e^2)$ where $\tau_w$ is the wall-shear stress, $\rho$ is the density and $U_e$ is the velocity at the boundary layer edge. The Clauser pressure gradient parameter, $\beta_{C}$, is defined as $\beta_{C} = (\delta^* / (\rho u_\tau^2)) (dP_e/dx)$, where $\delta^*$ is the displacement thickness, $u_\tau$ is the friction velocity, and $P_e$ is the pressure at the edge of the boundary layer.}
	\label{tab::cases}
	\begin{tabularx}{\textwidth}{@{\extracolsep{\fill}}rrrrrrrrr@{}}
		\hline \hline
		Position & $H$ & $Re_{\theta}$ & $Re_{\tau}$ & $Re_{\lambda}$  & $C_{f} \times 10^{-3}$ & $\beta_{ZS}$ & $\beta_{C}$ \\
		\hline

		DNS22-APG1  & 1.65  & 2985  & 626   & 57 - 78      	& 2.1932 & 0.33      	 & 4.51 \\
		DNS22-APG2  & 2.63  & 5769  & 455   & 86 - 110     	& 0.3619 & 0.13     	 & 42.48 \\
		DNS23-APG1  & 1.60  & 5335  & 1010  & 70 - 99       & 2.0506 & 0.27      	 & 4.42 \\
		DNS23-APG2  & 2.78  & 11727 & 850   & 129 - 175     & 0.2961 & 0.07        & 31.88 \\
		DNS23-FPG   & 1.60  & 12287 & 2156  & 153 - 218    	& 1.8612 & -0.15       & -2.27 \\
		ZPG         & 1.35 	& 6500  & 1990  & 85 - 116      & 2.7063 & $\approx$ 0 & $\approx$ 0 \\
		HST1        & N/A  	& N/A   & N/A	& 104   	    & N/A    & N/A         & N/A  \\
		HST2        & N/A   & N/A   & N/A 	& 248   	    & N/A    & N/A         & N/A  \\
		\hline \hline
	\end{tabularx}
\end{table*}

\section{DNS Databases and description of the flows}

The first PG TBL database, referred to as DNS22 and described in \citet{gungor2022}, represents a non-equilibrium APG TBL with a Reynolds number based on momentum thickness, $Re_\theta$, reaching 8650. The second database, referred to as DNS23 and detailed in \citet{gungor2024}, corresponds to a non-equilibrium TBL subjected to an APG, followed by a FPG region, with $Re_\theta$ reaching 13,000. A brief summary of the domain properties of both cases is given in Table ~\ref{tab::domain}. In addition to these databases, we employ two homogeneous shear turbulence (HST) databases with Taylor microscale Reynolds numbers of 104 and 248, as described in detail by \citet{dong}, to better understand the influence of mean shear in the outer layer. They represent a free shear flow where turbulence is influenced solely by mean shear.

\begin{figure}[h!]
	\centering
	
	\begin{tikzpicture}
		\node[anchor=north west] at (-3.5, 0) {\hspace{-1.8cm}\includegraphics[width=.95\linewidth]{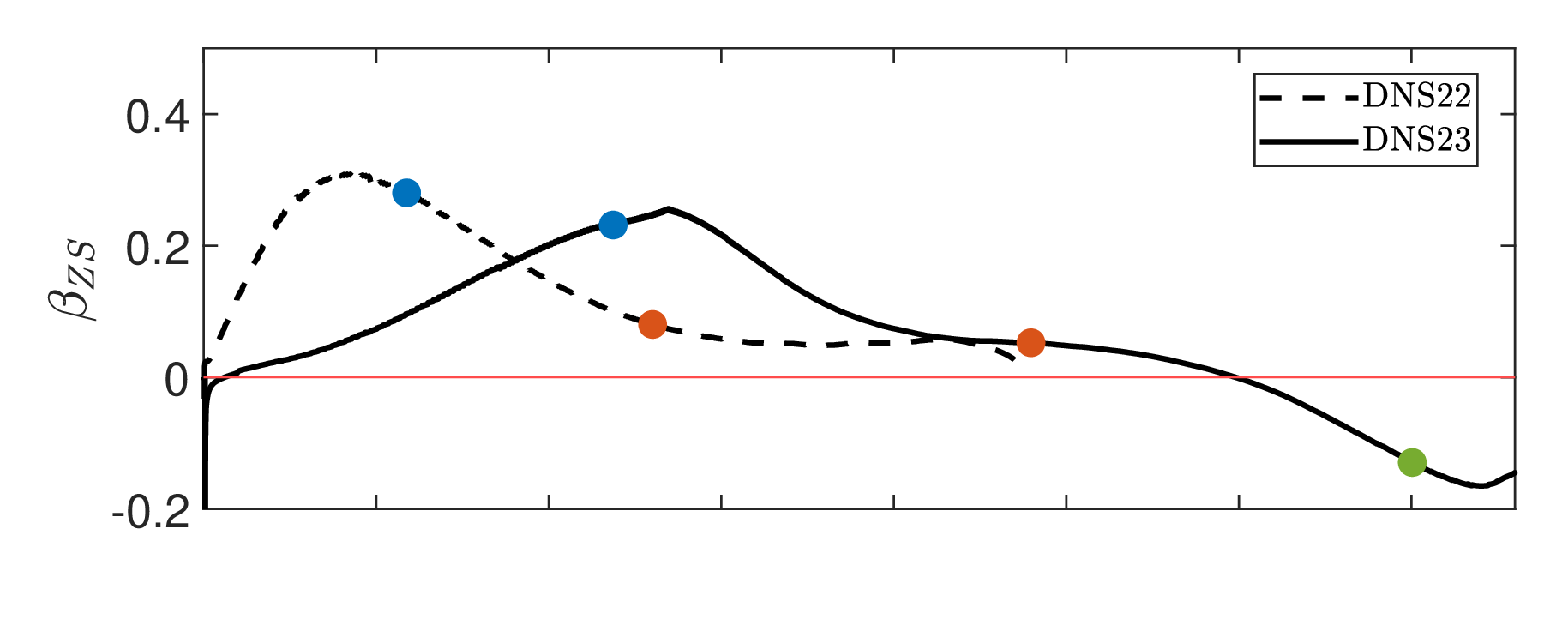}};
		\node[anchor=north west, xshift=-3em, yshift=-.8em] at (-4.3, 0) {$(a)$};
	\end{tikzpicture}
	
	\vspace{-.88cm}
	\begin{tikzpicture}
		\node[anchor=north west] at (-3.5, 10.1) {\hspace{-1.8cm}\includegraphics[width=.95\linewidth]{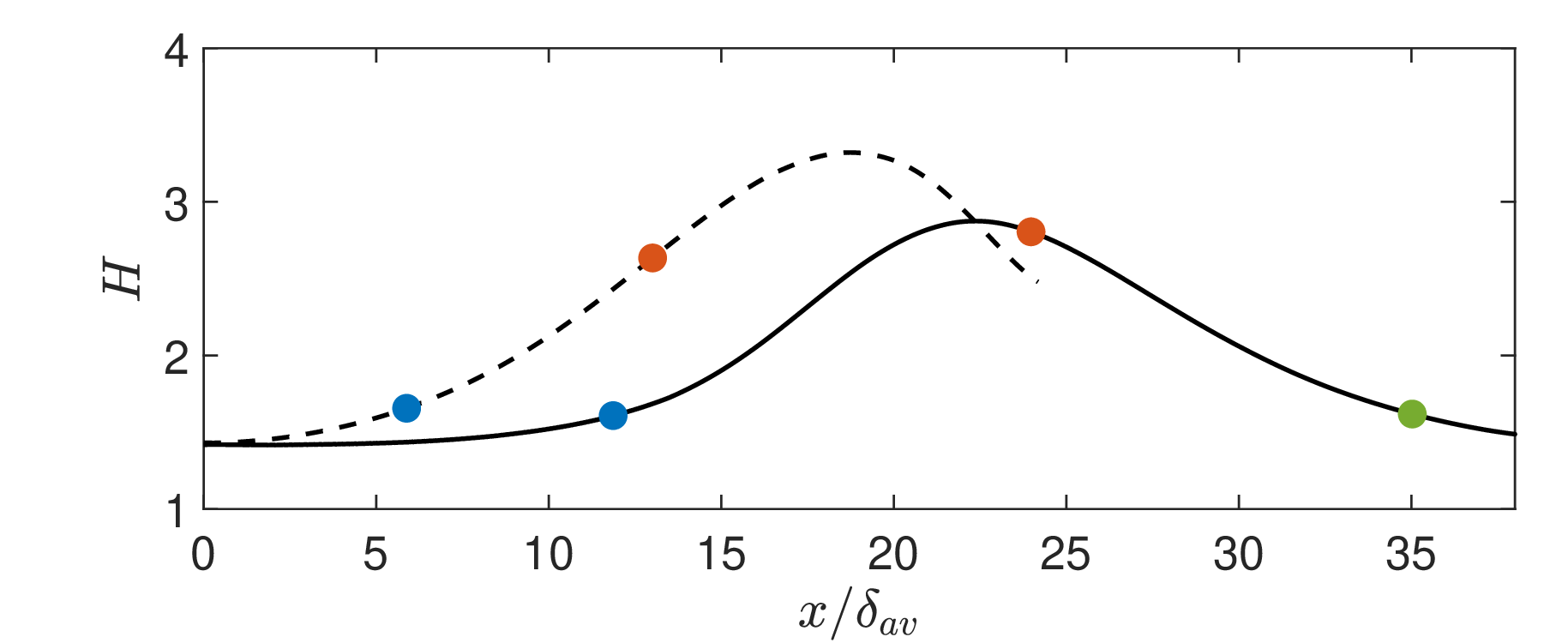}};
		\node[anchor=north west, xshift=-3em, yshift=-.7em] at (-4.3, 10) {$(b)$};
	\end{tikzpicture}
	
	\caption{The $\beta_{Z S}$ and $H$ distribution of the TBL cases.}
	\label{fig::dns22-23diff}
\end{figure}

Before analyzing the coherent structures, the key parameters and characteristics of the two PG TBLs are summarized. Fig.~\ref{fig::dns22-23diff}a shows the spatial evolution of the outer pressure gradient parameter based on Zagarola-Smits velocity, $\beta_{Z S}=\left(\delta / \rho U_{Z S}^{2}\right)\left(d p_{e} / d x\right)$, for the two PG TBLs, where $U_{ZS}=U_e\delta^*/\delta $, $\delta$ is the local boundary layer thickness, $p_{e}$ is the pressure at the edge of the boundary layer and $\delta^*$ is the displacement thickness. The streamwise length of the domain is normalized by $\delta_{av}$, the average boundary layer thickness within the useful range of the DNS domain \citep{gungor2022,gungor2024}. The pressure gradient parameter $\beta_{Z S}$ follows the ratio of pressure force to the turbulent force in the outer layer \citep{maciel2018}. Because of the initial increase of $\beta_{ZS}$, DNS23 evolves from a ZPG TBL to a TBL with a large velocity defect. Subsequently, it is subjected to a favorable pressure gradient, resulting in an FPG TBL with the flow history of an adverse pressure gradient TBL. DNS22 is a non-equilibrium APG TBL, which evolves from a ZPG TBL to a TBL on the verge of separation. The shape factor distribution in Fig.~\ref{fig::dns22-23diff}b gives an idea of the difference in mean velocity defect evolution between the two cases. DNS22 is in stronger disequilibrium than DNS23, whereas the Reynolds number is higher in DNS23 than in DNS22, as shown in Table ~\ref{tab::domain}.

For analyzing the sweeps and ejections in detail, we choose two streamwise positions from DNS22 and three positions from DNS23. These positions are indicated by filled circles in Fig.~\ref{fig::dns22-23diff}. Table~\ref{tab::cases} summarizes the main parameters of these streamwise positions. The first position of each flow has a small velocity defect $(H=1.65$ and $1.60$ for DNS22 and DNS23, respectively) while the second one has a large velocity defect ($H=2.63$ and $2.78$, for DNS22 and DNS23, respectively). The third position of DNS23 has the defect of the first position $(H=1.60)$ but the flow is under the effect of a FPG and a long non-equilibrium history. The ZPG TBL of \citet{sillero2013} at Reynolds number based on friction velocity, $Re_{\tau}=1990$ is also used as a reference case.  In addition to these parameters, the Reynolds number based on the Taylor microscale, $Re_\lambda$, is also presented in Table ~\ref{tab::cases}. Here, the Taylor microscale Reynolds number, $Re_\lambda$, is defined as $Re_\lambda = q^2(5/3\nu\varepsilon)^{1/2}$, where $q^2 = \langle u_i u_i \rangle$ represents turbulent kinetic energy, and the Taylor microscale is given by $\lambda = q \sqrt{5\nu / \varepsilon}$. The turbulent dissipation rate, $ \varepsilon = 2\nu  \left\langle s_{ij}s_{ij} \right\rangle$, is computed directly. $Re_\lambda$ is applicable to wall-bounded flows but not as a global parameter, as it varies with $y$. Therefore, it is presented as the minimum and maximum values within the region of interest in this study, $0.3\delta < y < 0.8\delta$.

\begin{figure}[h!]
	\centering
	\begin{tikzpicture}

		\node at (-1, 0) {\hspace{-0.3cm} \includegraphics[width=.74\linewidth]{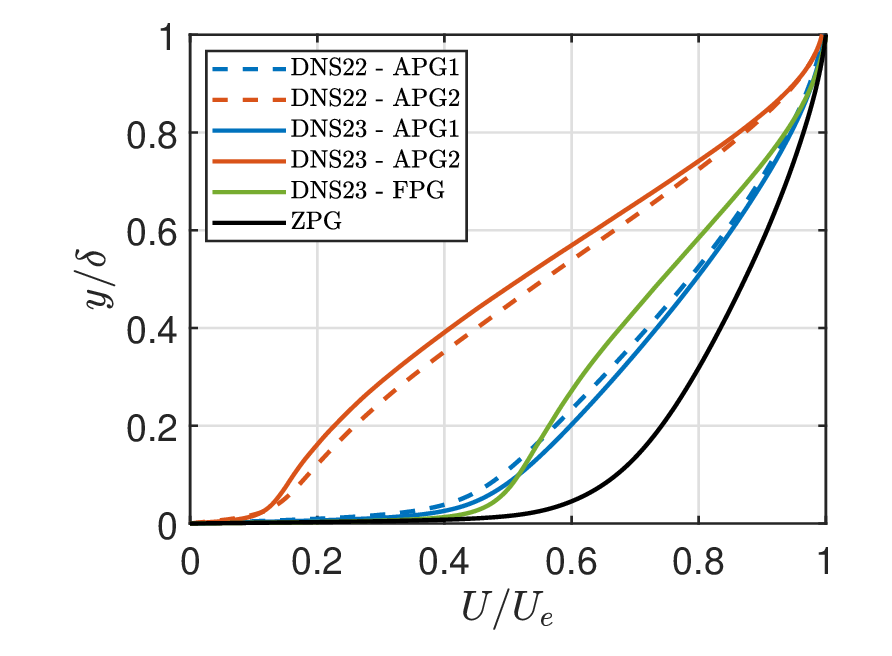}};
		\node[above left, inner sep=5pt, xshift=-5pt, yshift=0pt] at (-3.5, 1.85) {$(a)$}; 
		
		\node at (-1.,-4.7) {\hspace{-0.3cm} \includegraphics[width=.74\linewidth]{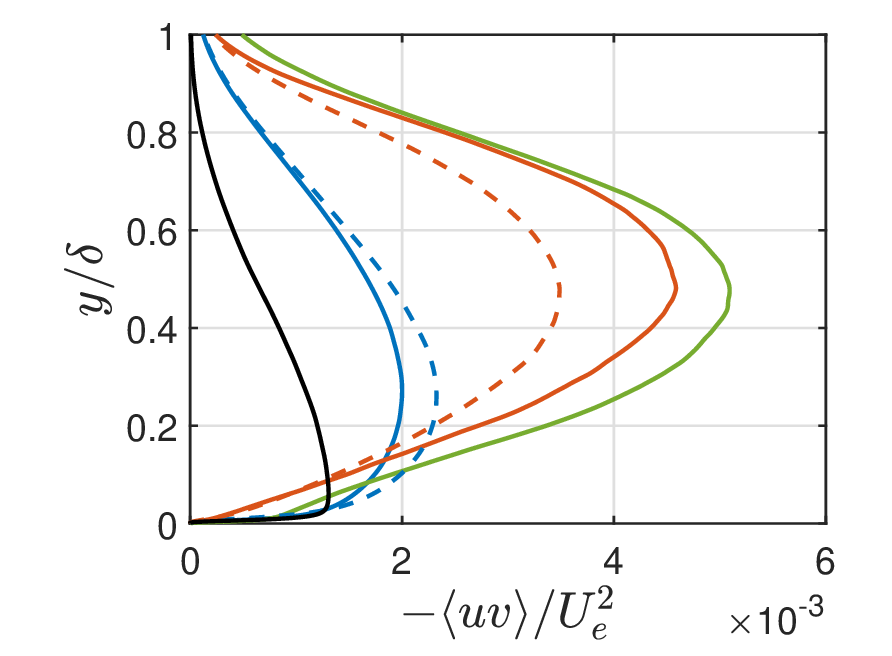}};
		\node[above left, inner sep=5pt, xshift=-5pt, yshift=15pt] at (-3.5, -3.5) {$(b)$}; 
		
		\node at (-1, -9.5) {\hspace{-0.3cm} \includegraphics[width=.74\linewidth]{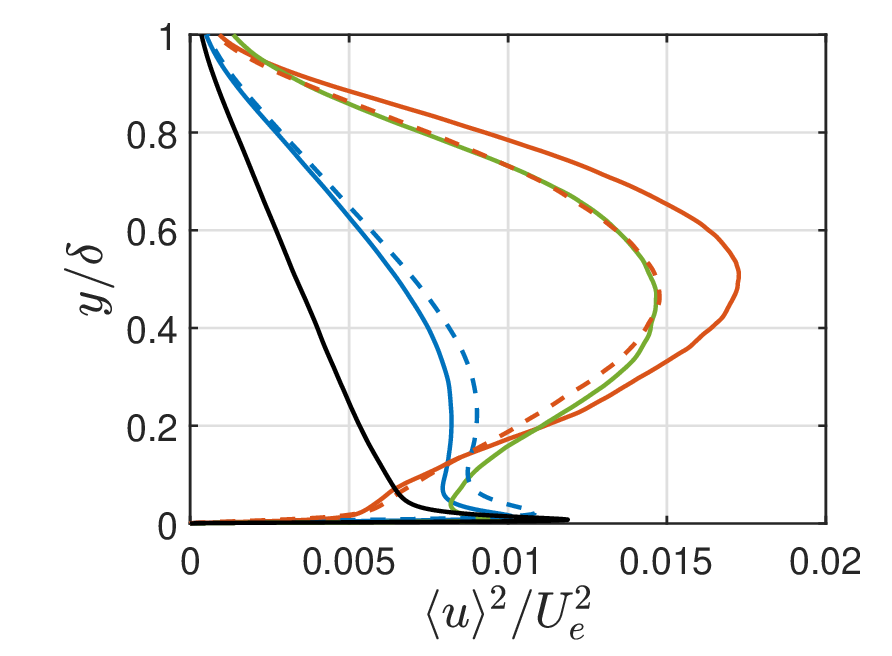}};
		\node[above left, inner sep=5pt, xshift=-5pt, yshift=-5pt] at (-3.5, -7.6) {$(c)$}; 
	\end{tikzpicture}
	
	\caption{The mean velocity (a), Reynolds shear stress $-\left\langle uv\right\rangle$ (b), and Reynolds normal stress $\langle u^{2}\rangle$ (c) profiles of the selected flow cases as a function of $y / \delta$.}
	
	\label{fig2}
\end{figure}
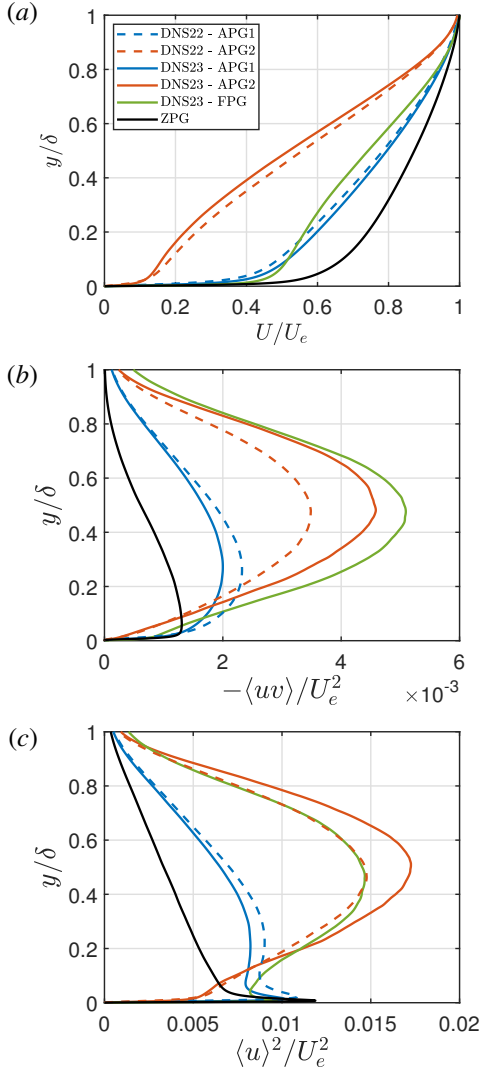

Fig.~\ref{fig2} presents the mean velocity, $-\left\langle uv\right\rangle$ and $\langle u^{2}\rangle$ profiles as a function of $y / \delta$ for the TBL cases. In both the small and large-defect APG cases, the two flows exhibit similar mean velocity profiles at comparable $H$ values. However, differences arise due to small variations in $H$ values, distinct flow histories, and Reynolds numbers. The FPG case exhibits an unusually significant velocity defect for an FPG TBL, comparable to the small-defect APG cases, due to its long and strong APG flow history. However, the FPG profile differs from the small-defect APG cases by gaining momentum more rapidly near the wall than in the outer region. Nonetheless, the mean shear is similar to that of the small-defect APG cases for $y>0.3\delta$. The mean shear decreases as the velocity defect diminishes in the outer layer, typically resulting in turbulence decay. However, as illustrated in Fig.~\ref{fig2}b, the Reynolds shear stress increases from APG2 (orange line) to FPG (green line). This increase is attributed to the delayed response of turbulence to changes in the pressure gradient and mean velocity \citep{gungor2024}.

\begin{figure}[hbtp!]
	\centering
	\begin{tikzpicture}
		\pgfdeclarelayer{background}
		\pgfdeclarelayer{foreground}
		\pgfsetlayers{background,main,foreground}
		
		\node at (-1., 0) {\includegraphics[width=.75\linewidth]{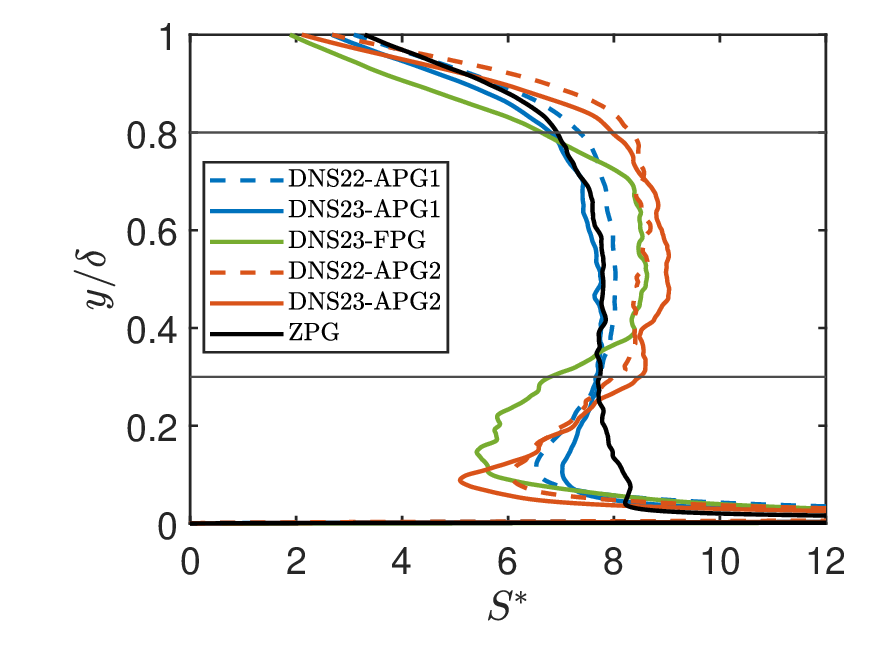}};
		\node[above left] at (-3.5, 1.85) {$(a)$};
		
		\node at (-1,-4.8) {\includegraphics[width=.75\linewidth]{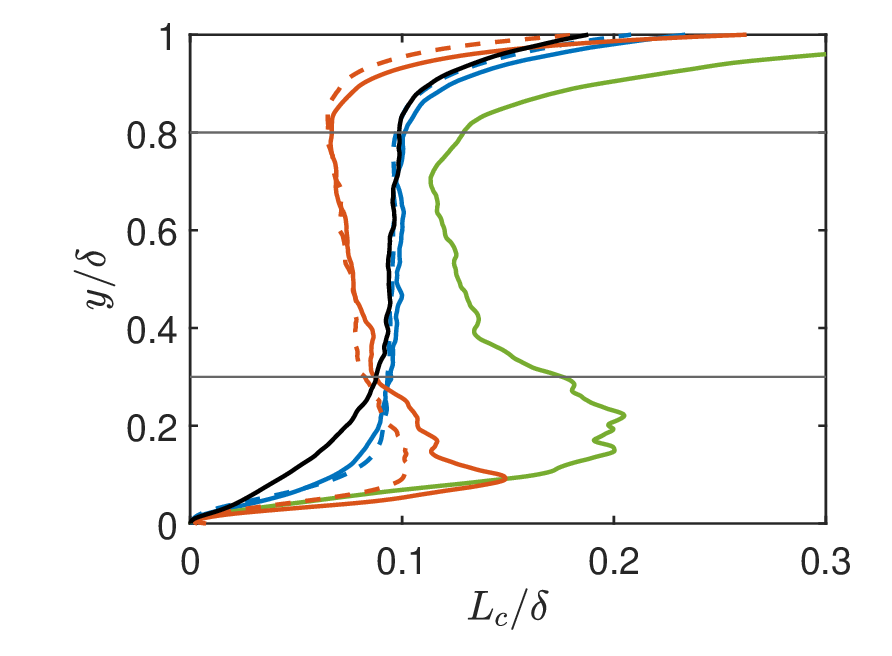}};
		\node[above left] at (-3.5, -2.95) {$(b)$}; 
	\end{tikzpicture}
	
	\caption{The Corrsin shear parameter (a) and Corrsin length scale (b) for the TBL cases as a function of $y / \delta$.}
	\label{fig::corrsin}
\end{figure}

\begin{figure*}[ht!]
	\centering
	\begin{tikzpicture}
		\node (top) at (-2, 4) {\includegraphics[width=\linewidth]{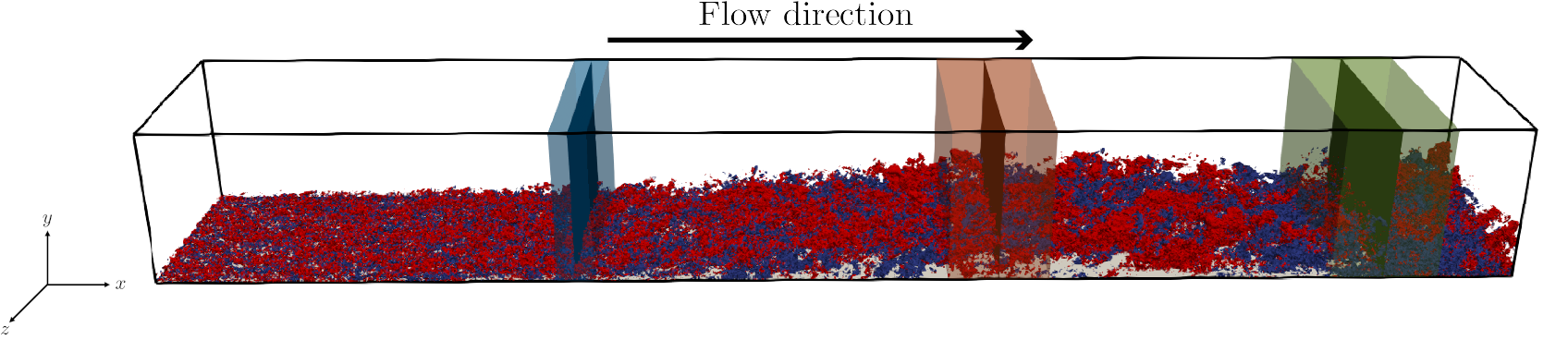}};
		
		\node (figax) at (-8, 1.8) {\includegraphics[width=.1\linewidth]{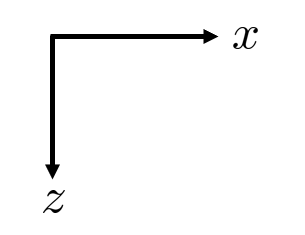}};  
		
		\node (figcol) at (-9, -1.8) [align=left] {
			\textcolor[rgb]{0.74,0,0}{Q2 in red} \\
			\textcolor[rgb]{0.1920,0.2470,0.5840}{Q4 in blue}
		};
		
		\node (fig1) at (-4.5, -.9) {\includegraphics[width=.21\linewidth]{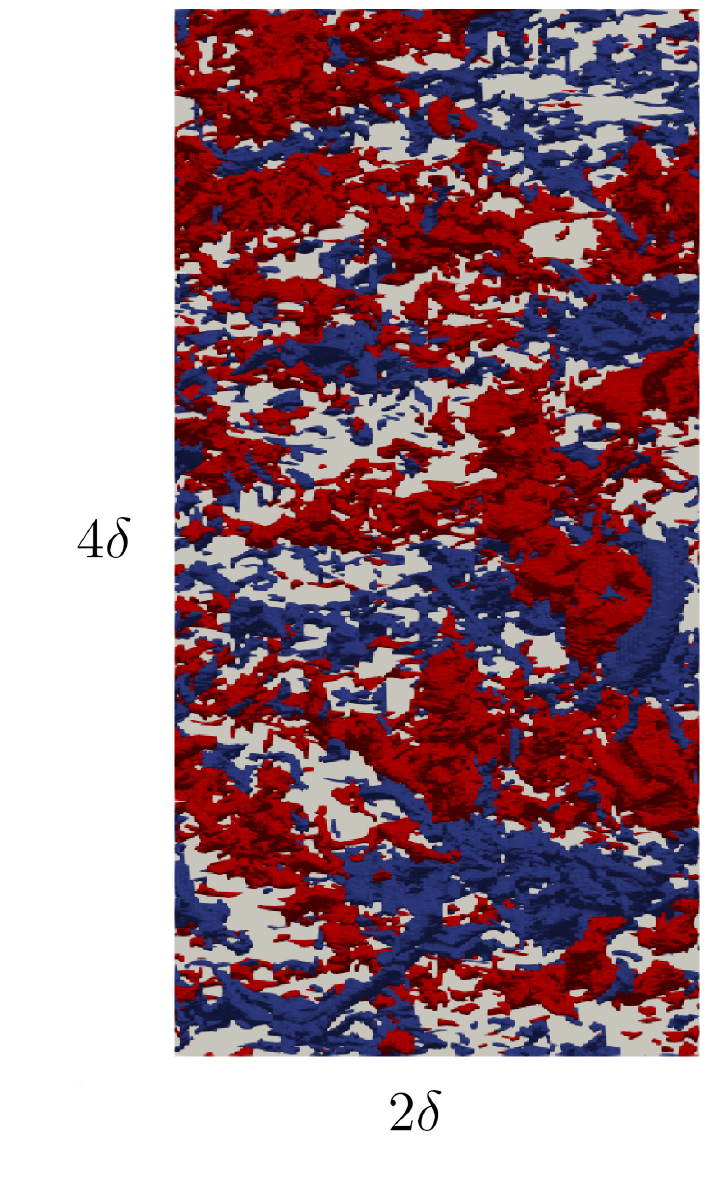}};  
		\node (fig2) at (0, -.9) {\includegraphics[width=.21\linewidth]{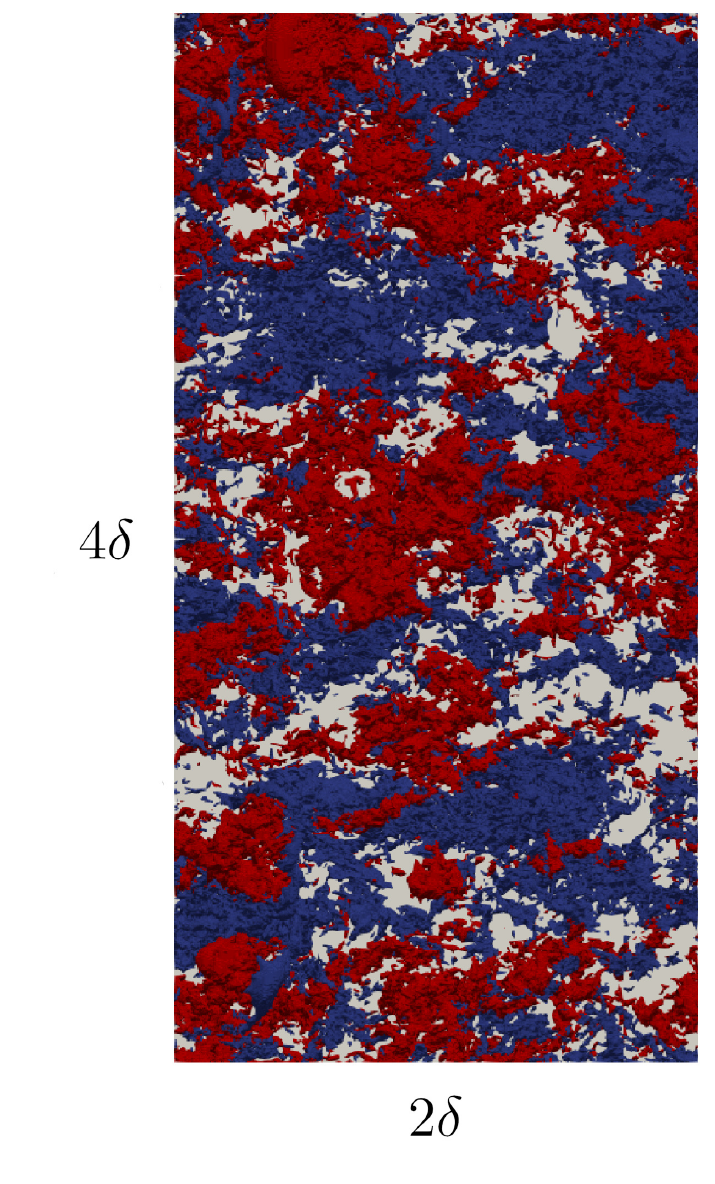}}; 
		\node (fig3) at (4.5, -.9) {\includegraphics[width=.21\linewidth]{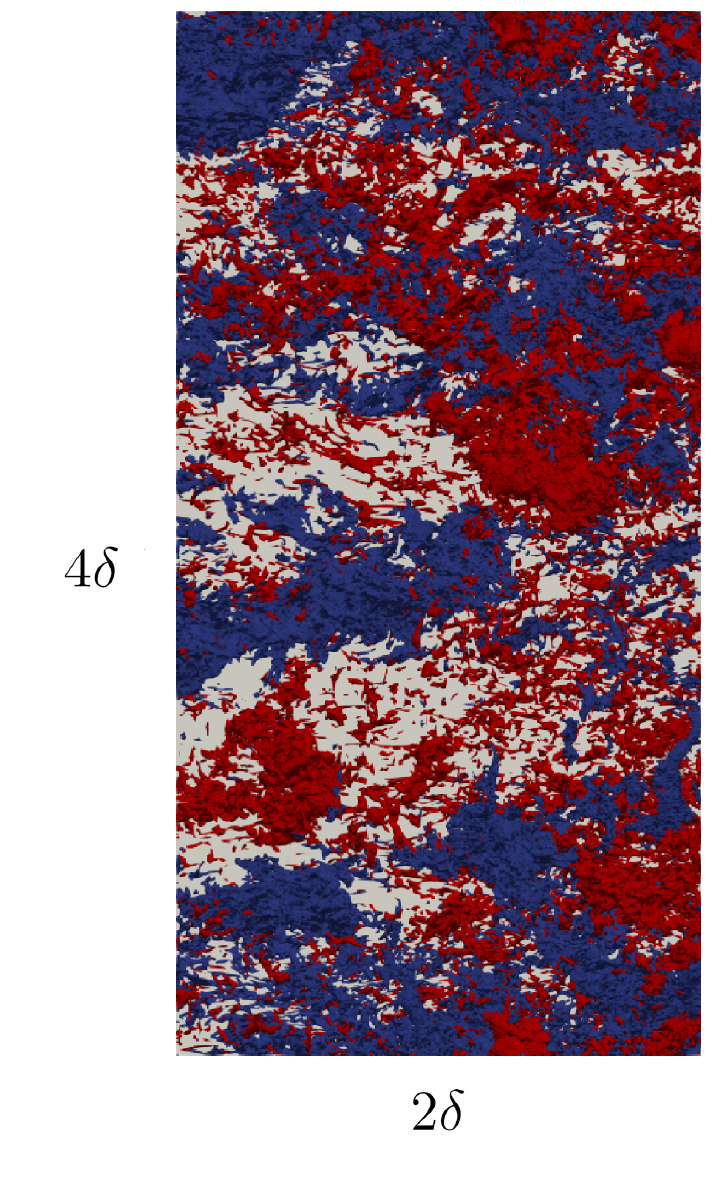}};

		\node[anchor=north west] at (top.north west) {\textbf{$(a)$}};
		\node[anchor=north west,xshift=-5cm,yshift=-0.2cm] at (fig1.north west) {\textbf{$(b)$}};
		
		\draw[thick, color={rgb,255:red,0; green,114; blue,189}] (-4.62,2.83) -- (-5.46,2.22);
		\draw[thick, color={rgb,255:red,0; green,114; blue,189}] (-4.2,2.84) -- (-2.68,2.22);
		
		\draw[thick, color={rgb,255:red,218; green,83; blue,25}] (-.05,2.85) -- (-1.0,2.22);
		\draw[thick, color={rgb,255:red,218; green,83; blue,25}] (1.15,2.86) -- (1.82,2.22);
		
		\draw[thick, color={rgb,255:red,119; green,172; blue,48}] (4.5,2.87)  -- (3.52,2.22);
		\draw[thick, color={rgb,255:red,119; green,172; blue,48}] (5.62,2.88)  -- (6.35,2.22);
	\end{tikzpicture}

	\caption{Spatial evolution of Q2 (red) and Q4 (blue) structures for DNS23. The flow direction is from left to right. In (a), the opaque colored planes indicate the three stations where spatio-temporal data were collected, while the translucent boxes represent the volumes used for the analysis with the fully-spatial data: blue (APG1), orange (APG2), and green (FPG). Figure (b) presents a top view of the three volumes, maintaining identical dimensions relative to $\delta$, while restricting the spanwise width to $4\delta$.}
	\label{fig:flowvis3d}
\end{figure*}
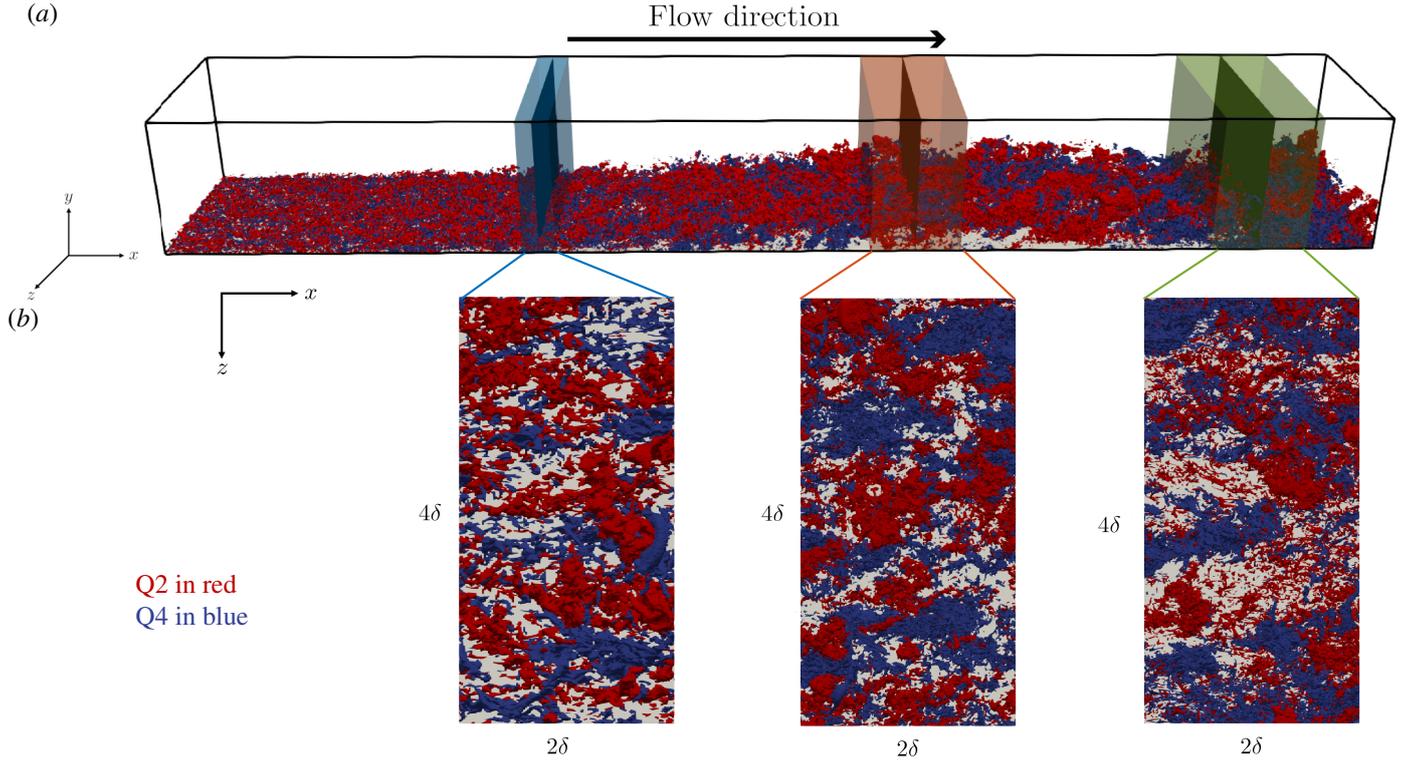

Fig.~\ref{fig2}c illustrates, as widely reported in the literature, that outer layer turbulence becomes dominant as the defect increases. In the small-defect APG cases (blue lines), the higher levels of $\langle u^{2}\rangle$ observed in DNS22 align with the larger $H$ and mean shear, while the contribution of flow history remains relatively small at these locations. Conversely, in the large-defect APG cases (orange lines), the larger mean shear and more gradual increase of the APG result in higher levels of $\langle u^{2}\rangle$ for DNS23. Regarding the $\langle u^{2}\rangle$ profile of the FPG case, although it has recovered the near-wall peak, its outer level is much higher than those of the small-defect APG TBLs and resembles those of the large-defect cases. This indicates again a delayed response of outer turbulence. An equilibrium FPG TBL with the same $\beta_{ZS}$ value would exhibit lower outer turbulence levels than the ZPG TBL. Thus, we have a set of flow cases that combine both similar and very different PG TBLs.

To understand the importance of the mean shear for the outer layer, we examine the Corrsin shear parameter $\left(S^{*}\right)$ in all cases, defined as $S^{*}=S q^{2} / \varepsilon$, where $S$ is the mean shear. It indicates the importance of the interaction between the shear and energy-carrying structures. When $S^{*}$ is much greater than one, these structures are primarily influenced by the local mean shear. Conversely, when $S^{*}$ is less than or approximately equal to one, turbulence becomes disconnected from the mean shear \citep{jimenez_nearwall}. Fig.~\ref{fig::corrsin}a presents $S^{*}$ as a function of $y / \delta$ for the TBL cases. The parameter $S^{*}$ remains fairly constant between $y / \delta=0.3$ and 0.8 for all cases except DNS23-FPG where it is constant between $y / \delta=0.4$ and 0.7. This suggests a similar interaction between the mean shear and the energy-carrying structures across all flow cases in the bulk of the outer region.

In association with the Corrsin shear parameter, we employ the Corrsin length scale $\left(L_{c}\right)$ to scale the coherent structures size, where $L_c = \left(\varepsilon / S^{3}\right)^{1 / 2}$. The Corrsin length scale is an intermediate length scale that represents the size of the smallest structures interacting with the mean shear. When $L_{c}<1$, turbulent structures are decoupled from mean shear and become isotropic in size. Fig.~\ref{fig::corrsin}b shows the Corrsin length scale $\left(L_{c}\right)$, normalized with $\delta$, and plotted as a function of $y / \delta$. One striking result is that the $L_c/\delta$ profiles exhibit very similar behavior for each defect case (ZPG/APG1 with a small velocity defect, APG2, and FPG) in the region $0.3 < y/\delta<0.8$. The smallest structures interacting with the mean shear are smaller relative to $\delta$ in the large-defect cases.

\section{Structure Identification Method}

We now describe the method used to identify structures responsible for carrying Reynolds shear stress. First, all $uv$ structures are classified into four groups based on their positions in the $u-v$ plane quadrants: outward interactions $(Q 1, u>0$ and $v>0)$, ejections $(Q 2, u<0$ and $v>0)$, inward interactions $(Q 3, u<0$ and $v<0)$ and sweeps $(Q 4, u>0$ and $v<0$). Although all Q structures are identified, the analysis primarily focuses on intense Q2 and Q4 structures, as they predominantly contribute to the Reynolds shear stress. To identify the intense Q structures, we employ the clustering technique where structures are defined as connected regions satisfying the following condition 
 \citep{lozano2012,maciel2017b}:

\begin{equation}
	|u(\mathbf{x}) v(\mathbf{x})|>H^{*} \sigma_{u} \sigma_{v} 
\end{equation}

\noindent
Here, $H^{*}$ is the threshold constant and $\sigma$ is the root-mean-square of the velocity fluctuation denoted by the index. Connectivity is defined using the six orthogonal neighbours in the DNS mesh. The value of $H^{*}$ was determined to be 1.75 through percolation analysis by \citet{lozano2012} for channel flows and by \citet{maciel2017b} for various APG TBLs. This threshold value was also shown to be valid for homogeneous shear turbulence by \citet{dong}. Because of its consistent value and for the sake of comparison, we also adopt $H^{*}=1.75$.

In this study, Q structures are identified using two types of three-dimensional velocity fields: fully spatial fields and spatio-temporal fields using Taylor's frozen  turbulence hypothesis. In the fully spatial case, the instantaneous three-dimensional spatial data is used to identify and analyze structures within a rectangular volume of streamwise length $2\delta$ at each streamwise position, using independent snapshots. For the fully-spatial data, structures with a volume smaller than $(3\Delta x)^3$ and structures with a wall-normal length smaller than $0.005\delta$ are discarded to avoid resolution issues. Additionally, structures with a maximum wall distance ($y_{max}$) greater than $1.2\delta$ are discarded. In the spatio-temporal case, time histories in $y-z$ planes are recorded and Taylor's frozen turbulence hypothesis is applied to each Q structure individually to convert time into the streamwise length $x$, utilizing the volume averaged instantaneous velocity as the convection velocity.

\begin{table*}[ht!]
	\centering
	\caption{Parameters of the attached/detached structures. $N_{i}$ and $V_{i}$ are the percentages of each type of $Q^-$ in terms of their number and volume for fully-spatial and spatio-temporal data, with parameters normalized by the total value for all attached and detached $Q^-$ structures.}
	\begin{tabular}{lrrrrrrrr}
		\hline \hline
		\multirow{2}{*}{Position} & \multicolumn{2}{c}{NQ2} & \multicolumn{2}{c}{NQ4} & \multicolumn{2}{c}{VQ2} & \multicolumn{2}{c}{VQ4} \\
		\cline{2-9}
		& Attached & Detached & Attached & Detached & Attached & Detached & Attached & Detached \\
		\hline
		\multicolumn{9}{l}{\textbf{Fully-spatial Data}} \\
		DNS23-APG1                & 22.7     & 30.7     & 19.8     & 26.8     & 49.4    & 16.1     & 23.4     & 11.1     \\
		DNS23-APG2                & 5.1      & 44.1     & 6.2      & 44.6     & 31.6    & 24.1     & 29.1     & 15.2     \\
		DNS23-FPG                 & 17.1     & 36.3     & 13.2     & 33.4     & 49.3    & 8.2     & 29.8     & 12.7     \\
		\hline
		\multicolumn{9}{l}{\textbf{Spatio-temporal Data}} \\
		DNS22-APG1                & 16.9     & 31.5     & 18.4     & 33.2     & 34.9     & 25.2     & 13.3     & 26.7     \\
		DNS22-APG2                & 2.9      & 42.2     & 4.5      & 50.4     & 31.2     & 29.6     & 15.9     & 23.3     \\
		DNS23-APG1                & 18.2     & 31.2     & 19.1     & 31.6     & 42.6     & 20.5     & 15.7     & 21.3     \\
		DNS23-APG2                & 2.0      & 42.6     & 3.4      & 52.0     & 26.7     & 31.2     & 19.9     & 22.3     \\
		DNS23-FPG                 & 13.2     & 35.0     & 14.3     & 37.5     & 51.3     & 11.7     & 23.5     & 13.6     \\
		\hline \hline
	\end{tabular}
	\label{tab2}
\end{table*}

\section{Analysis of Reynolds shear stress structures}

Fig.~\ref{fig:flowvis3d}a illustrates the spatial evolution of the ejections (red) and sweeps (blue) at a single time instant in the case of DNS23, using the fully-spatial data. As the boundary layer thickens in the streamwise direction, sweeps and ejections also increase in size. Fig.~\ref{fig:flowvis3d}b presents a top view of sweeps and ejections within the three selected volumes for spatial analysis. The plot dimensions are scaled by $\delta$, highlighting the relative size of the structures compared to the boundary layer thickness and enabling comparisons across streamwise positions. While differences in structures between flow cases are not immediately apparent in this visualization, they do exist and will be revealed through the subsequent statistical analysis. The most notable difference in Fig.~\ref{fig:flowvis3d}b is that the structures become more finely corrugated as the Reynolds number increases, due to the greater scale separation between large- and small-scale structures.

\begin{figure}[h!]
	\centering
	\begin{tikzpicture}
		\pgfdeclarelayer{background}
		\pgfdeclarelayer{foreground}
		\pgfsetlayers{background,main,foreground}
		
		\begin{pgfonlayer}{foreground}
			\node at (-1, 0) {\includegraphics[width=.75\linewidth]{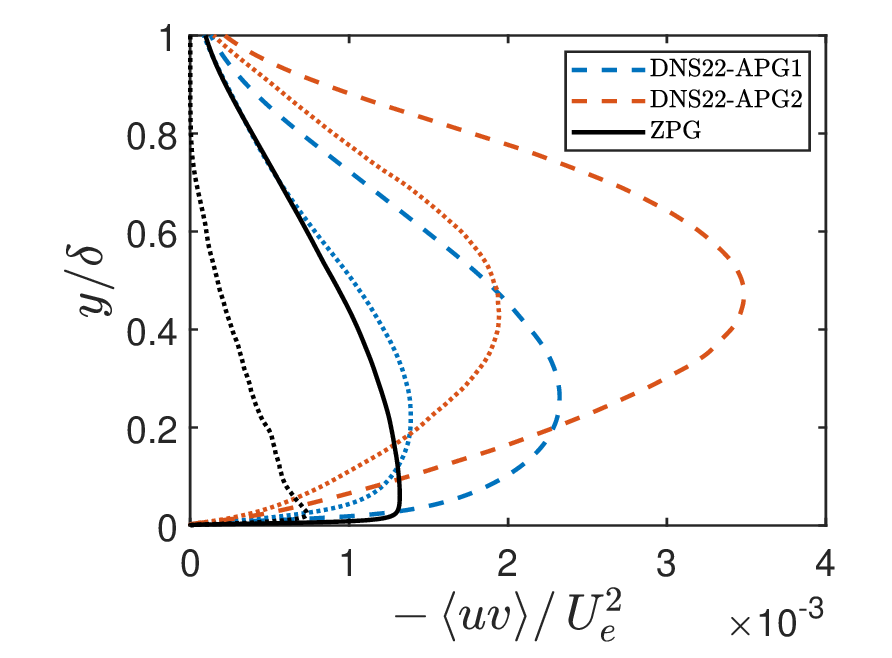}};
			\node[above left] at (-3.5, 1.75) {$(a)$}; 
		\end{pgfonlayer}
		
		\begin{pgfonlayer}{foreground}
			\node at (-1,-4.9) {\includegraphics[width=.75\linewidth]{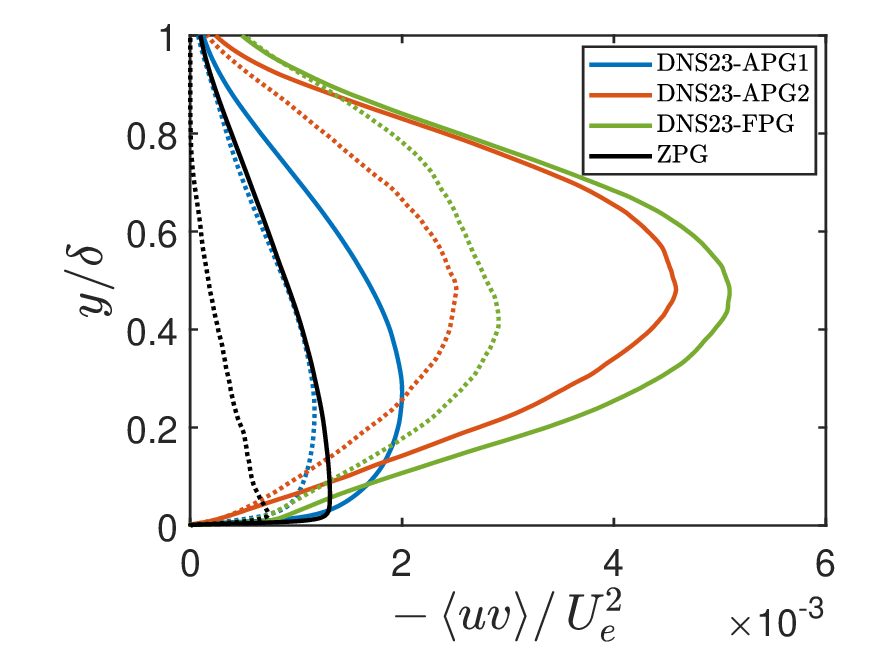}};
			\node[above left] at (-3.5, -3.1) {$(b)$}; 
		\end{pgfonlayer}
	\end{tikzpicture}
	
	\caption{Reynolds shear stress profiles compared with the Reynolds shear stress carried by the identified intense $Q^{-}$ events (dotted lines) for DNS22 (a) and DNS23 (b) obtained using spatio-temporal data along with the ZPG TBL results of \citet{maciel2017a}. }
	\label{fig4}
\end{figure}

Fig.~\ref{fig4} compares the amount of Reynolds shear stress carried by the identified $Q^{-}$ structures, where $Q^{-}$stands for combined intense Q2s and Q4s, to the total Reynolds shear stress as a function of $y / \delta$ for all cases along with the ZPG TBL results of \citet{maciel2017a}. With the chosen extraction threshold value $\left(H^{*}=1.75\right)$, the identified $Q^{-}$ structures carry approximately half of the total amount of Reynolds shear stress in all cases.

We distinguish between wall-attached and wall-detached Q structures due to their distinct properties and dynamic significance. In canonical wall flows, wall-attached Q structures are larger and predominantly carry the Reynolds shear stress in the overlap layer \citep{lozano2012,maciel2017a,maciel2017b}. However, in large-defect APG TBLs, detached Q structures become more numerous and contribute equally to the Reynolds shear stress as attached ones \citep{maciel2017a}. As in~\citet{maciel2017a,maciel2017b}, attached structures are defined as those whose bottom boundary is located at a minimum distance of less than $0.05\delta$ from the wall, while the remaining structures are considered detached. The joint probability density functions of the minimum and maximum wall distances of the Qs (not shown) confirm that this boundary at $y=0.05 \delta$ adequately separates both types of structures in all flow cases.

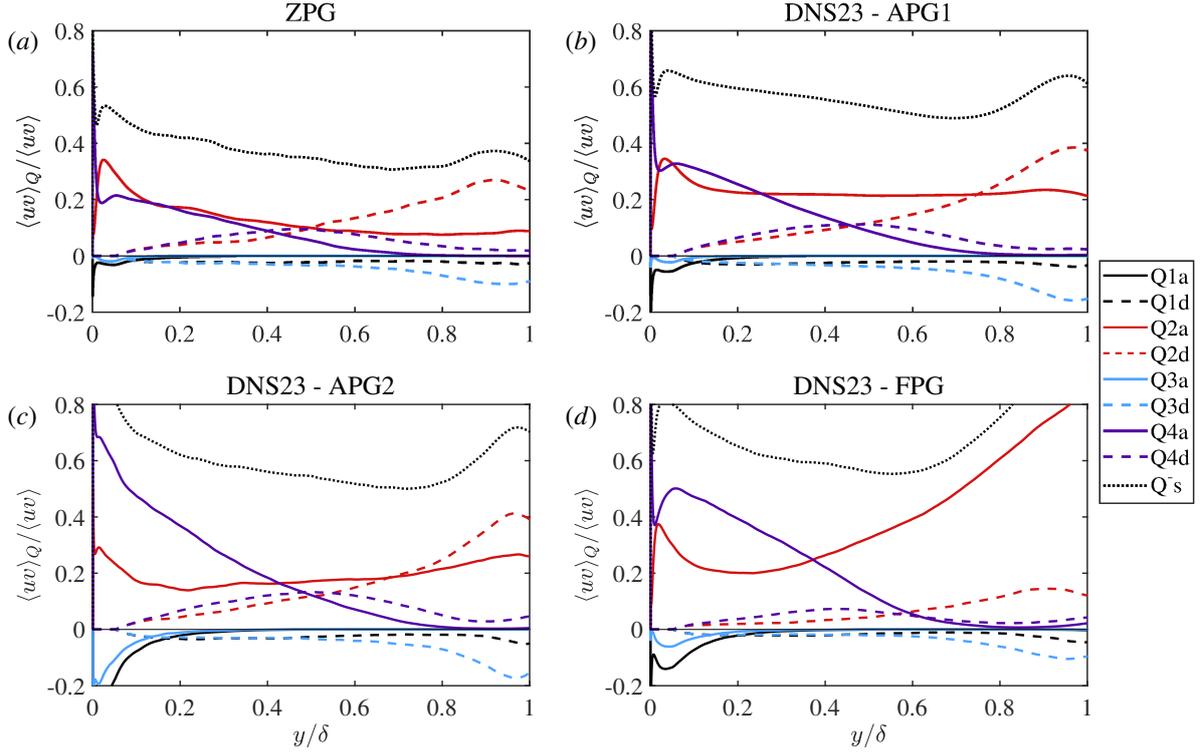
\begin{figure*}[htbp!]
	\centering
	\begin{tikzpicture}
		\node[anchor=south west] (a) at (0.05\linewidth, 0) {\includegraphics[width=0.37\linewidth]{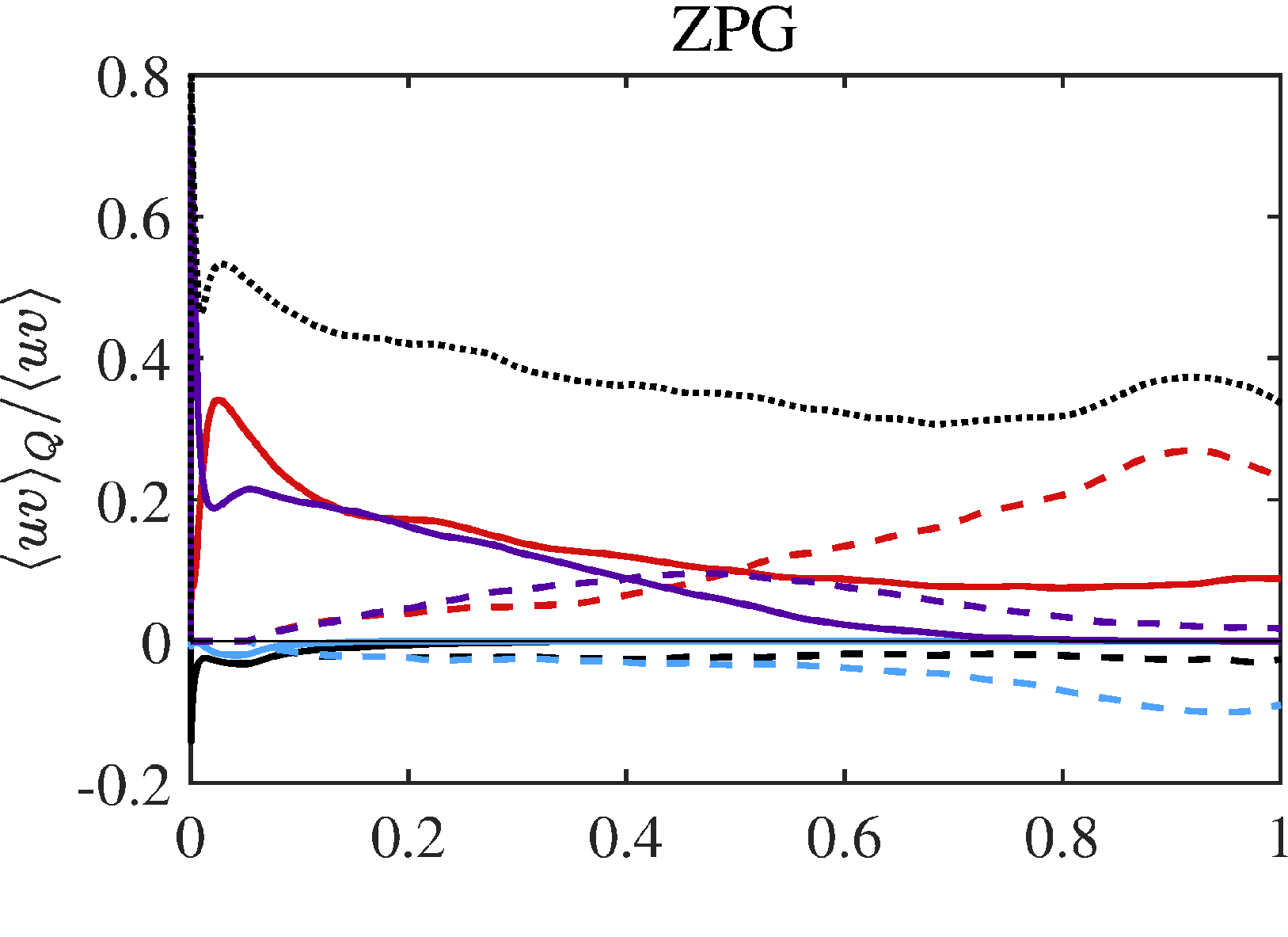}};
		\node[anchor=north west, inner sep=5pt, xshift=-5pt, yshift=-10pt] at (a.north west) { $(a)$};
		
		\node[anchor=south west] (b) at (0.45\linewidth, 0) {\includegraphics[width=0.37\linewidth]{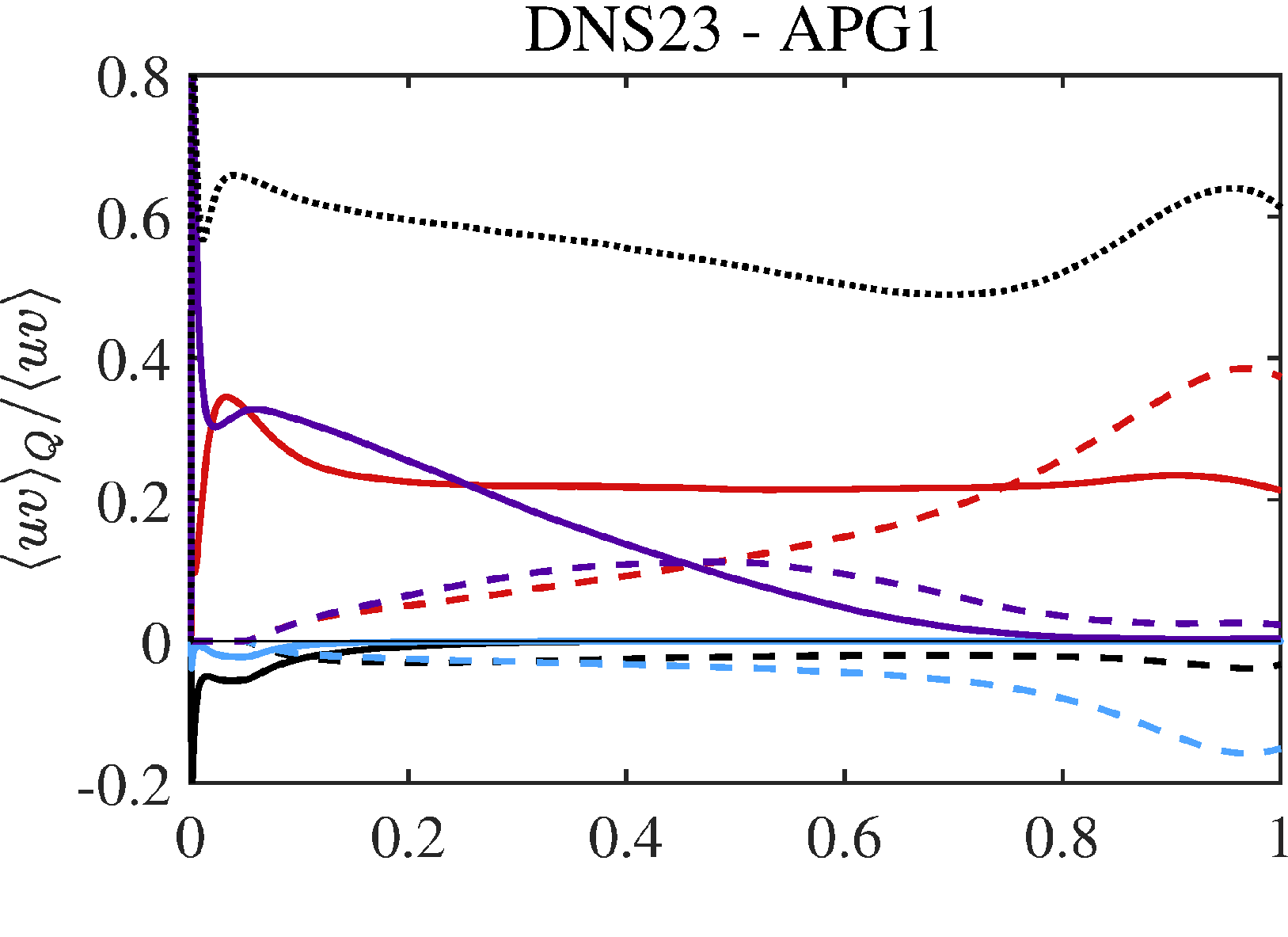}};
		\node[anchor=north west, inner sep=5pt, xshift=-5pt, yshift=-10pt] at (b.north west) { $(b)$};
		
		\node[anchor=south west] (c) at (0.05\linewidth, -0.27\linewidth) {\includegraphics[width=0.37\linewidth]{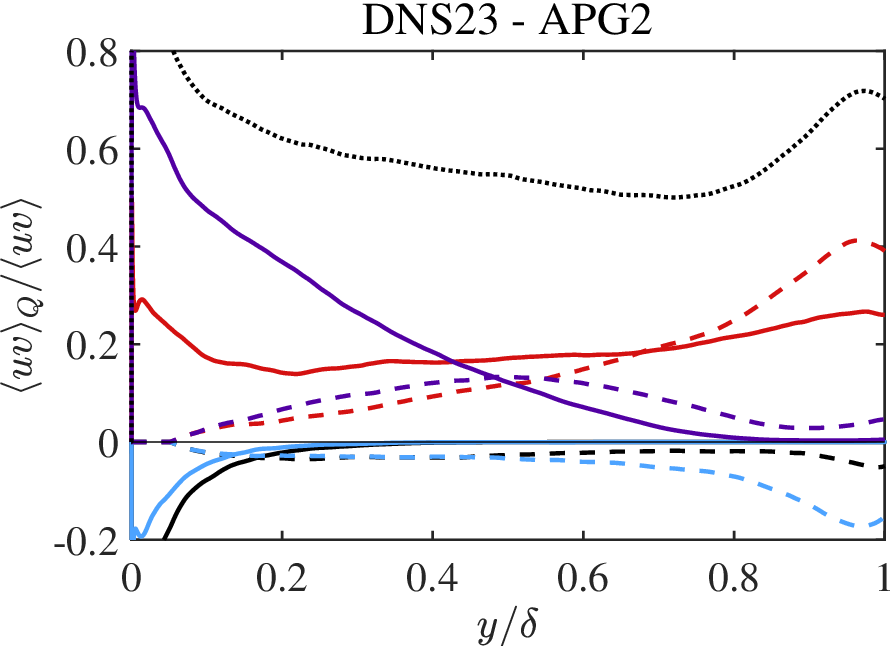}};
		\node[anchor=north west, inner sep=5pt, xshift=-5pt, yshift=-10pt] at (c.north west) { $(c)$};
		
		\node[anchor=south west] (d) at (0.45\linewidth, -0.27\linewidth) {\includegraphics[width=0.37\linewidth]{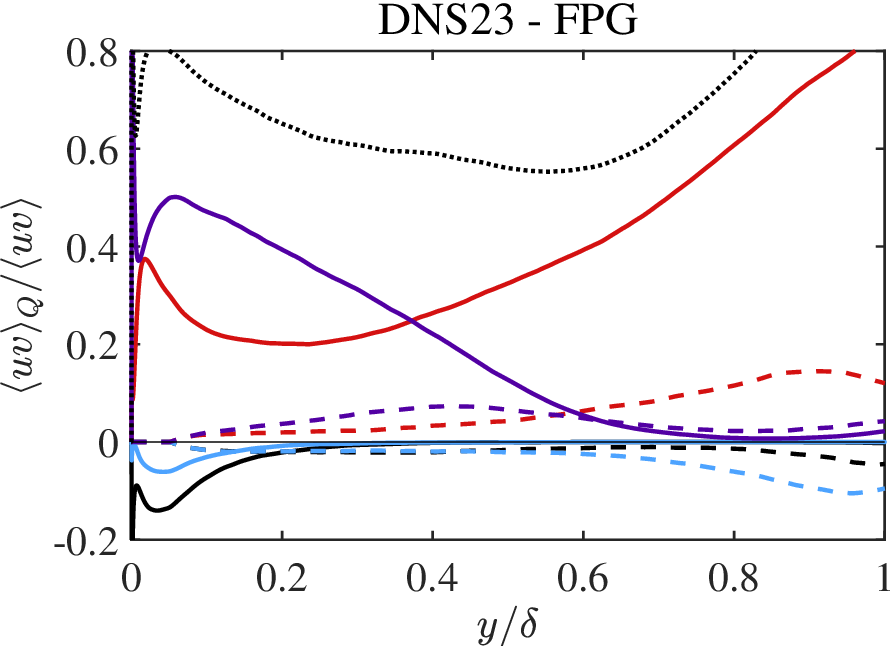}};
		\node[anchor=north west, inner sep=5pt, xshift=-5pt, yshift=-10pt] at (d.north west) { $(d)$};
		
		\node[anchor=west] at (0.825\linewidth, 0*\linewidth) {\includegraphics[width=0.07\linewidth]{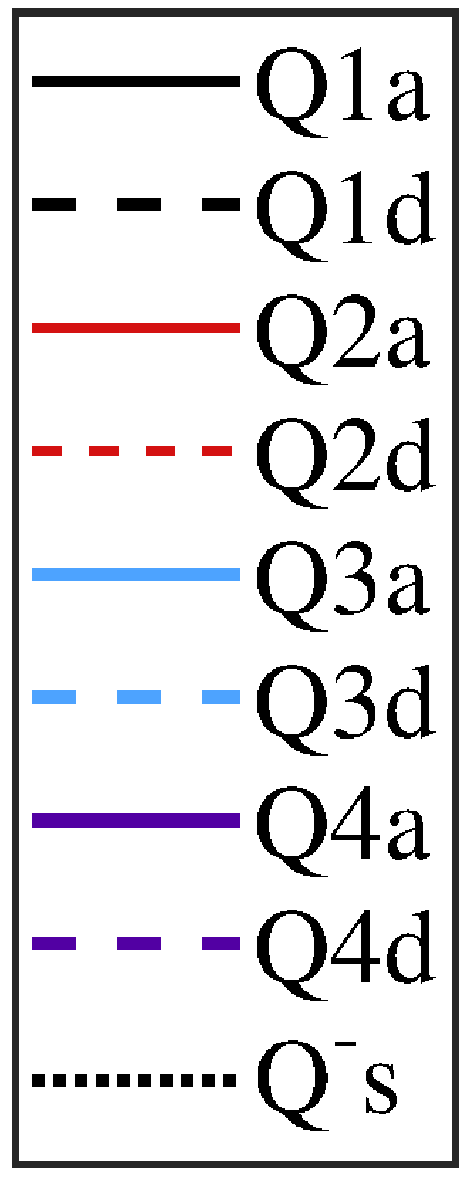}};
	\end{tikzpicture}
	\caption{Fractional contributions to $\langle u v\rangle$ from attached Qs (solid), and detached Qs (dashed) for ZPG (a), APG1 (b), APG2 (c), and FPG (d) of DNS23 for spatio-temporal data. The dotted line is the total fractional contribution of all $Q^-$s (Q2s and Q4s).}
	\label{fig::qcontrib}
\end{figure*}

Table ~\ref{tab2} shows the percentage of attached and detached $Q^{-} \mathrm{s}$ in terms of their number and volume for the present flow cases. The volume $V$, defined as the volume of connected points within a structure, and the number of structures are normalized by the total volume and total number of all attached and detached $Q^-$ structures, respectively, for both fully-spatial and spatio-temporal data. In all flow cases, detached structures are more numerous than attached ones, with the difference being particularly pronounced in the large defect cases (APG2). However, despite their greater number, detached structures generally occupy a smaller volume than attached structures. In the small-defect APG TBLs (APG1), attached Q2s occupy nearly twice the volume of detached ones, similar to observations in canonical wall flows \citep{maciel2017a}. In the large-defect APG TBLs (APG2), focusing on the more reliable fully-spatial data, the volume difference becomes less pronounced, although attached $Q^-$ structures still remain larger on average. In the FPG TBL, attached $Q^{-} \mathrm{s}$ dominate the total volume ($\sim80\%$), far exceeding detached structures ($\sim20\%$). This dominance may be related to the slight downward advection of the structures, which increases the space occupied by attached structures and potentially their number. This aligns with observations by \citet{joshi2014}, who found that an FPG reduces outward migration of near-wall turbulence.

\begin{figure*}[h!]
	\centering
	\begin{tikzpicture}
		
		\node (top) at (0, 0) {\includegraphics[width=0.8\linewidth]{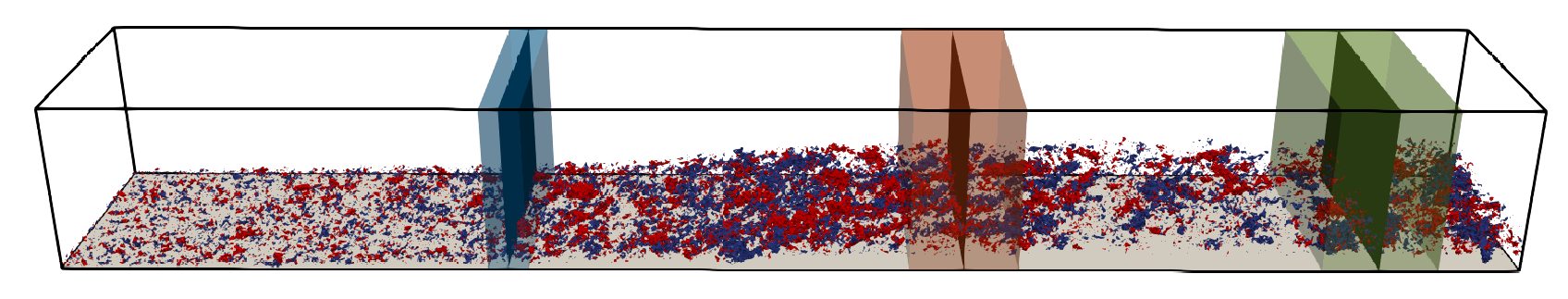}};
		
		\node (figax) at (-8, -1.) {\includegraphics[width=.1\linewidth]{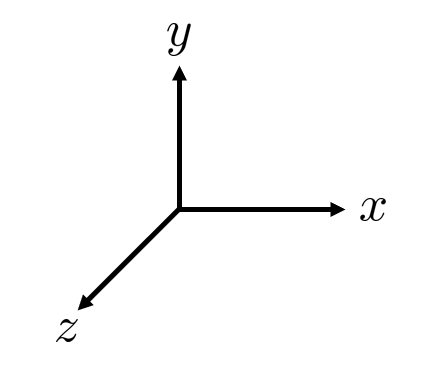}};  
		
		\node (figax) at (-8, -2.8) {\includegraphics[width=.1\linewidth]{./figures/xzaxes.png}};  
		\node (figax) at (-8, -10.3) {\includegraphics[width=.1\linewidth]{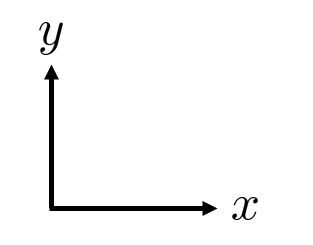}}; 
		
		\node (figcol) at (-9, -5.8) [align=left] {
			\textcolor[rgb]{0.74,0,0}{Q2 in red} \\
			\textcolor[rgb]{0.1920,0.2470,0.5840}{Q4 in blue}
		};

		\node (fig1) at (-5, -5.5) {\includegraphics[width=0.21\linewidth]{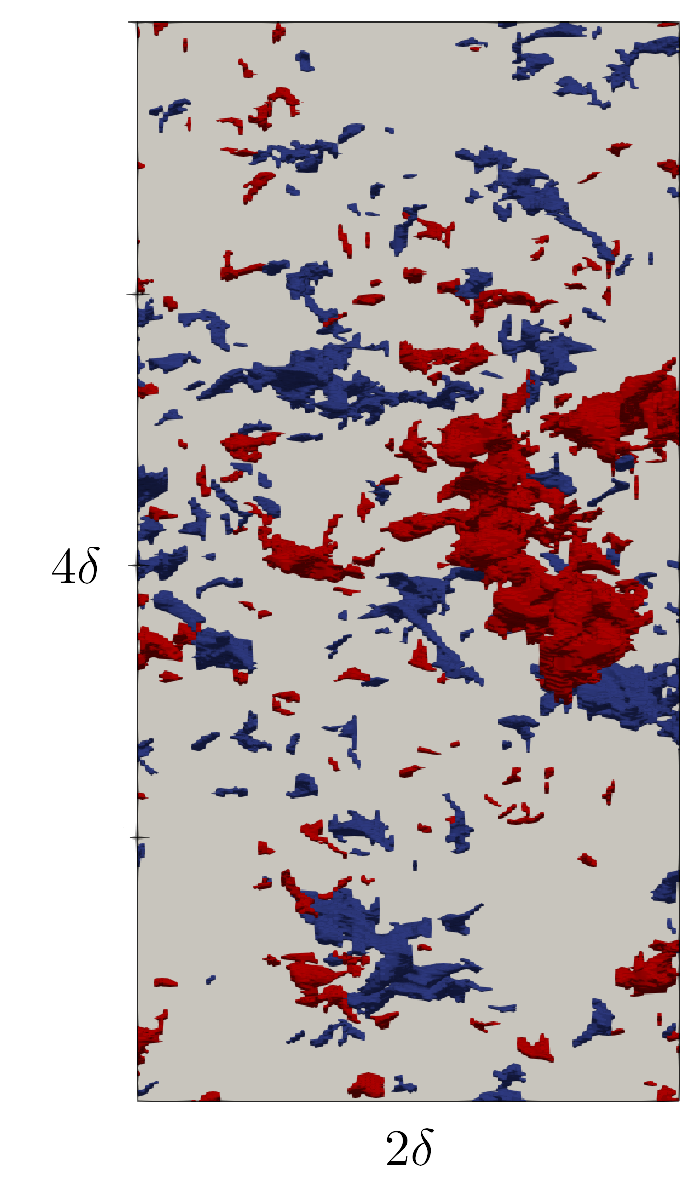}};
		\node (fig2) at (0, -5.5) {\includegraphics[width=0.21\linewidth]{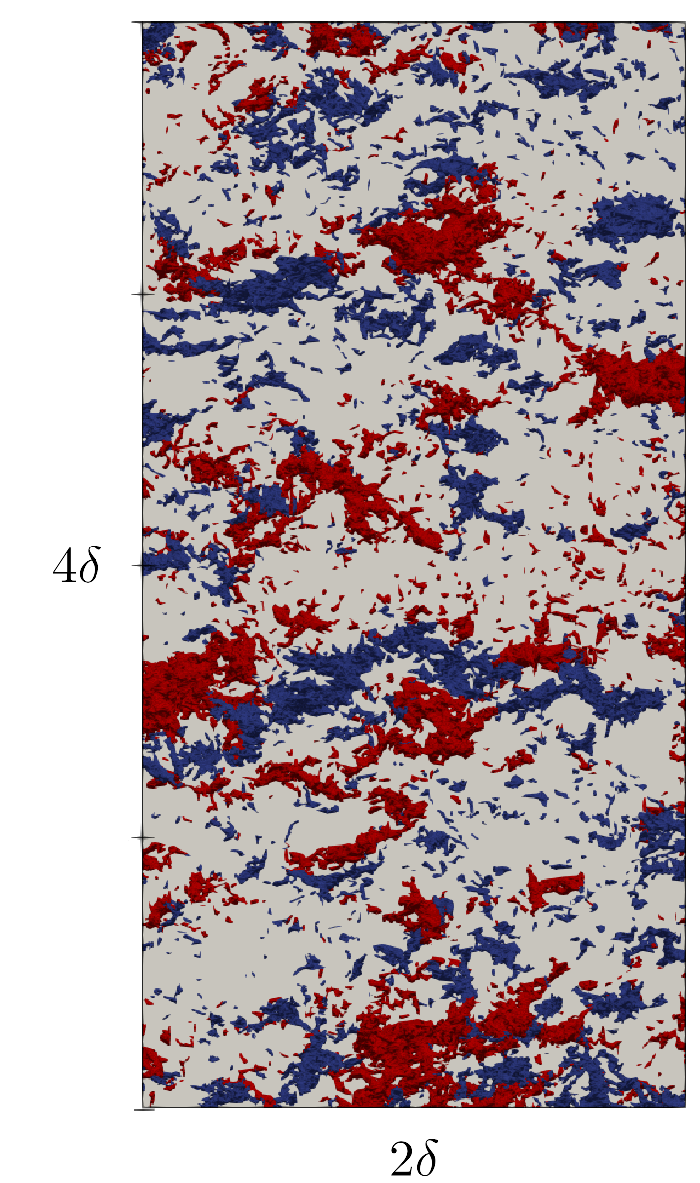}};
		\node (fig3) at (5, -5.5) {\includegraphics[width=0.21\linewidth]{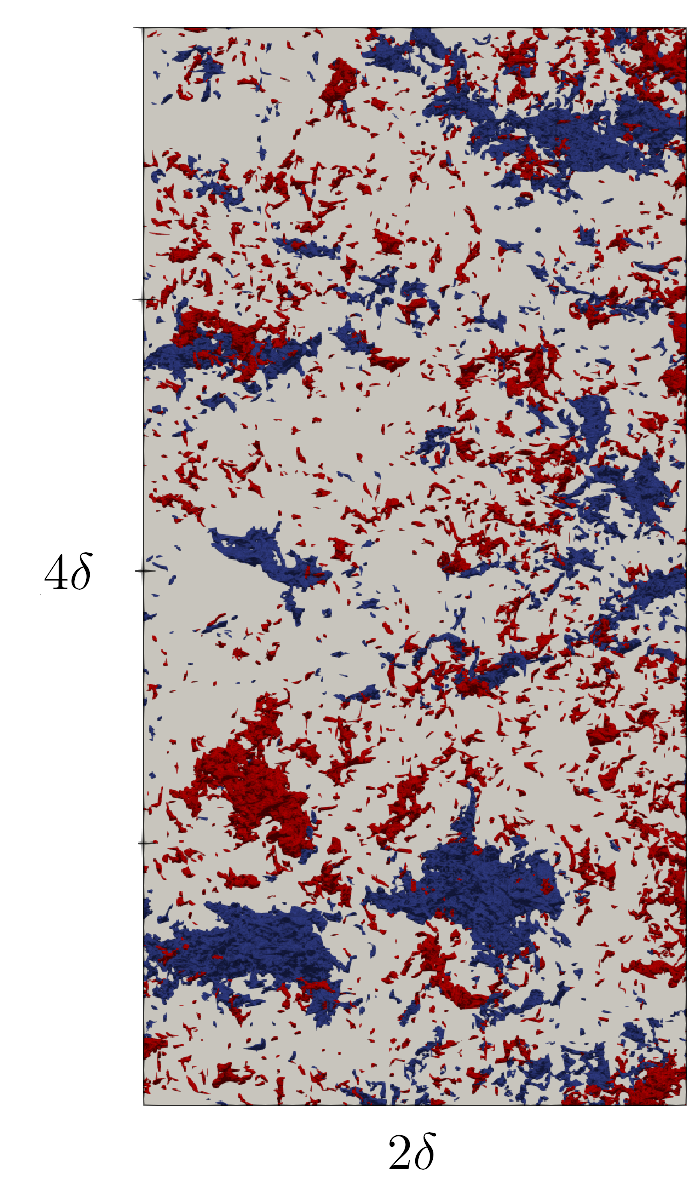}};

		\node (fig4) at (-5, -10) {\includegraphics[width=0.21\linewidth]{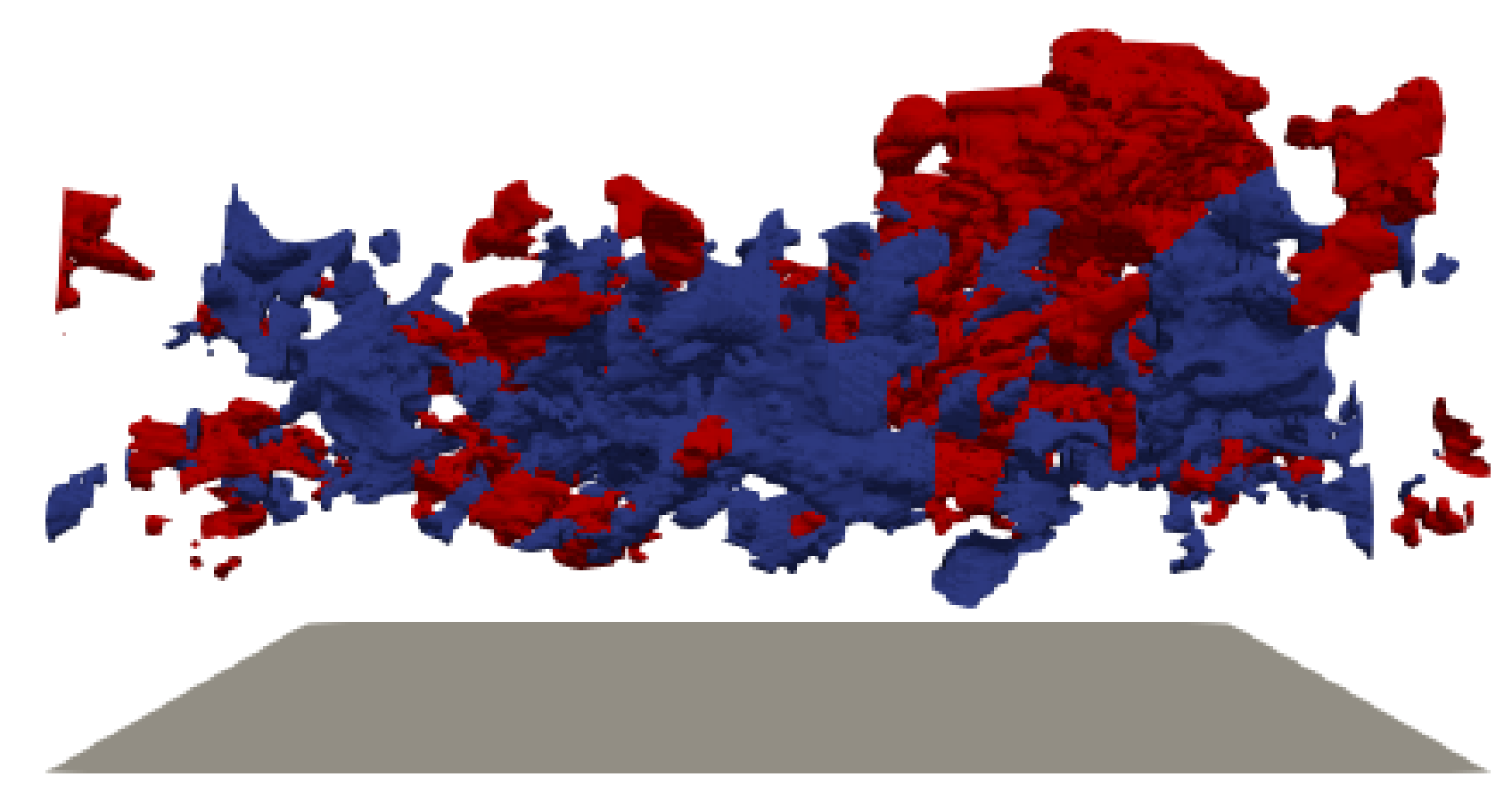}};
		\node (fig5) at (0, -10) {\includegraphics[width=0.21\linewidth]{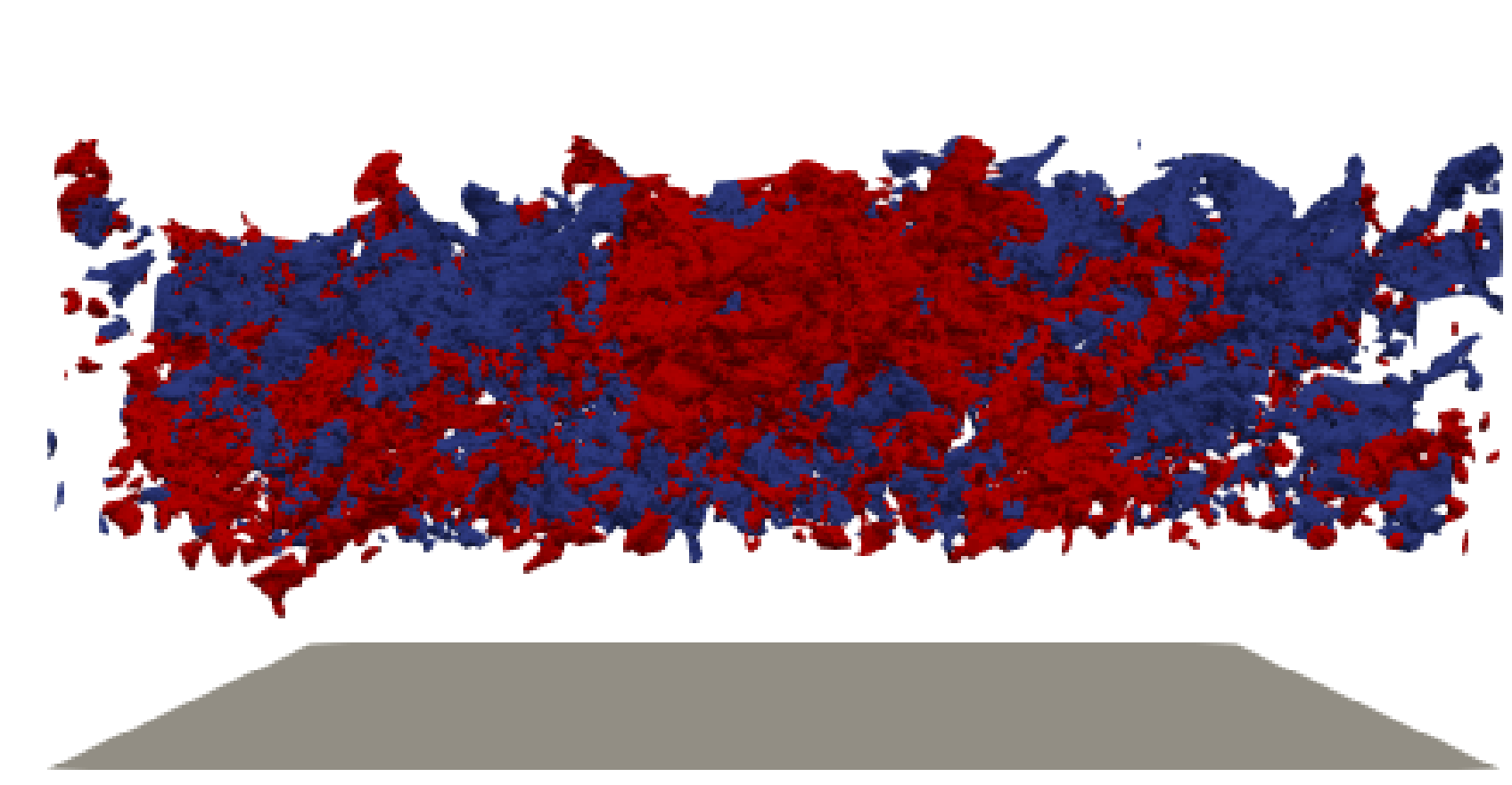}};
		\node (fig6) at (5, -10) {\includegraphics[width=0.21\linewidth]{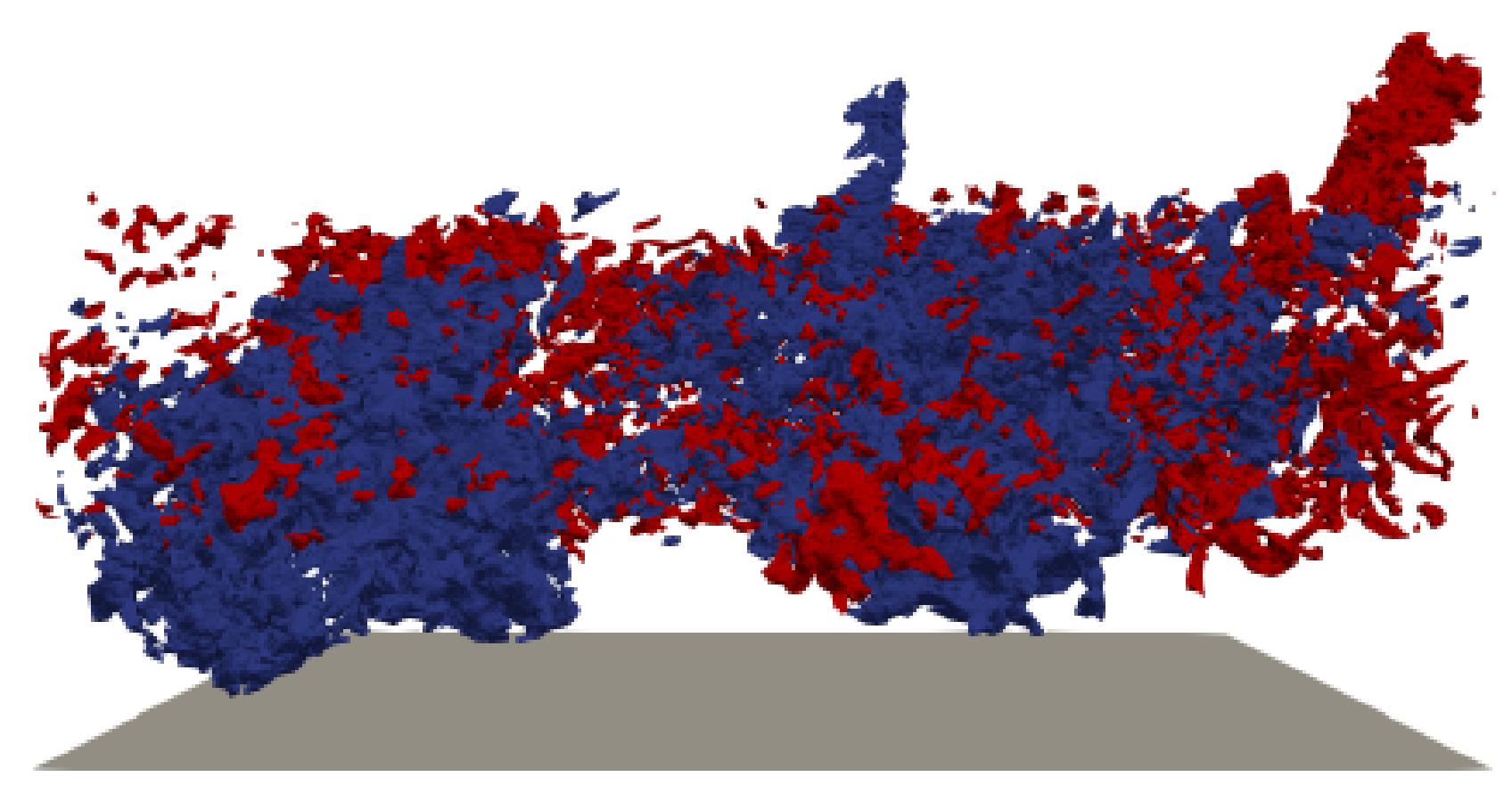}};
		
		\node[anchor=north west, xshift=-2.5cm, yshift=-5pt] at (top.north west) {\textbf{$(a)$}};
		\node[anchor=north west, xshift=-3cm, yshift=-5pt] at (fig1.north west) {\textbf{$(b)$}};
		\node[anchor=north west, xshift=-3cm, yshift=-5pt] at (fig4.north west) {\textbf{$(c)$}};

	\end{tikzpicture}
	
	\caption{Spatial evolution of detached structures with their centers, $y_c$, located within the range $0.3\delta < y_c < 0.8\delta$, for Q2 (red) and Q4 (blue) structures in DNS23. The flow direction is from left to right. In (a), the opaque colored planes indicate the three stations where spatio-temporal data were collected, while the translucent boxes represent the volumes used for the analysis with the fully-spatial data: blue (APG1), orange (APG2), and green (FPG). (b) presents a top view of the three volumes, maintaining identical dimensions relative to $\delta$, while restricting the spanwise width to $4\delta$. (c) presents the side view of the three volumes.}

	\label{fig:flowvis3ddetached}
\end{figure*}

\subsection{Reynolds Shear Stress Contributions}

Having focused on Q2 and Q4 structures, we now investigate the contributions of all Q types, including Q1 and Q3 structures, to the Reynolds shear stress. Fig.~\ref{fig::qcontrib} illustrates the Reynolds shear stress contributions from intense $Q$ structures separated both in quadrant type and attached/detached type, as a function of $y / \delta$ for DNS23 and the ZPG TBL case of \citet{sillero2013}. It is important to stress that the contributions are normalized with the local value of $\langle u v\rangle$ and therefore it may give the impression that the Reynolds shear stress increases towards the boundary layer edge, which is not the case as shown in Fig.~\ref{fig4}.

The results shows that the behavior of intense Qs exhibit similarities in all cases. In the inner region of the boundary layer, attached structures, particularly Q2 and Q4 motions, dominate the momentum transfer, highlighting their critical role near the wall. Conversely, in the outer region, the contribution from detached structures becomes increasingly significant. Moreover, the contribution of Q2 structures increases with $y$ in the outer region, while that of Q4 structures decreases.

Despite these similarities, the pressure gradient alters the structures significantly. In the lower part of the boundary layer, the contribution of $Q^{+} \mathrm{s}$ $(\mathrm{Q} 1 \mathrm{s}$ and $\mathrm{Q} 3 \mathrm{s}$) becomes more relevant for the large-defect case APG2. Their intensity matches that of Q2 structures in APG2, whereas it remains small in APG1. In the near-wall region of canonical wall flows, it is well-established that sweeps dominate below the inner peak of $\langle u^2 \rangle$, while ejections dominate above it \citep{KimMoinMoser1987,moser1999,jimenez2010}.
This behavior is evident in Fig.~\ref{fig::qcontrib}a for the ZPG case, but also in Fig.~\ref{fig::qcontrib}b for APG1. Notably, in APG1 and APG2, this behavior also occurs near the outer maximum of $\langle u^2 \rangle$, at approximately $y = 0.25\delta$ for APG1 and $0.5\delta$ for APG2, suggesting a similar turbulence mechanism in the outer region as in the inner region. These findings are consistent with those reported by \citet{maciel2017a} and \citet{deshpande2024}. This phenomenon persists in the FPG case, where the outer peak of $\langle u^2 \rangle$ remains pronounced at $y \approx 0.45\delta$ due to the flow's history. One significant difference between the FPG case and the other three cases is that attached structures dominate the flow even in the upper part of the boundary layer. This may be partially attributed to the downward ($v<0$) advection of structures in the FPG region.

\subsection{Size and shape of the structures}

\begin{figure*}[h!]
	\centering
	\begin{tikzpicture}

		\def\gridwidth{0.25\linewidth} 
		\def\gridheight{0.25\linewidth} 
		\def\hspacing{-0.7cm} 
		\def\vspacing{-4.0cm} 
		\def\figspacing{0.5cm} 

		\pgfdeclarelayer{background}
		\pgfdeclarelayer{middle}
		\pgfdeclarelayer{foreground}
		\pgfdeclarelayer{title}
		\pgfsetlayers{background,middle,foreground,title}

		\begin{pgfonlayer}{title}
			\node[anchor=north] at (2.1, 4.35) {\includegraphics[width=.08\linewidth]{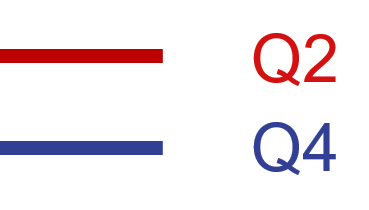}};
		\end{pgfonlayer}
		
		\begin{pgfonlayer}{foreground}
			\node[anchor=south west] at (0, 0) {\includegraphics[width=\gridwidth, height=\gridheight]{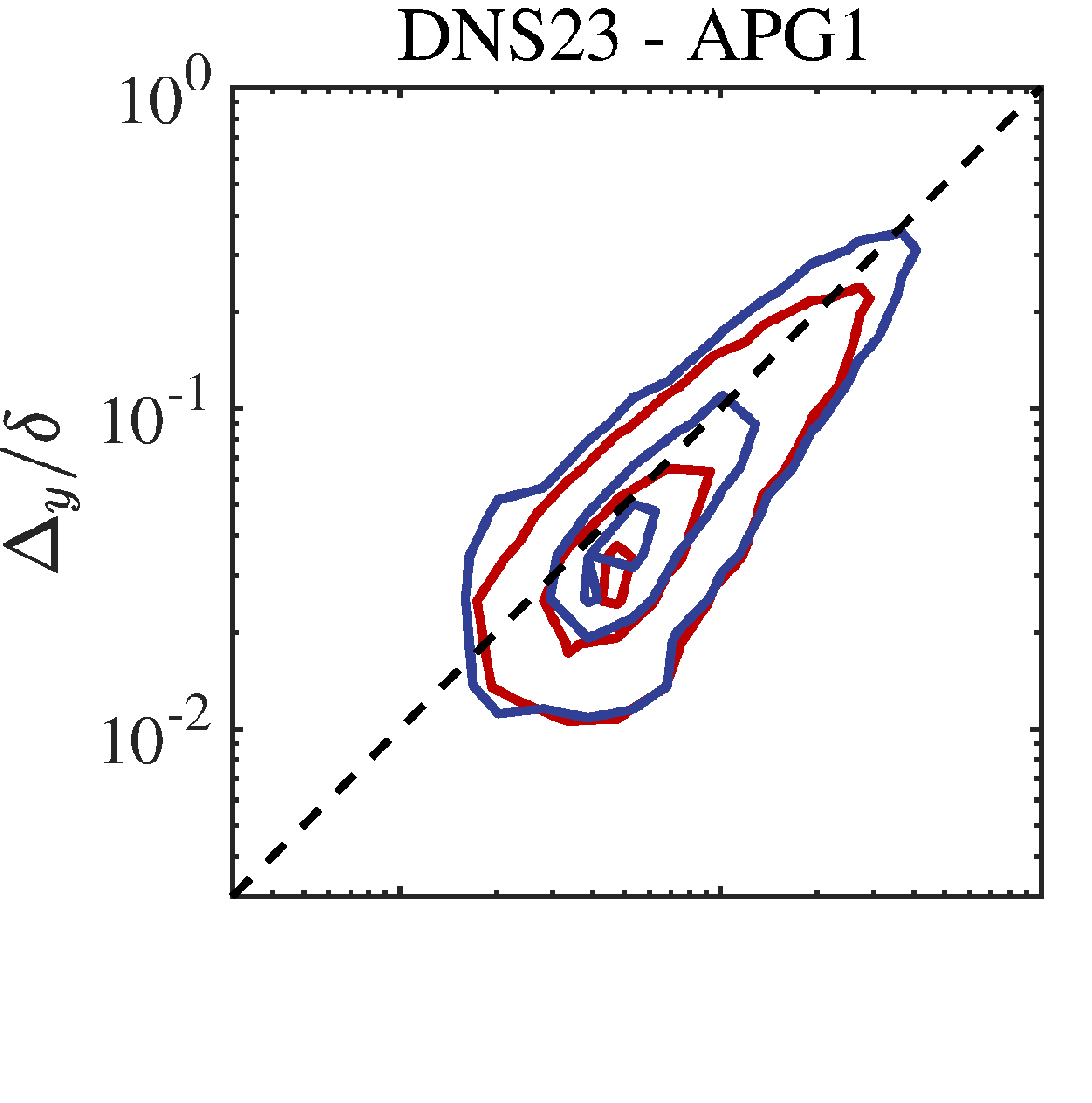}};
		\end{pgfonlayer}
		\begin{pgfonlayer}{middle}
			\node[anchor=south west] at (\gridwidth + \hspacing, 0) {\includegraphics[width=\gridwidth, height=\gridheight]{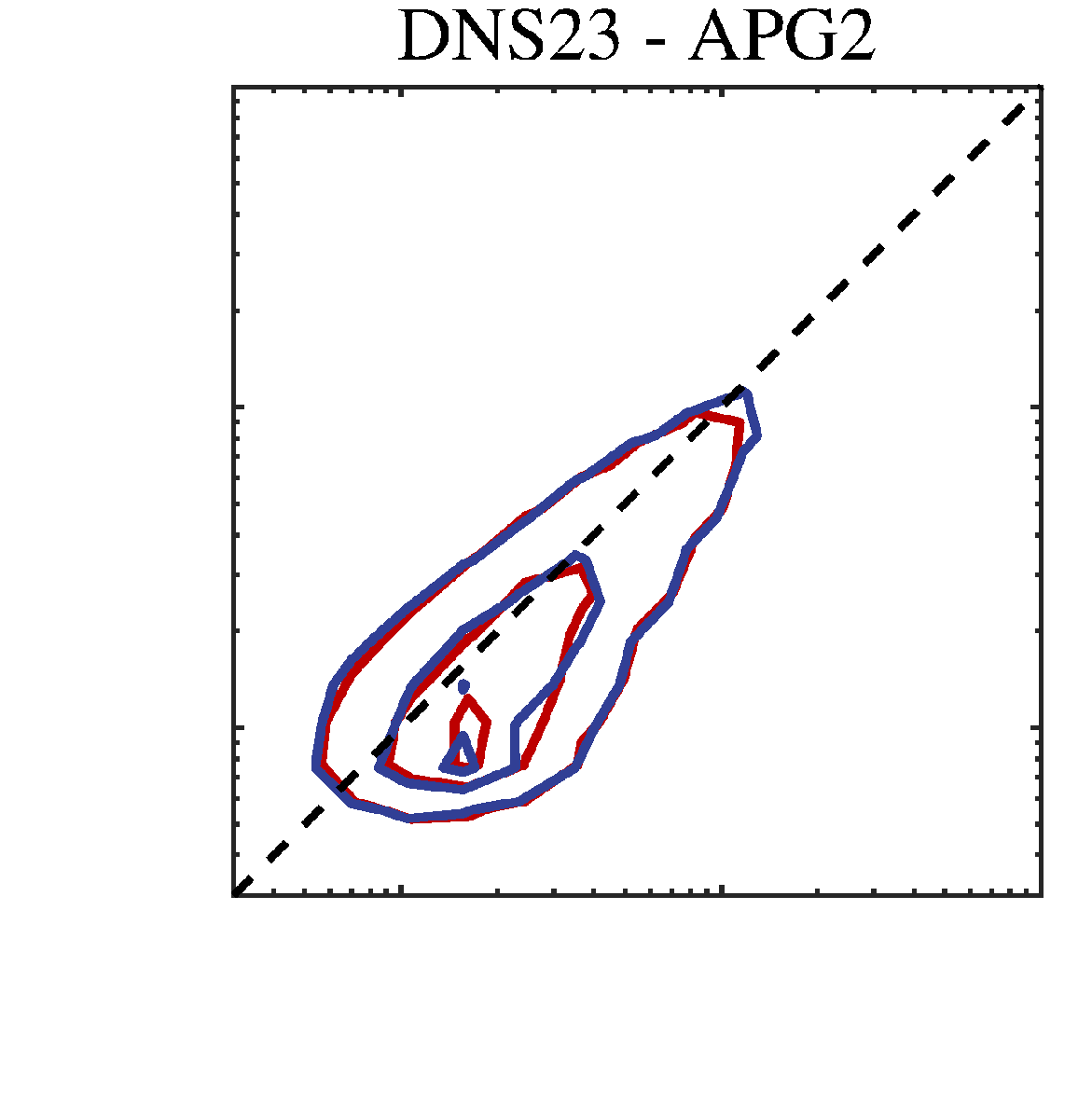}};
		\end{pgfonlayer}
		\begin{pgfonlayer}{background}
			\node[anchor=south west] at (2 * \gridwidth + 2 * \hspacing, 0) {\includegraphics[width=\gridwidth, height=\gridheight]{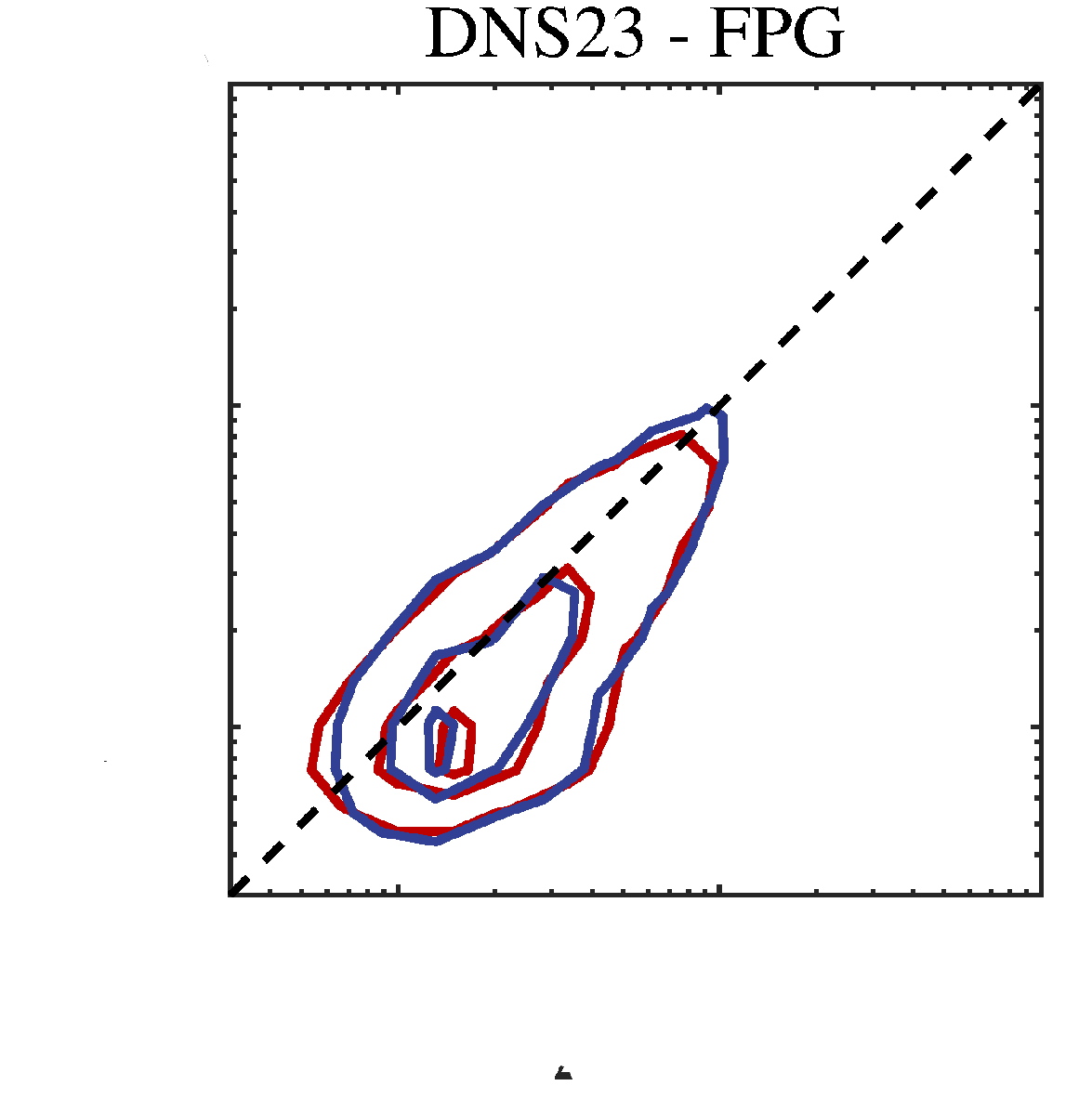}};
		\end{pgfonlayer}

		\begin{pgfonlayer}{foreground}
			\node[anchor=east] at (-0.1, \gridheight/2) {{Fully-spatial}};
		\end{pgfonlayer}

		\begin{pgfonlayer}{foreground}
			\node[anchor=south west] at (0, \vspacing) {\includegraphics[width=\gridwidth, height=\gridheight]{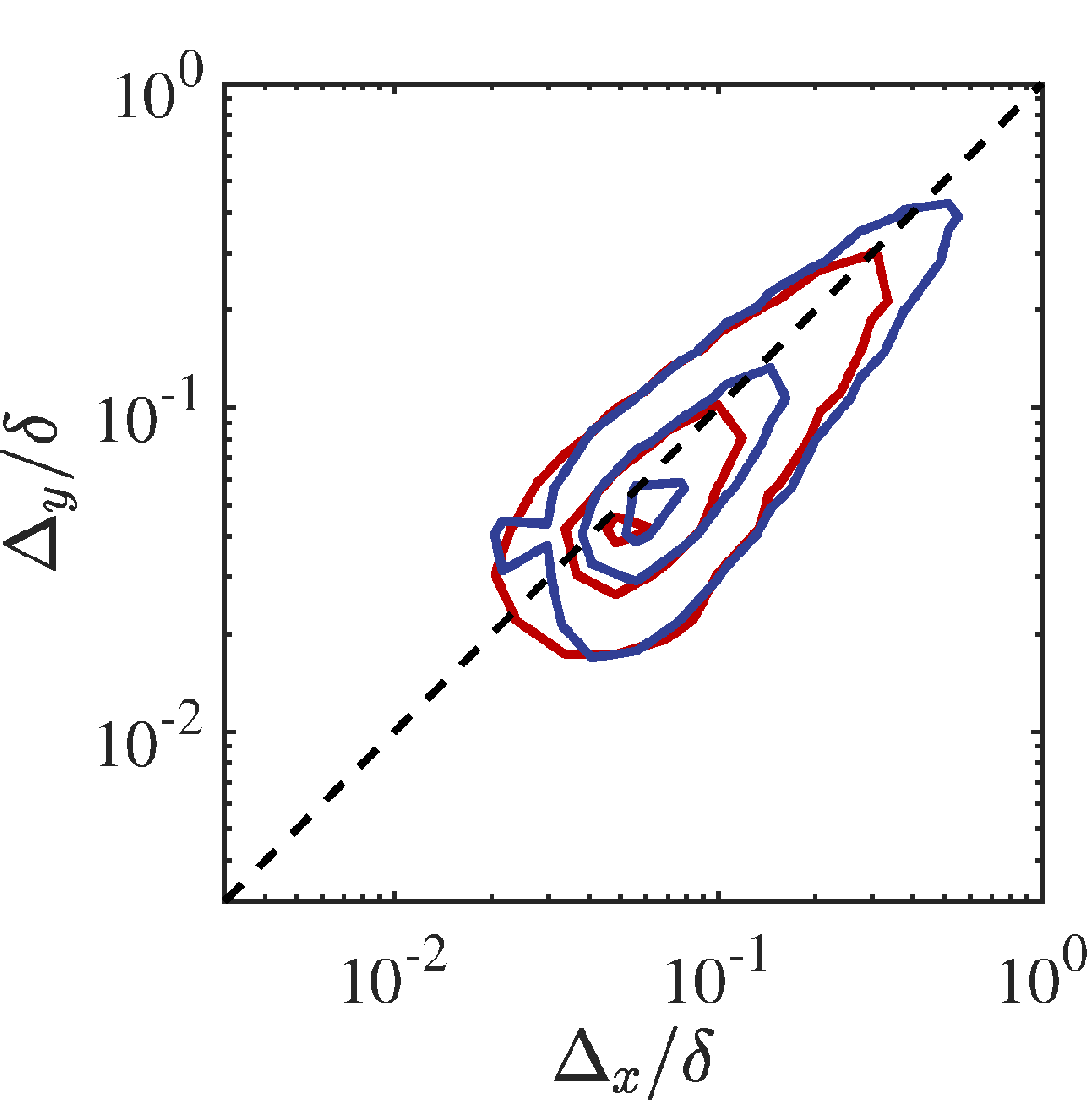}};
		\end{pgfonlayer}
		\begin{pgfonlayer}{middle}
			\node[anchor=south west] at (\gridwidth + \hspacing, \vspacing) {\includegraphics[width=\gridwidth, height=\gridheight]{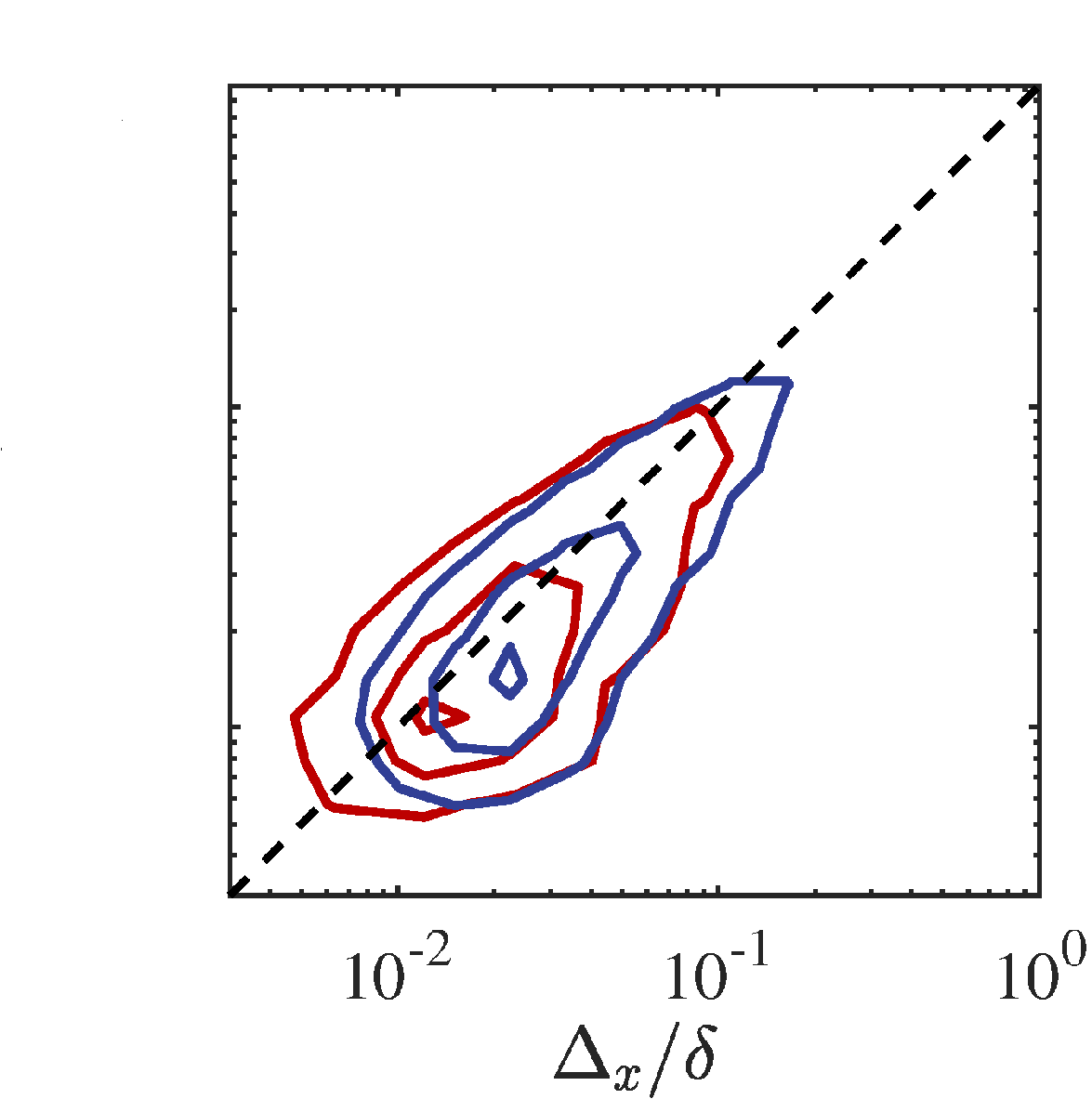}};
		\end{pgfonlayer}
		\begin{pgfonlayer}{background}
			\node[anchor=south west] at (2 * \gridwidth + 2 * \hspacing, \vspacing) {\includegraphics[width=\gridwidth, height=\gridheight]{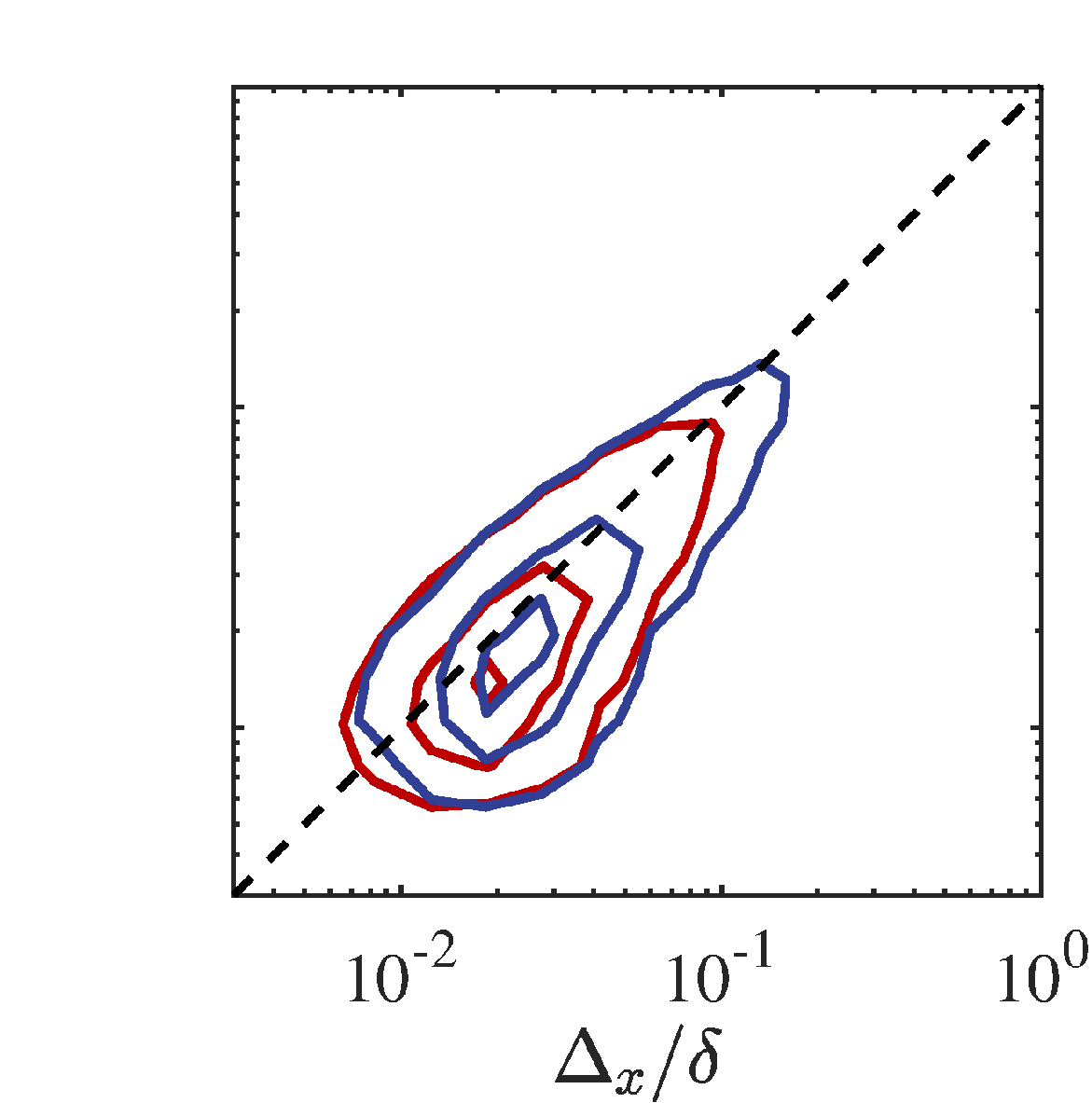}};
		\end{pgfonlayer}

		\begin{pgfonlayer}{foreground}
			\node[anchor=east] at (-0.1, \vspacing + \gridheight/2) {{Spatio-temporal}};
		\end{pgfonlayer}

		\begin{pgfonlayer}{foreground}
			\node[anchor=south west] at (-0.5, \gridheight) {$(a)$};
		\end{pgfonlayer}

		\begin{pgfonlayer}{foreground}
			\node[anchor=south west] at (0, 2.35 * \vspacing) {\includegraphics[width=\gridwidth, height=\gridheight]{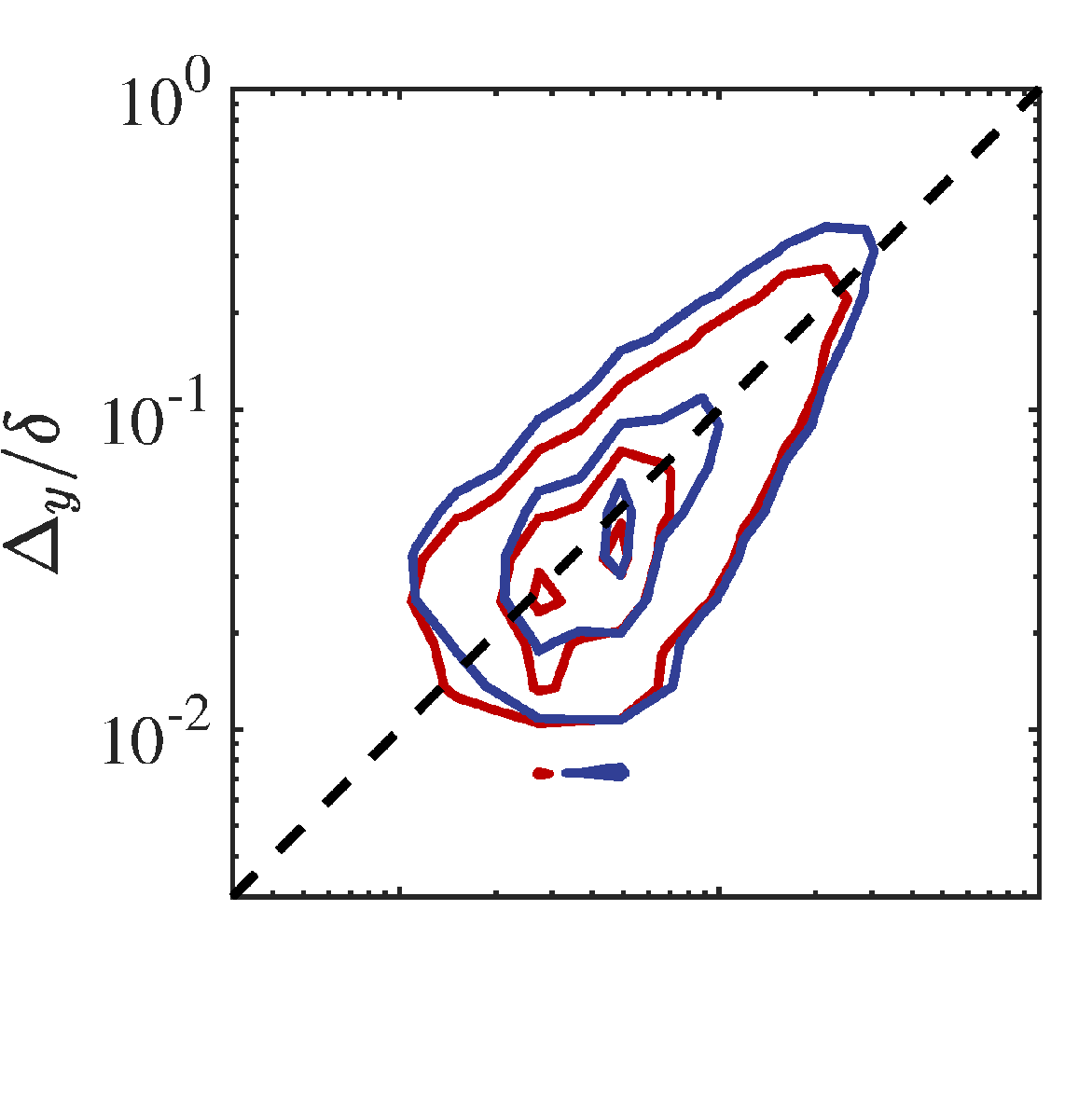}};
		\end{pgfonlayer}
		\begin{pgfonlayer}{middle}
			\node[anchor=south west] at (\gridwidth + \hspacing, 2.35 * \vspacing) {\includegraphics[width=\gridwidth, height=\gridheight]{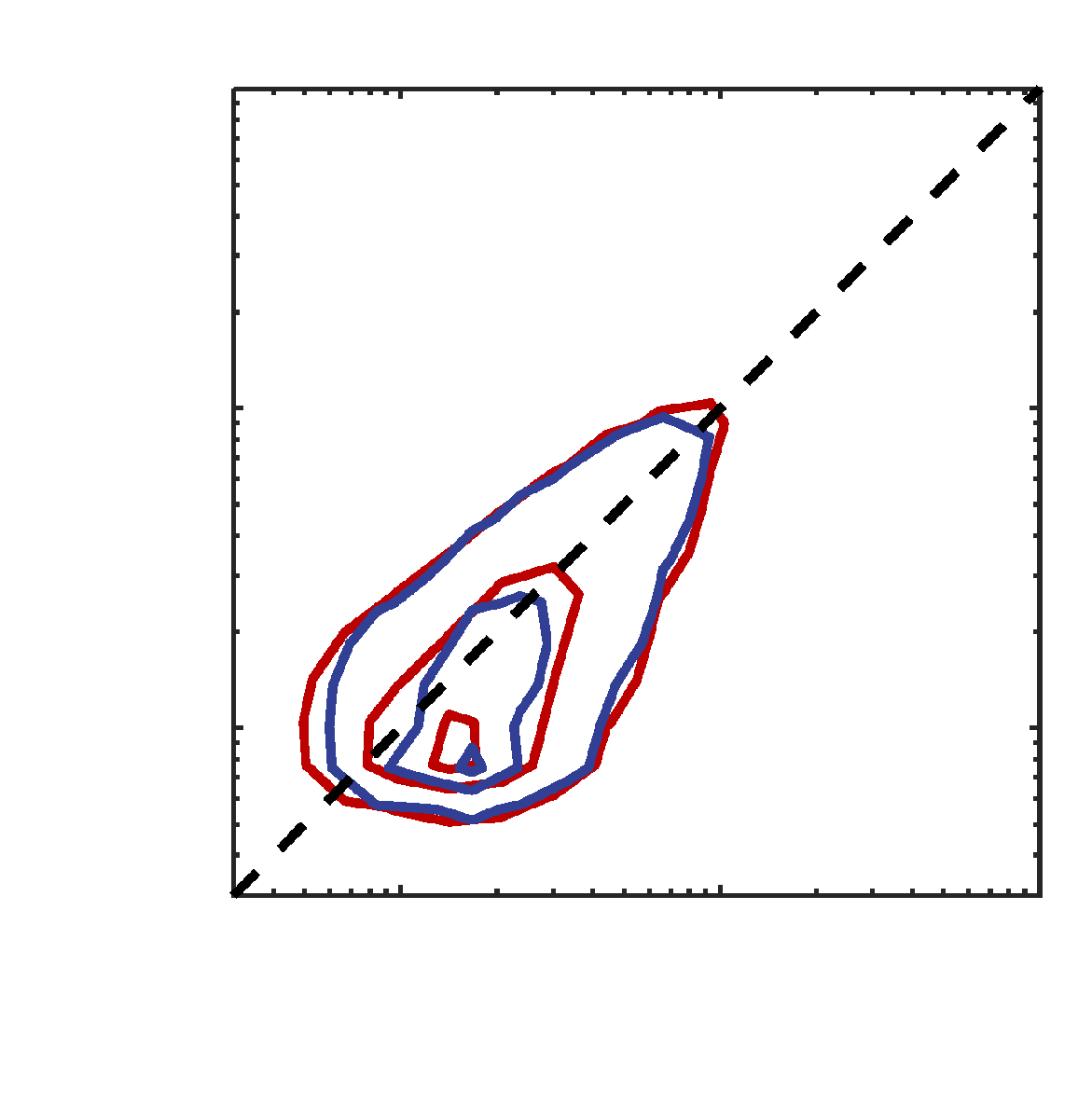}};
		\end{pgfonlayer}
		\begin{pgfonlayer}{background}
			\node[anchor=south west] at (2 * \gridwidth + 2 * \hspacing, 2.35 * \vspacing) {\includegraphics[width=\gridwidth, height=\gridheight]{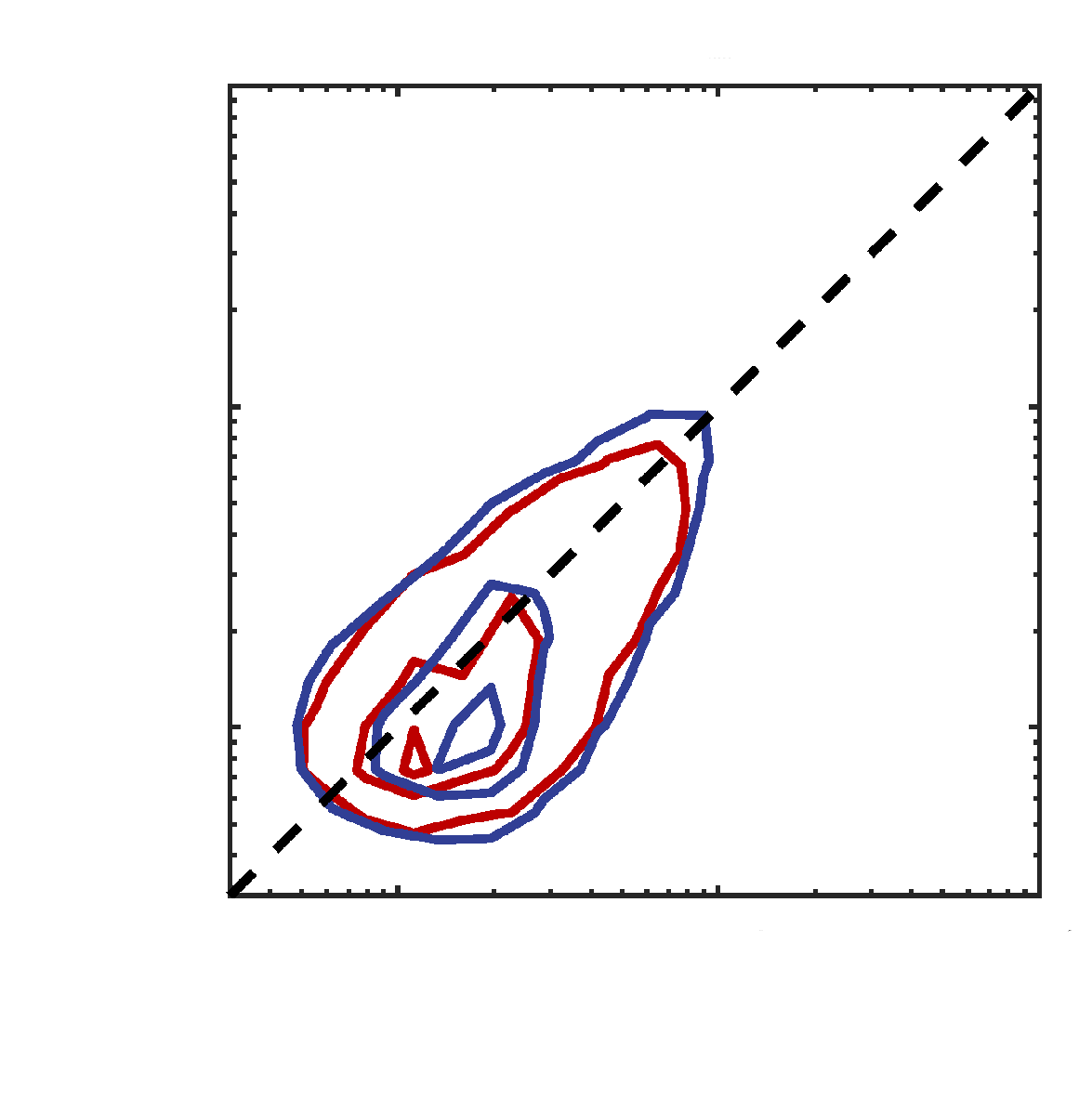}};
		\end{pgfonlayer}

		\begin{pgfonlayer}{foreground}
			\node[anchor=east] at (-0.1, 2 * \vspacing + \gridheight/2) {{Fully-spatial}};
		\end{pgfonlayer}

		\begin{pgfonlayer}{foreground}
			\node[anchor=south west] at (0, 3.35 * \vspacing) {\includegraphics[width=\gridwidth, height=\gridheight]{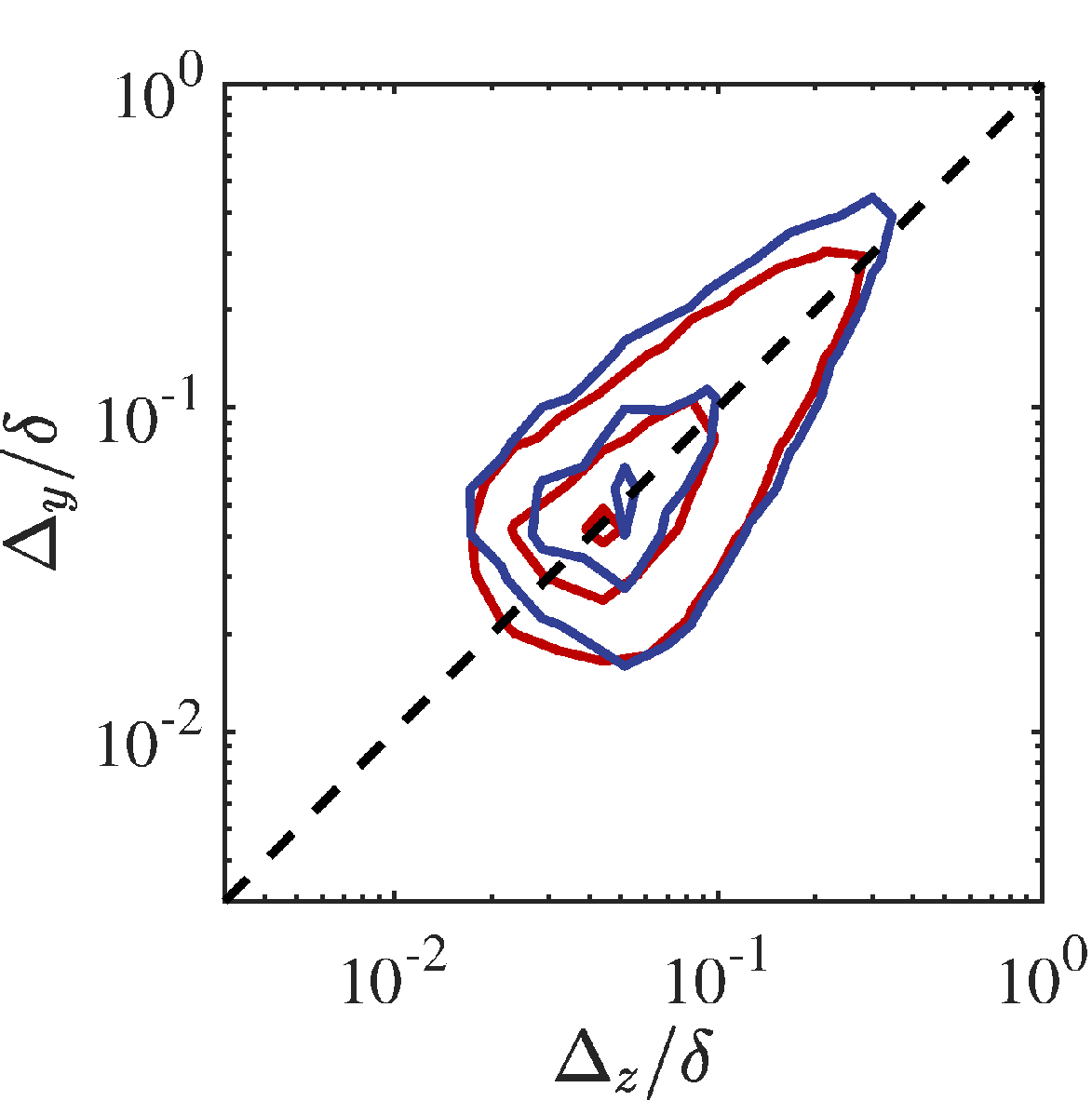}};
		\end{pgfonlayer}
		\begin{pgfonlayer}{middle}
			\node[anchor=south west] at (\gridwidth + \hspacing, 3.35 * \vspacing) {\includegraphics[width=\gridwidth, height=\gridheight]{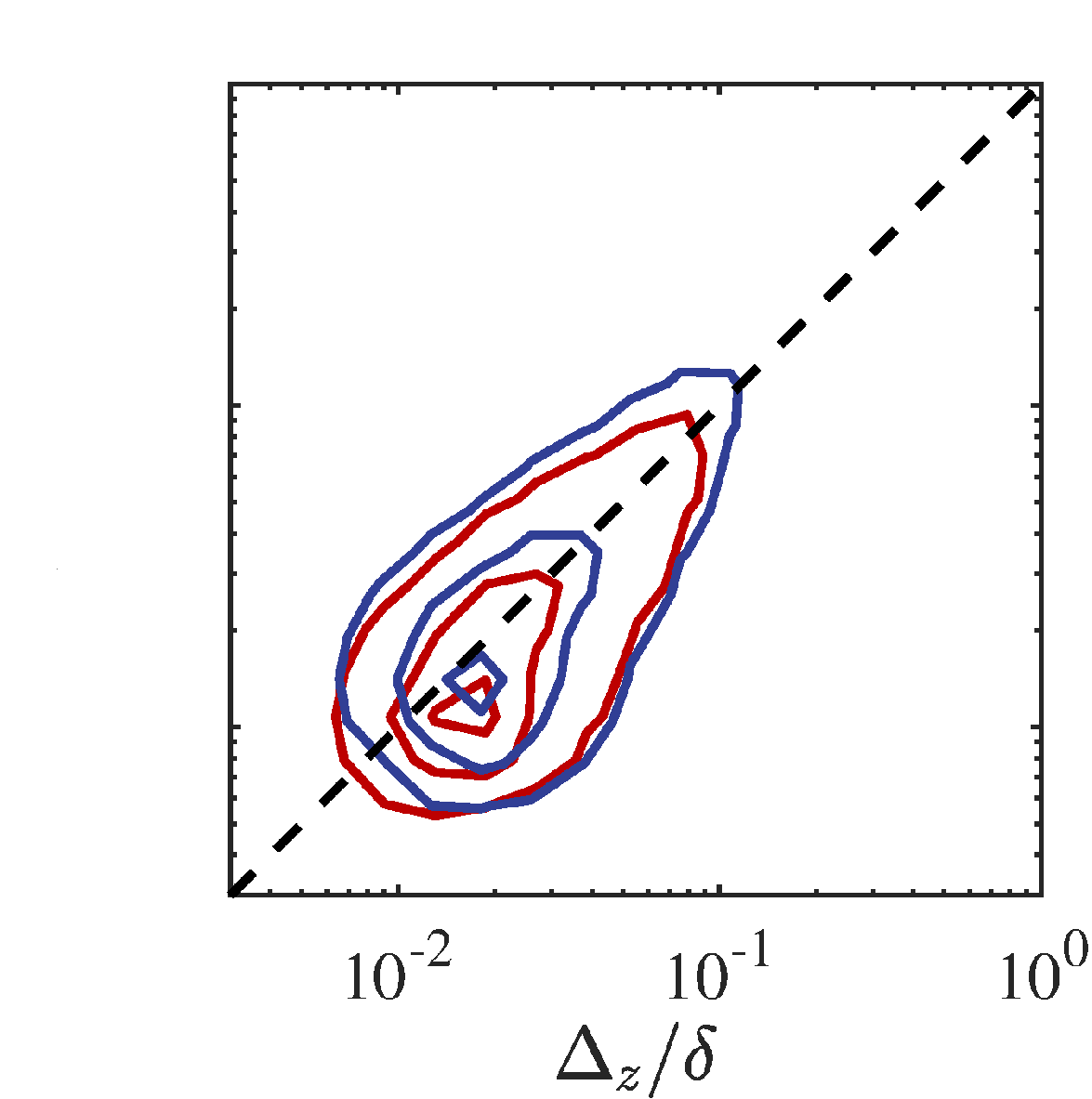}};
		\end{pgfonlayer}
		\begin{pgfonlayer}{background}
			\node[anchor=south west] at (2 * \gridwidth + 2 * \hspacing, 3.35 * \vspacing) {\includegraphics[width=\gridwidth, height=\gridheight]{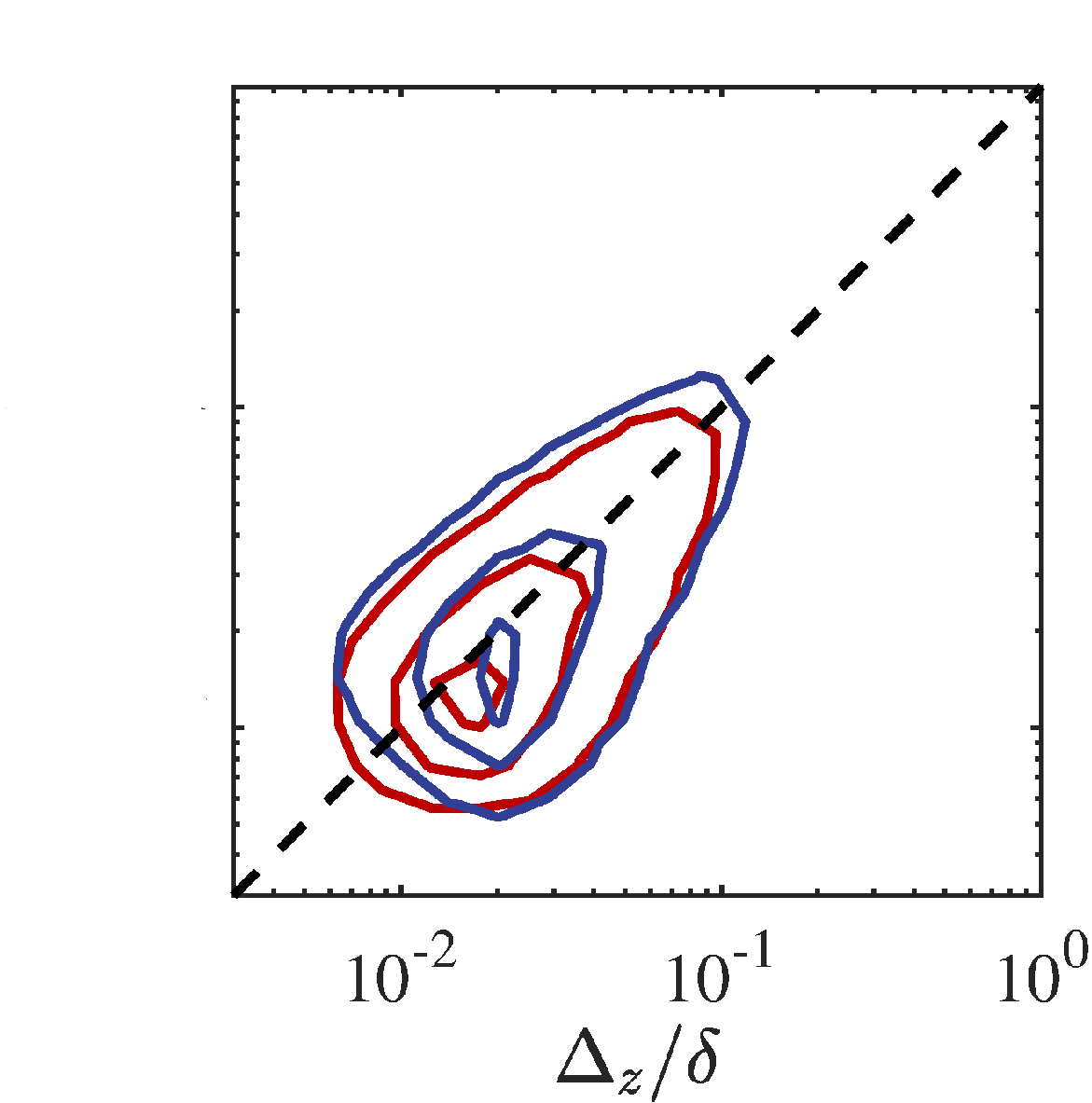}};
		\end{pgfonlayer}

		\begin{pgfonlayer}{foreground}
			\node[anchor=east] at (-0.1, 3 * \vspacing + \gridheight/2) {{Spatio-temporal}};
		\end{pgfonlayer}

		\begin{pgfonlayer}{foreground}
			\node[anchor=south west] at (-0.5,  -\gridheight - 0*\vspacing) {$(b)$};
		\end{pgfonlayer}
	\end{tikzpicture}
	
	\caption{Joint PDFs of (a) $\Delta y /\delta$ and $\Delta x /\delta$ and (b) $\Delta y /\delta$ and $\Delta z /\delta$. The first row is obtained with fully-spatial data, and the second row is obtained with spatio-temporal data for Q2 (red) and Q4 (blue). The contour levels are [0.1 0.5 0.9] of the maximum. The dimensions are normalized with local boundary layer thickness. The dashed line indicates $\Delta x /\delta = \Delta y /\delta$ in (a) and $\Delta y /\delta = \Delta z /\delta$ in (b).}
	\label{fig:jpdfs_combined}
\end{figure*}

\begin{figure*}[h!]
	\centering
	\begin{tikzpicture}

		\def\gridwidth{0.25\linewidth} 
		\def\gridheight{0.25\linewidth} 
		\def\hspacing{-0.5cm} 
		\def\vspacing{-3.95cm} 

		\pgfdeclarelayer{background}
		\pgfdeclarelayer{middle}
		\pgfdeclarelayer{foreground}
		\pgfdeclarelayer{title}
		\pgfsetlayers{background,middle,foreground,title}
		
			\begin{pgfonlayer}{title}
			\node[anchor=north] at (2.1, 4.35) {\includegraphics[width=.08\linewidth]{./figures/color.png}};
		\end{pgfonlayer}

		\begin{pgfonlayer}{foreground} 
			\node[anchor=south west] at (0, 0) {\includegraphics[width=\gridwidth, height=\gridheight]{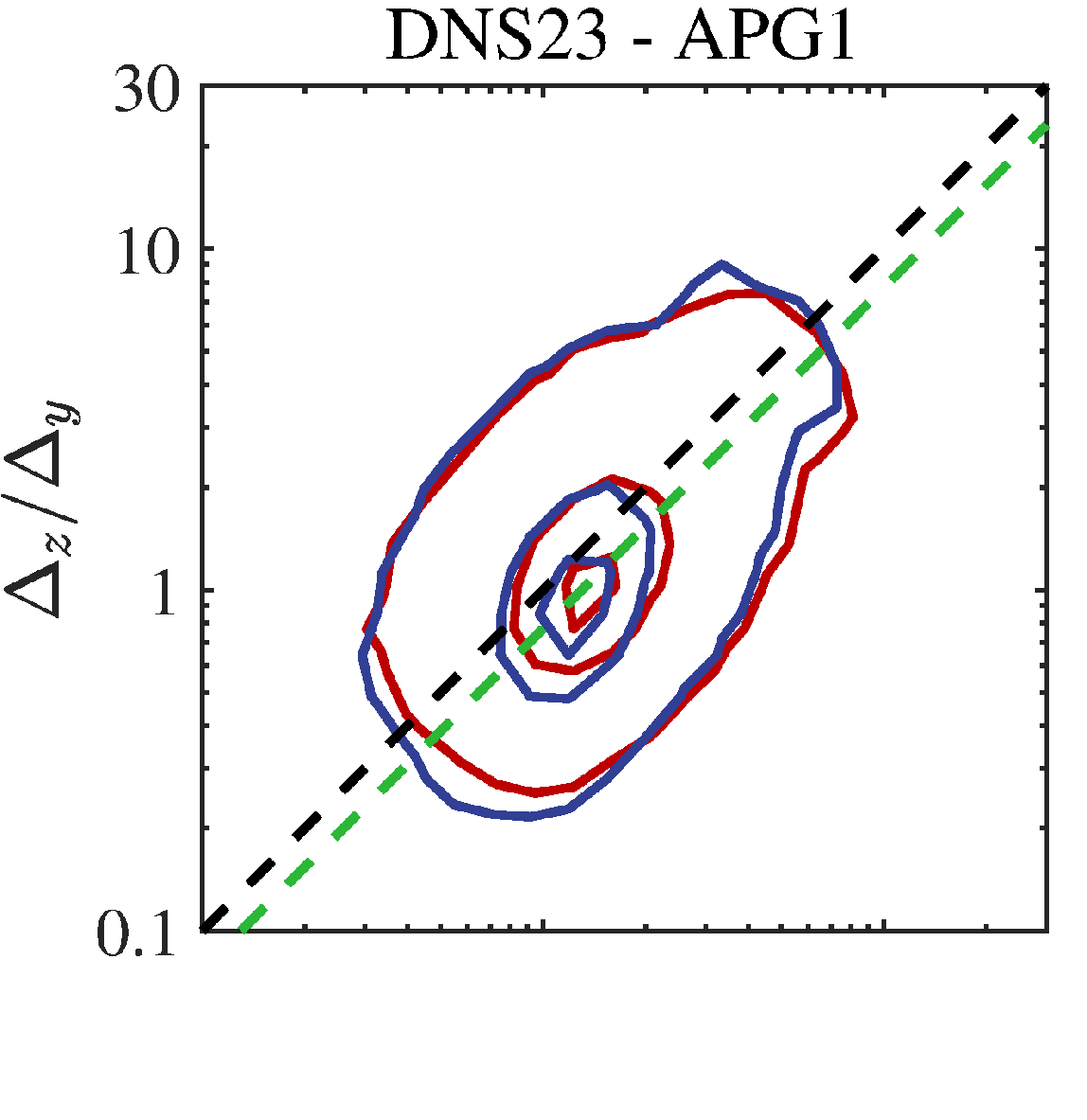}};
		\end{pgfonlayer}
		\begin{pgfonlayer}{middle} 
			\node[anchor=south west] at (\gridwidth + \hspacing, 0) {\includegraphics[width=\gridwidth, height=\gridheight]{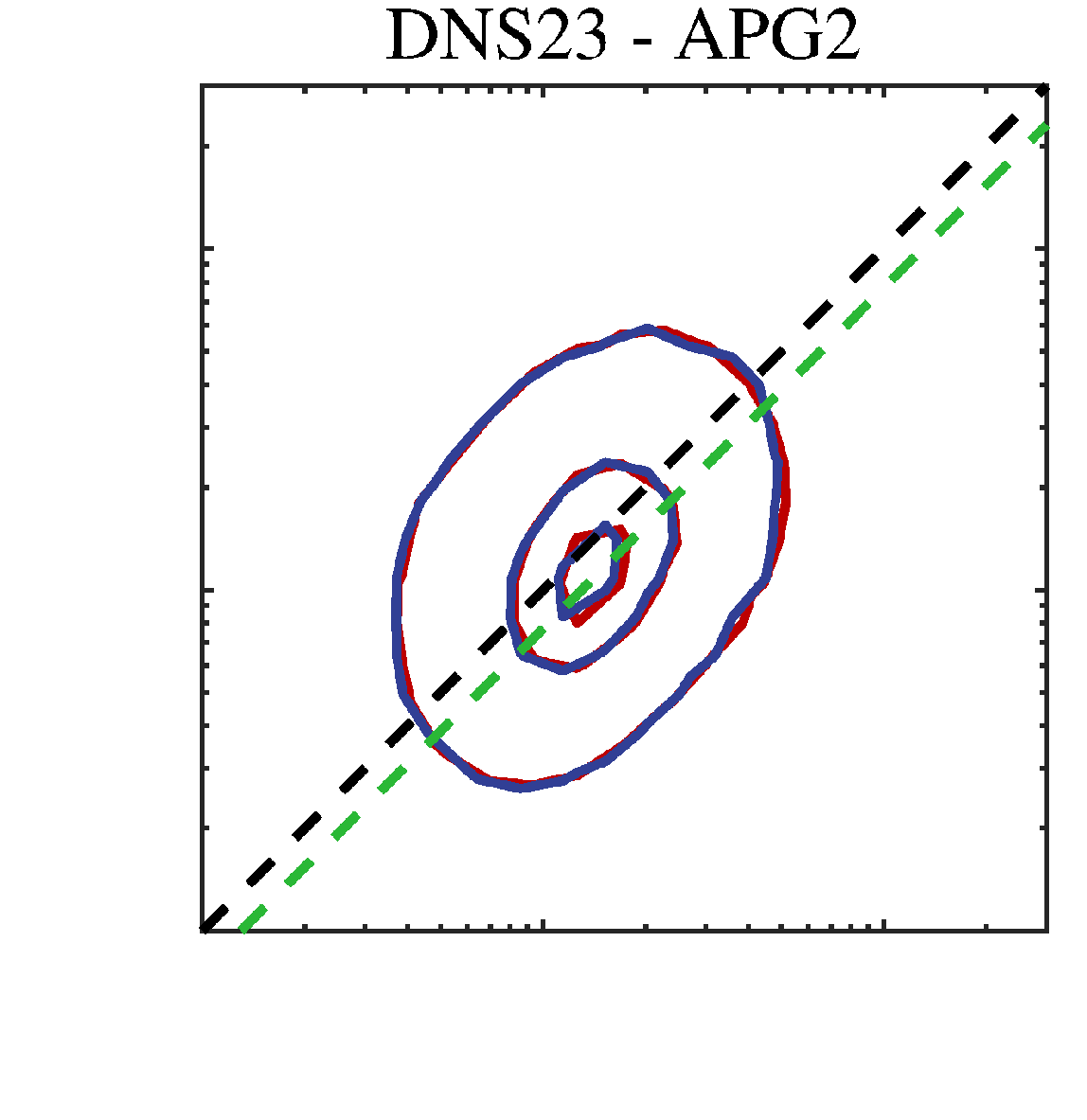}};
		\end{pgfonlayer}
		\begin{pgfonlayer}{background} 
			\node[anchor=south west] at (2 * \gridwidth + 2 * \hspacing, 0) {\includegraphics[width=\gridwidth, height=\gridheight]{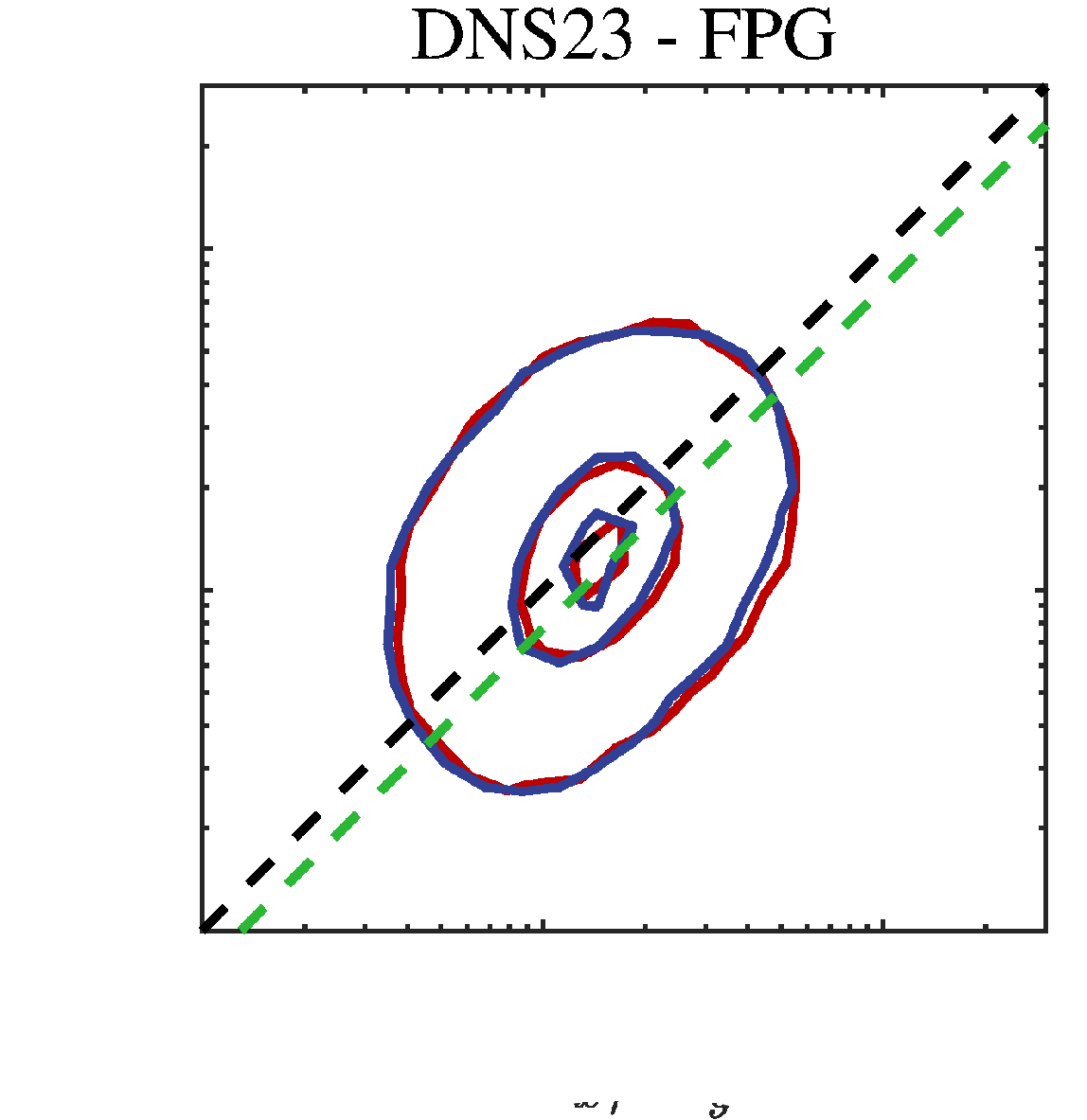}};
		\end{pgfonlayer}

		\begin{pgfonlayer}{foreground}
			\node[anchor=east] at (-0.1, \gridheight/2) {{Fully-spatial}};
		\end{pgfonlayer}

		\begin{pgfonlayer}{foreground} 
			\node[anchor=south west] at (0, \vspacing) {\includegraphics[width=\gridwidth, height=\gridheight]{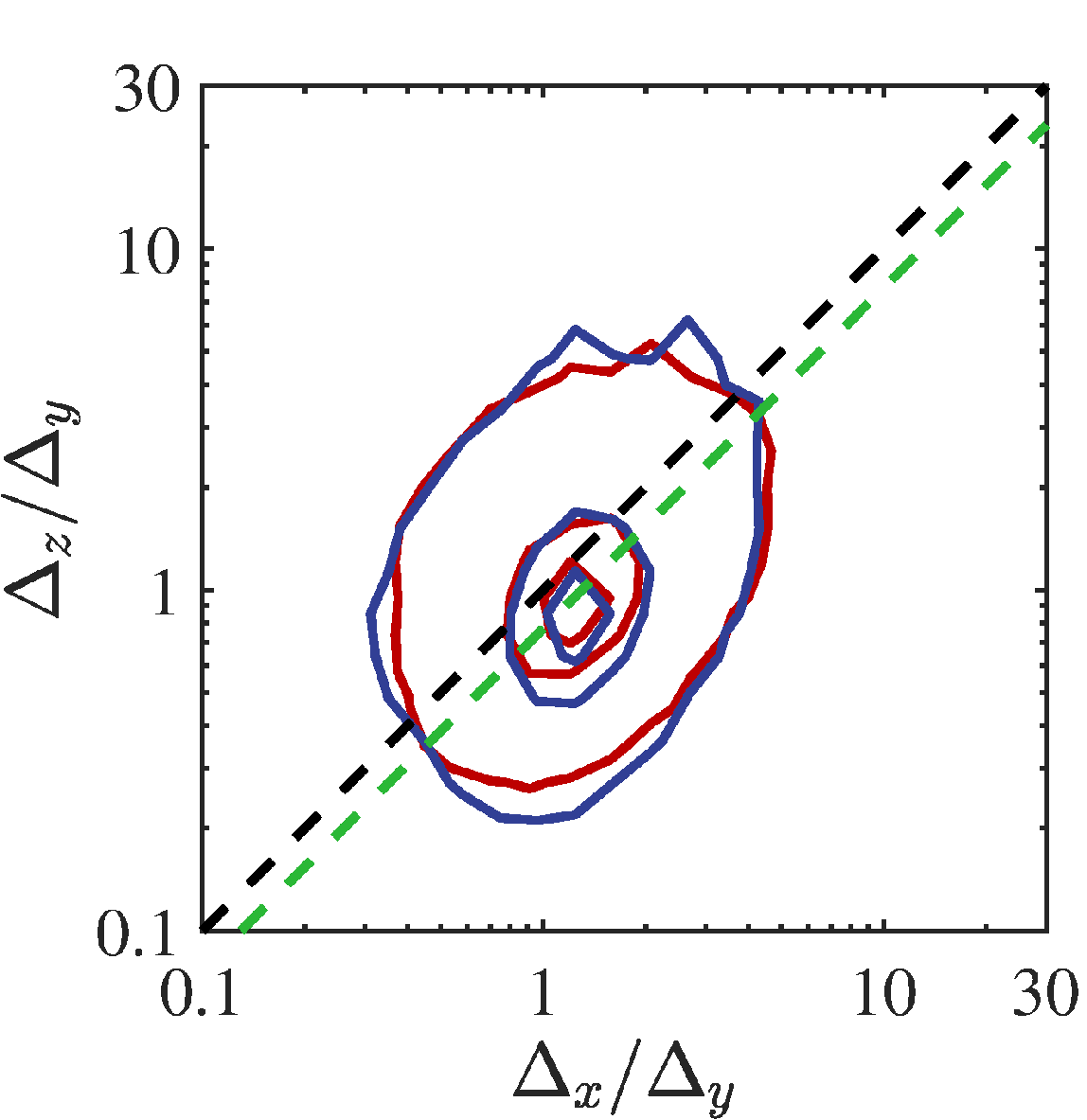}};
		\end{pgfonlayer}
		\begin{pgfonlayer}{middle} 
			\node[anchor=south west] at (\gridwidth + \hspacing, \vspacing) {\includegraphics[width=\gridwidth, height=\gridheight]{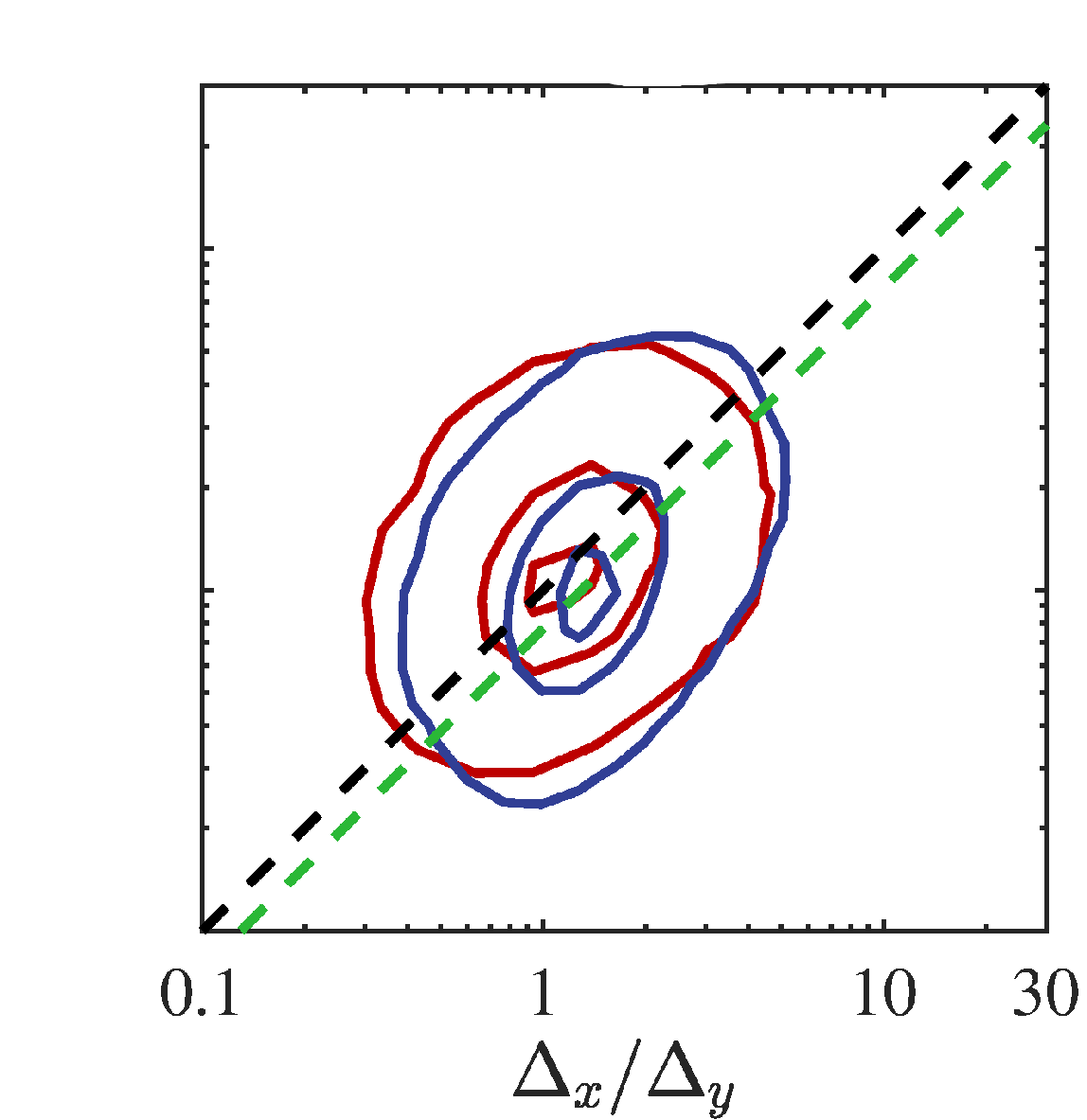}};
		\end{pgfonlayer}
		\begin{pgfonlayer}{background} 
			\node[anchor=south west] at (2 * \gridwidth + 2 * \hspacing, \vspacing) {\includegraphics[width=\gridwidth, height=\gridheight]{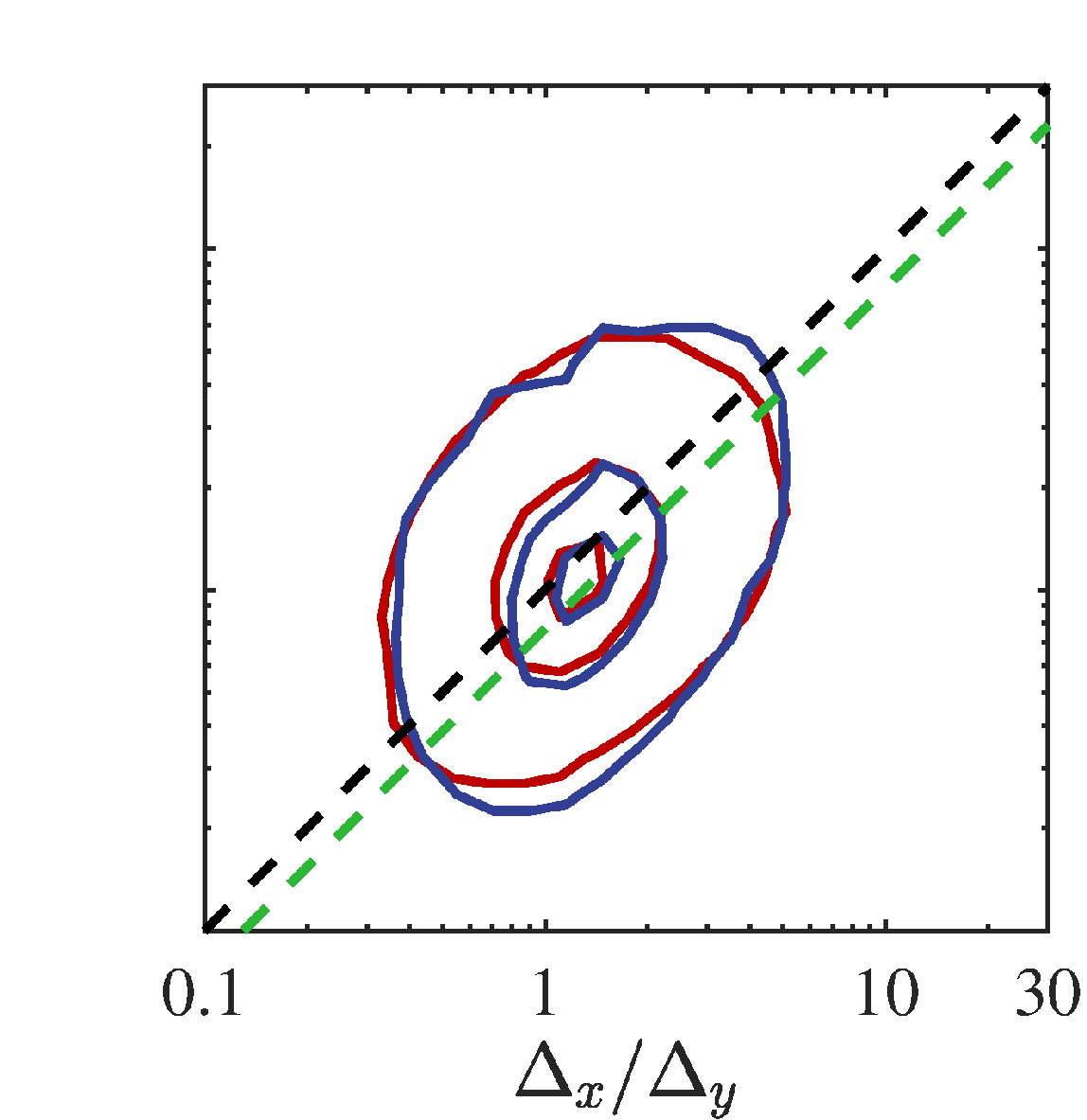}};
		\end{pgfonlayer}

		\begin{pgfonlayer}{foreground}
			\node[anchor=east] at (-0.1, \vspacing + \gridheight/2) {{Spatio-temporal}};
		\end{pgfonlayer}
	\end{tikzpicture}
	
	\caption{Joint PDFs of the logarithms of the aspect ratios of the boxes circumscribing Q2s (red) and Q4s (blue). The first row is obtained with fully-spatial data, and the second row is obtained with spatio-temporal data. The contour levels are [0.1 0.5 0.98] of the maximum. The dashed black line indicates $\Delta x = \Delta z$ and the green line indicates $\Delta x = 1.3\Delta z$.}
	\label{fig:jpdfaspectratio}
\end{figure*}

\begin{table}[htbp!]
	\centering
	\caption{Parameters of detached Q2s and Q4s with $0.3 \delta < y_{c} < 0.8 \delta$ and $y_{\text{min}} > 0.05 \delta$. $N_{i}$ and $V_{i}$ are the percentages of $Q^-$s in terms of their number and volume for fully-spatial and spatio-temporal data with parameters normalized by the total for all detached $Q^-$ structures.}
	\begin{tabular}{lrrrr}
		\hline \hline
		Position       & NQ2 & NQ{4} & VQ{2} & VQ{4} \\
		\hline

		\multicolumn{5}{l}{\textbf{Fully-spatial Data}} \\
		DNS23-APG1     & 51.9    & 48.1    & 56.1    & 43.9      \\
		DNS23-APG2     & 48.8    & 51.2    & 57.8    & 42.2      \\
		DNS23-FPG      & 53.1    & 46.9    & 38.1    & 61.9      \\
		\hline

		\multicolumn{5}{l}{\textbf{Spatio-temporal Data}} \\
		DNS22-APG1     & 46.4    & 53.7    & 62.8    & 37.2    \\
		DNS22-APG2     & 44.1    & 56.0    & 61.5    & 38.5    \\
		DNS23-APG1     & 47.7    & 52.3    & 59.3    & 40.7    \\
		DNS23-APG2     & 44.3    & 55.7    & 62.3    & 37.7    \\
		DNS23-FPG      & 49.2    & 50.8    & 45.5    & 54.5    \\
		\hline \hline
	\end{tabular}
	\label{tab:structure_parameters}
\end{table}

We proceed to examine the spatial features of Q2 and Q4 structures, with their centers, $y_c$, located within the region $0.3\delta<y_c<0.8\delta$. The location of the structure's center, $y_c$, is determined by calculating the arithmetic mean of its minimum and maximum $y$ coordinates.  Additionally, we only consider wall-detached structures for this analysis, because wall-attached structures can be geometrically affected by the wall and their spatial features can be very different from those of wall-detached ones \citep{dong}. By doing so, we aim to isolate the effect of the mean shear across different flow cases. Fig.~\ref{fig:flowvis3ddetached} presents a similar visualization of the sweeps and ejections as in Fig.~\ref{fig:flowvis3d}, for the same flow snapshot, but only for detached structures within the region $0.3\delta<y_c<0.8\delta$. A comparison between Fig.~\ref{fig:flowvis3d}b and \ref{fig:flowvis3ddetached}b shows that large-scale wall-attached structures are removed by the selection process. As mentioned for Fig.~\ref{fig:flowvis3d}, drawing conclusions about modifications in the shape and size of the detached structures is difficult from visualizations like those in Fig.~\ref{fig:flowvis3ddetached}. Table ~\ref{tab:structure_parameters} presents the number and volume of detached Q2 and Q4 structures in the region $0.3\delta<y_c<0.8\delta$. It is worth noting that the values of detached structures presented in Tables ~\ref{tab2} and ~\ref{tab:structure_parameters} differ because both the wall-normal range and parameter normalization are different. In Table \ref{tab2}, values correspond to attached and detached $Q^-$ structures across the boundary layer ($0 < y/\delta < 1.2$), with parameters normalized by the total value for all attached and detached $Q^-$ structures. In contrast, Table \ref{tab:structure_parameters} focuses exclusively on detached $Q^-$ structures ($y_{min} > 0.05\delta$) within the range $0.3\delta < y_c < 0.8\delta$, with parameters normalized by the total for detached $Q^-$ structures only.

Table~\ref{tab:structure_parameters} reveals that the number of detached Q2s is slightly less than detached Q4s in most cases. For their volumes, which are more important than their numbers, Q2 structures occupy a larger volume than Q4s in the APG TBL cases. The ratio of the volume occupied by Q2s to that occupied by Q4s is consistent across all APG cases.  Contrary to this, Q4s have a larger volume in the FPG case.

\setcounter{figure}{10}
\begin{figure*}[h!]
	\centering
	\begin{tikzpicture}

		\def\gridwidth{8cm} 
		\def\gridheight{2.5cm} 
		\def\hspacing{.5cm} 
		\def\vspacing{1.5cm} 
		
		\pgfdeclarelayer{background}
		\pgfdeclarelayer{middle}
		\pgfdeclarelayer{foreground}
		\pgfdeclarelayer{title}
		\pgfsetlayers{background,middle,foreground,title}
		
		\begin{pgfonlayer}{title}
			\node[anchor=south] at (0.25*\gridwidth , \gridheight+1.5cm) {Q2 + Q4}; 
			\node[anchor=south] at (0.2*\gridwidth , -1.2*\gridheight+3cm) {Q2}; 
			\node[anchor=south] at (0.2*\gridwidth , -1.65*\gridheight) {Q4}; 
		\end{pgfonlayer}

		\begin{pgfonlayer}{foreground}
			\node[anchor=south west] (a) at (0, 0) {\includegraphics[width=\gridwidth]{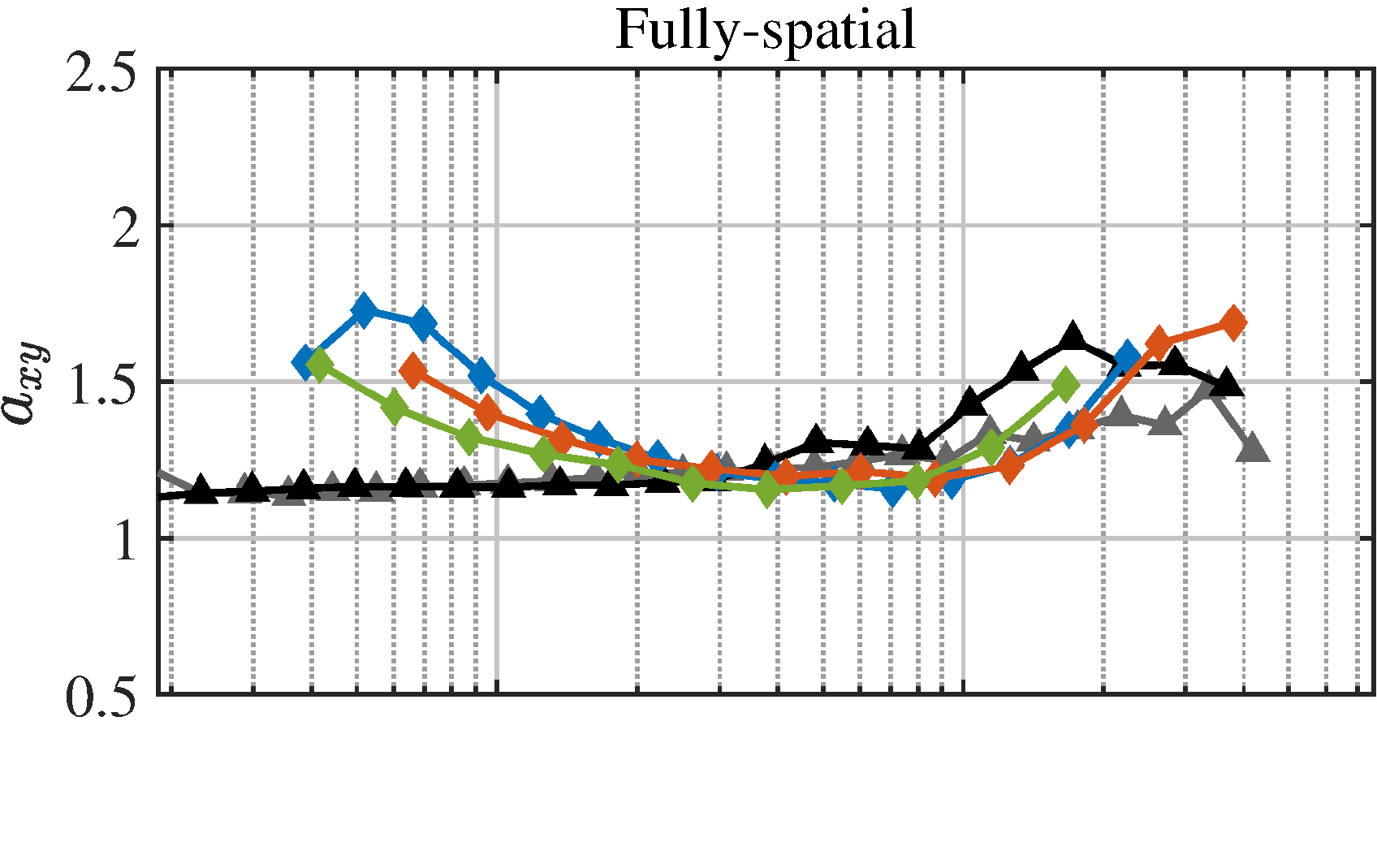}};
			\node[anchor=north west, inner sep=5pt, xshift=-0.25cm, yshift=-10pt] at (a.north west) { $(a)$};

			\node[anchor=south west] (b) at (\gridwidth + \hspacing, 0) {\includegraphics[width=\gridwidth]{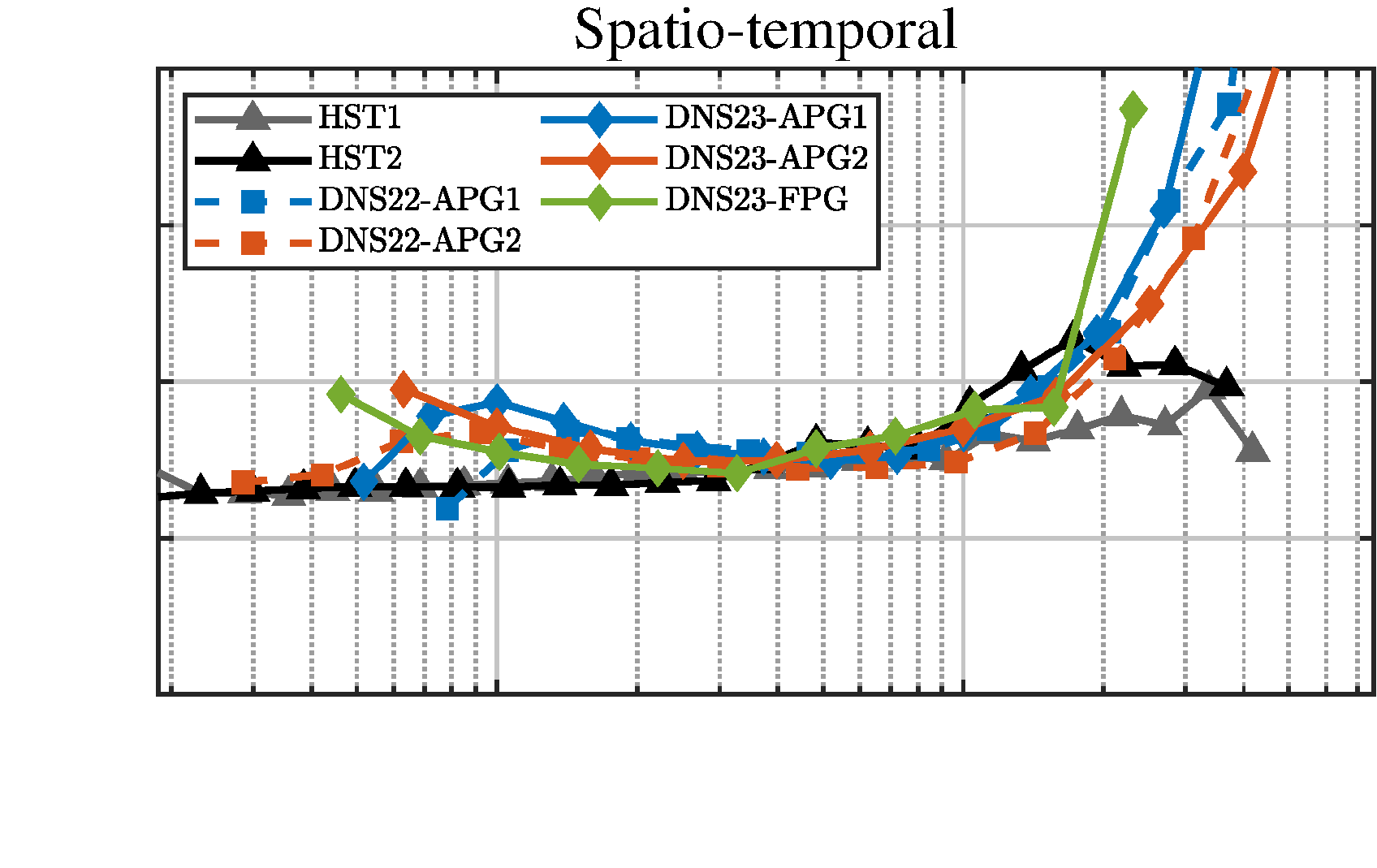}};
			\node[anchor=north west, inner sep=5pt, xshift=-0.1cm, yshift=-10pt] at (b.north west) { $(b)$};
			
			\node[anchor=south west] (c) at (0, -\gridheight - \vspacing) {\includegraphics[width=\gridwidth]{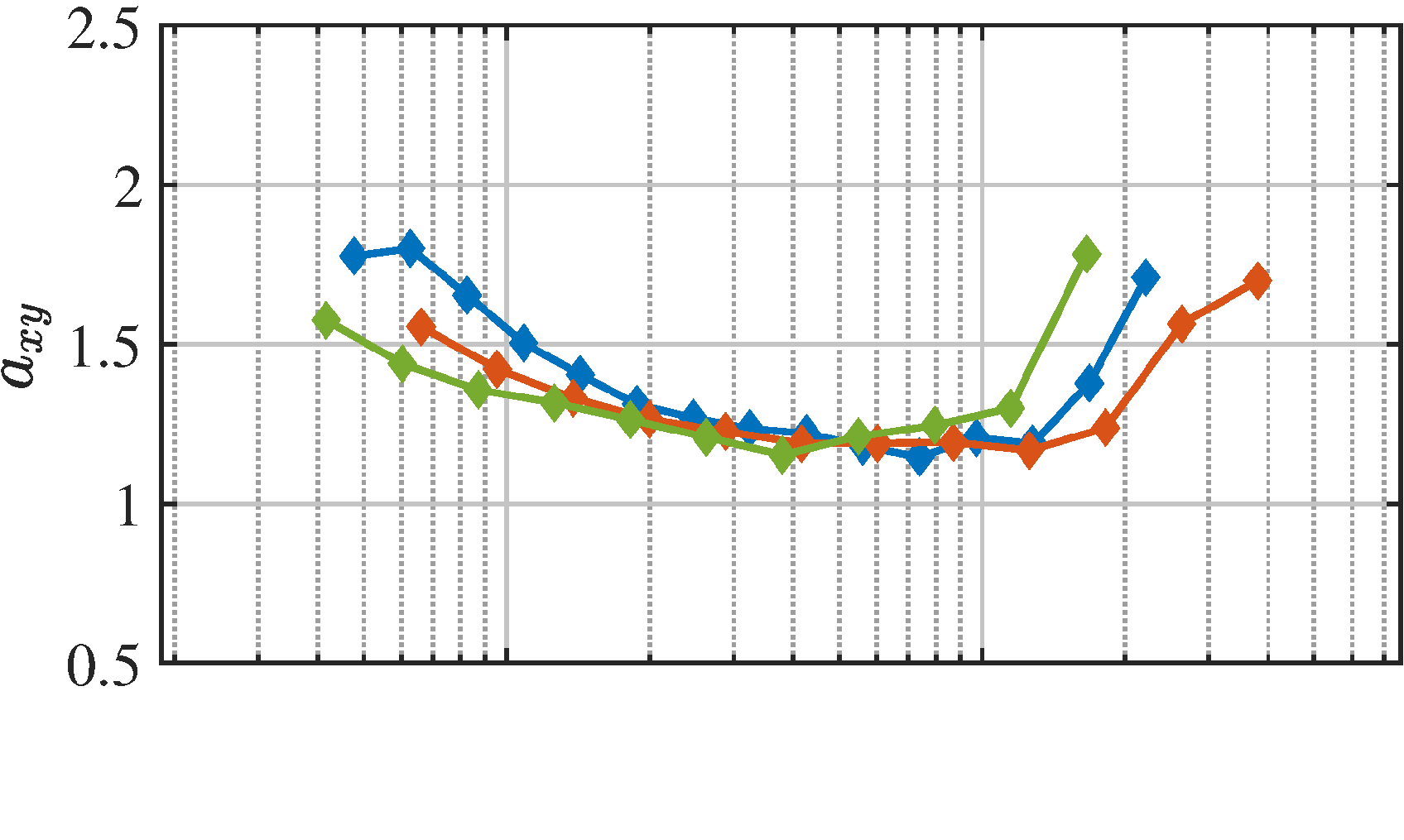}};
			\node[anchor=north west, inner sep=5pt, xshift=-0.25cm, yshift=-1pt] at (c.north west) { $(c)$};
			
			\node[anchor=south west] (d) at (\gridwidth + \hspacing, -\gridheight - \vspacing) {\includegraphics[width=\gridwidth]{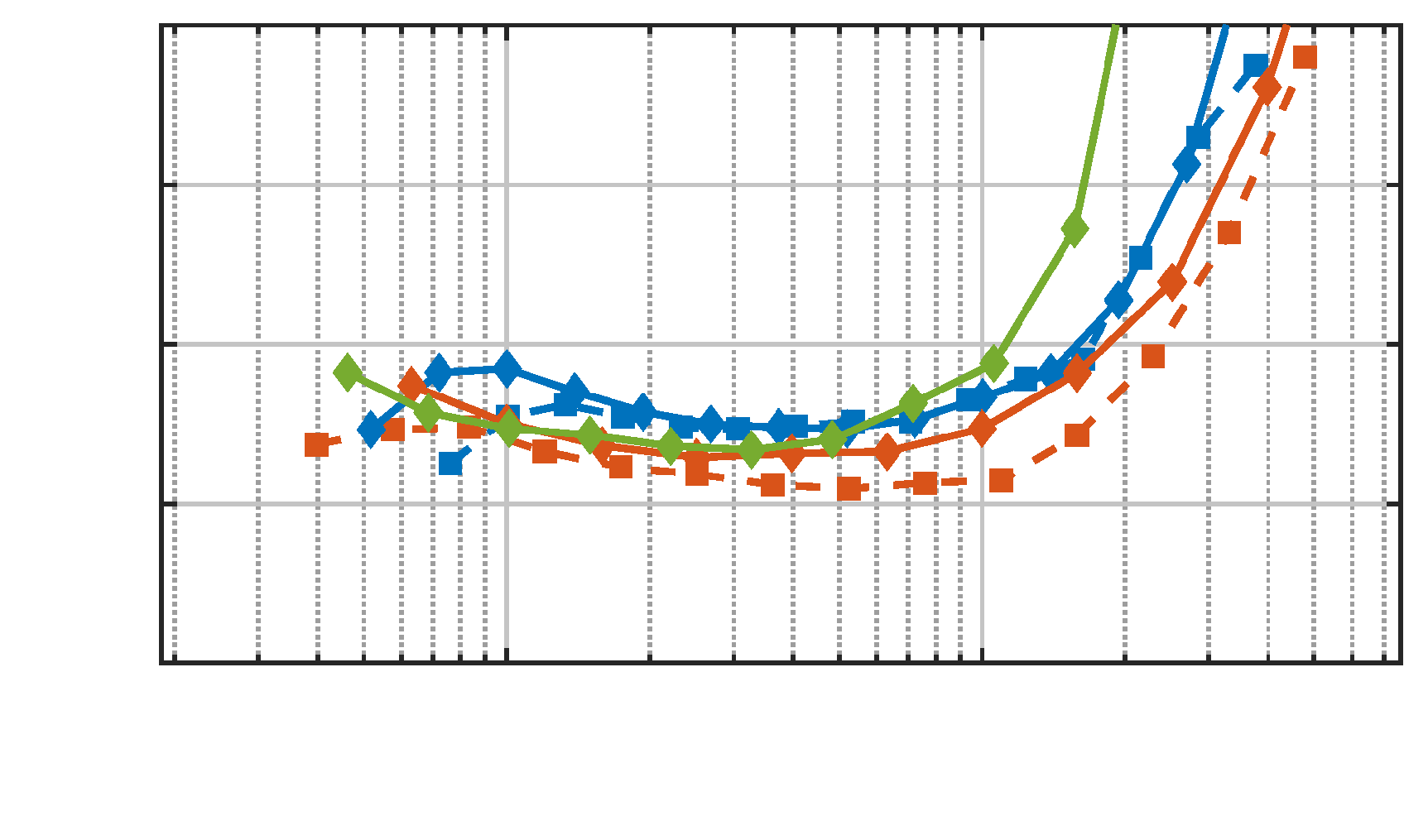}};
			\node[anchor=north west, inner sep=5pt, xshift=-0.1cm, yshift=-1pt] at (d.north west) { $(d)$};
			
			\node[anchor=south west] (e) at (0, -2 * \gridheight - 2 * \vspacing) {\includegraphics[width=\gridwidth]{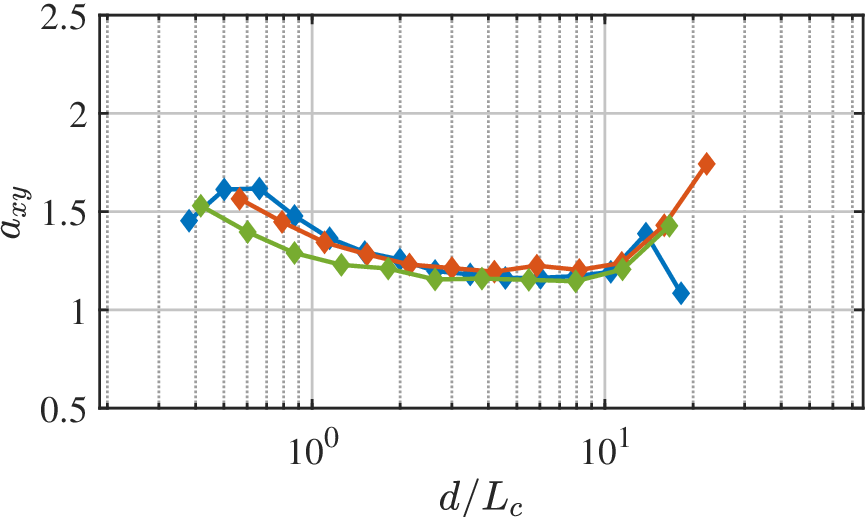}};
			\node[anchor=north west, inner sep=5pt, xshift=-0.25cm, yshift=-1pt] at (e.north west) { $(e)$};
			
			\node[anchor=south west] (f) at (\gridwidth + \hspacing, -2 * \gridheight - 2 * \vspacing) {\includegraphics[width=\gridwidth]{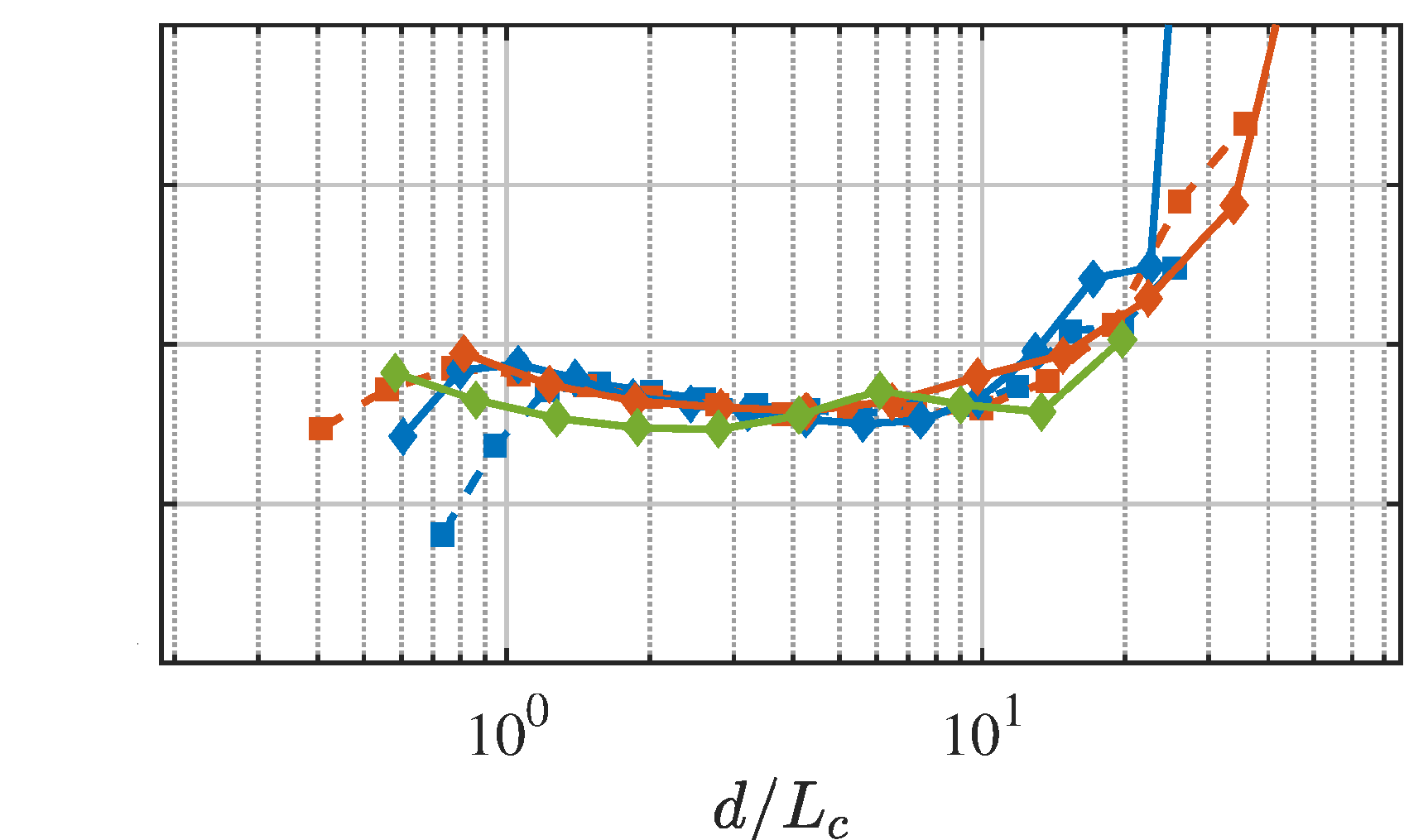}};
			\node[anchor=north west, inner sep=5pt, xshift=-0.1cm, yshift=-1pt] at (f.north west) { $(f)$}; 
		\end{pgfonlayer}
	\end{tikzpicture}
	\caption{Average aspect ratio, $a_{xy} =\Delta_x / \Delta_y$, of the circumscribing boxes for detached $Q^{-}$ structures whose center is between $y_c / \delta = 0.3$ and 0.8 as a function of the box diagonal of the structure for all $Q^{-} \mathrm{s}$ combined $(a, b), \mathrm{Q}2\mathrm{s}(c, d)$ and $\mathrm{Q}4\mathrm{s}(e, f)$. The box diagonal is normalized with $L_{c}$.}
	\label{fig::aspectratioxy}
\end{figure*}

We now examine the detached structures' lengths in all directions and their overall extent using a circumscribing box. To illustrate this, Fig.~\ref{fig::structure} depicts a single Q2 structure enclosed within its circumscribing box. The box dimensions $\Delta_x, \Delta_y$ and $\Delta_z$ are utilized to represent the dimensions of the structure. Fig.~\ref{fig:jpdfs_combined}a illustrates the joint probability density functions (PDFs) of the logarithms of streamwise and wall-normal sizes of detached Q2s and Q4s, while Fig.~\ref{fig:jpdfs_combined}b shows the joint PDFs of their spanwise and wall-normal lengths, with the first row representing joint PDFs obtained from fully-spatial data and the second row from spatio-temporal data. The dimensions are normalized by the local boundary layer thickness, $\delta$. Spatio-temporal and fully-spatial data exhibit similar behavior for the low-defect APG TBL (APG1), where Q2 and Q4 structures tend to be streamwise elongated. However, in the large-defect APG TBL (APG2), spatio-temporal data indicates that Q2 structures are elongated in the streamwise direction, while Q4 structures are less elongated. In contrast, fully-spatial data show that both Q2 and Q4 structures exhibit similar streamwise elongations. This discrepancy may arise because Taylor's hypothesis underestimates the streamwise length of Q4 structures in the spatio-temporal case for large-defect flows, even though the convection velocity was specifically chosen as the structure's volume-averaged $U$. In the FPG case, although the shape factor is similar to that of low-defect APGs, the structure sizes are more comparable to those in large-defect APG cases, possibly due to the flow history. The dimensions of Q2 and Q4 structures are generally similar across all flows, but the largest structures tend to be Q4s.

\setcounter{figure}{9}
\begin{figure}[H]
	\centering

	\includegraphics[width=0.85\linewidth]{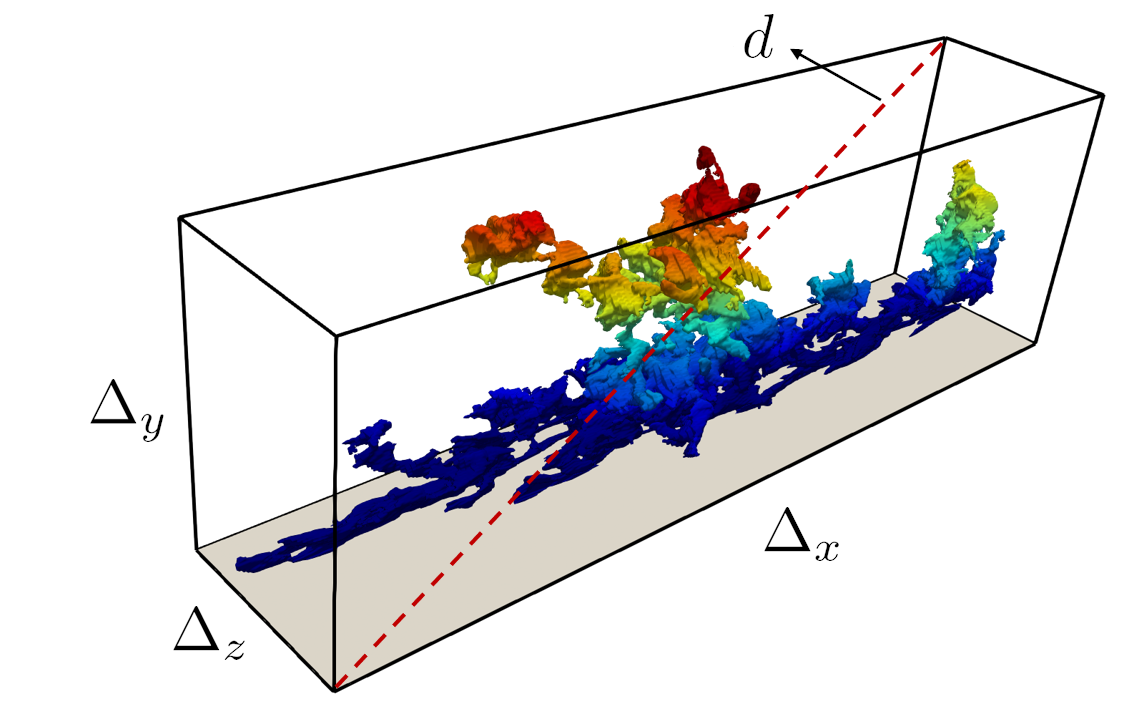}
	
	\caption{A Q2 structure with the circumscribing box used to indicate its length in all directions and its overall extent. The dashed line shows the diagonal, $d$.}
	\label{fig::structure}
\end{figure}

\setcounter{figure}{11}
\begin{figure*}[h!]
	\centering
	\begin{tikzpicture}

		\def\gridwidth{8cm} 
		\def\gridheight{2.5cm} 
		\def\hspacing{.5cm} 
		\def\vspacing{1.5cm} 

		\pgfdeclarelayer{background}
		\pgfdeclarelayer{middle}
		\pgfdeclarelayer{foreground}
		\pgfdeclarelayer{title}
		\pgfsetlayers{background,middle,foreground,title}
		
		\begin{pgfonlayer}{title}
			\node[anchor=south] at (0.25*\gridwidth , \gridheight+1.5cm) {Q2 + Q4}; 
			\node[anchor=south] at (0.2*\gridwidth , -1.2*\gridheight+3cm) {Q2}; 
			\node[anchor=south] at (0.2*\gridwidth , -1.65*\gridheight) {Q4}; 
		\end{pgfonlayer}

		\begin{pgfonlayer}{foreground}
			\node[anchor=south west] (a) at (0, 0) {\includegraphics[width=\gridwidth]{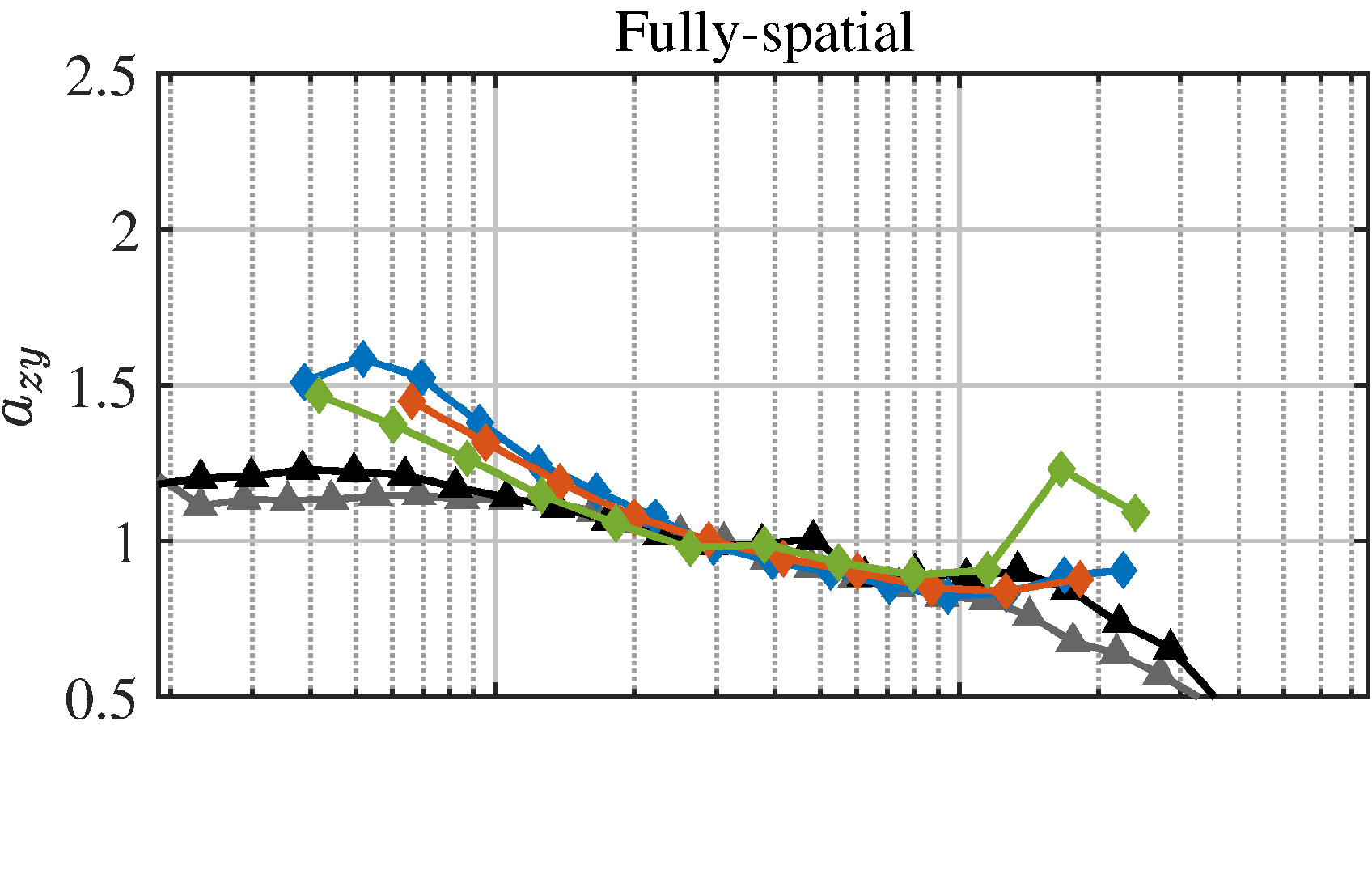}};
			\node[anchor=north west, inner sep=5pt, xshift=-0.25cm, yshift=-10pt] at (a.north west) { $(a)$};
			
			\node[anchor=south west] (b) at (\gridwidth + \hspacing, 0) {\includegraphics[width=\gridwidth]{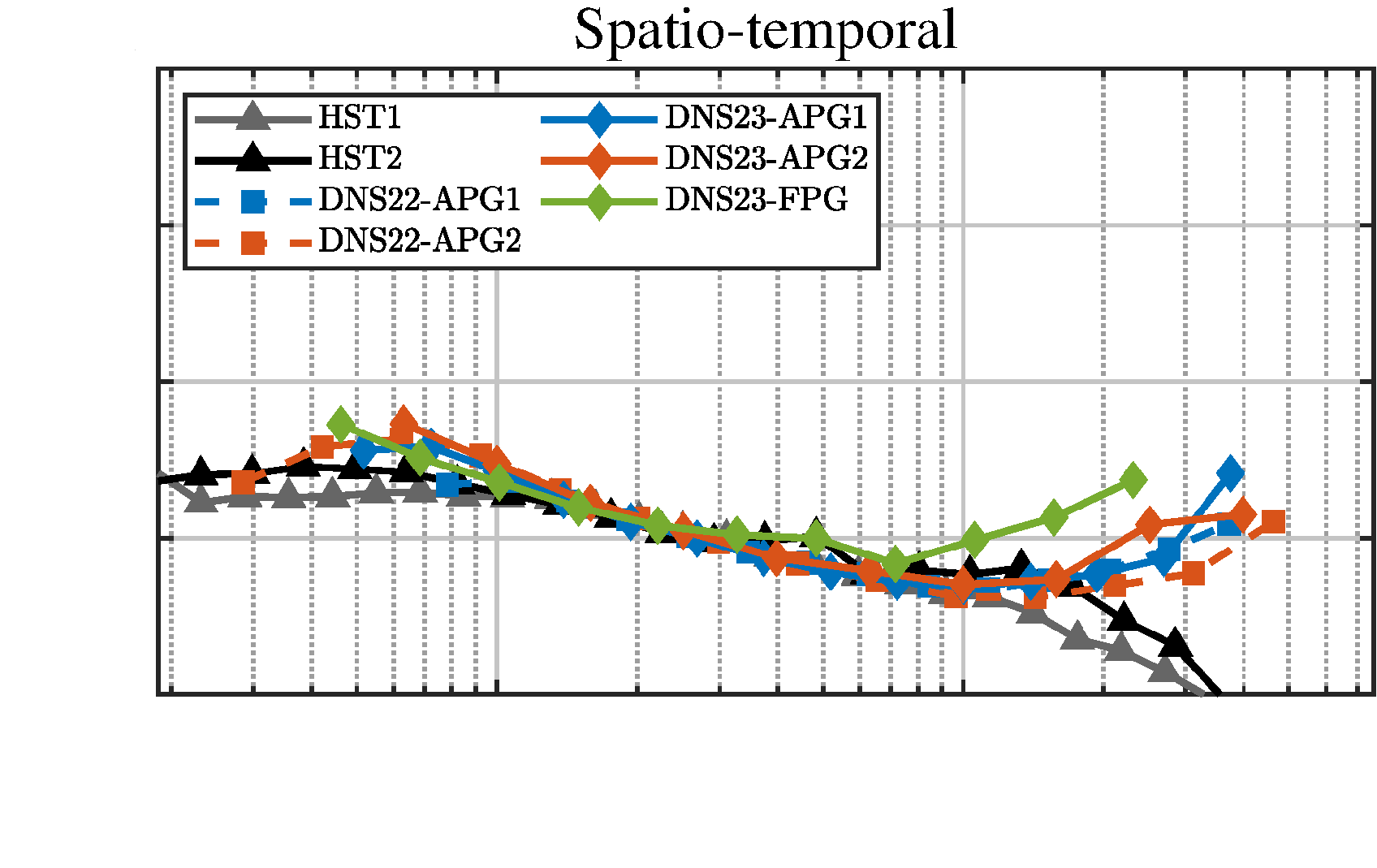}};
			\node[anchor=north west, inner sep=5pt, xshift=-0.1cm, yshift=-10pt] at (b.north west) { $(b)$}; 
			
			\node[anchor=south west] (c) at (0, -\gridheight - \vspacing) {\includegraphics[width=\gridwidth]{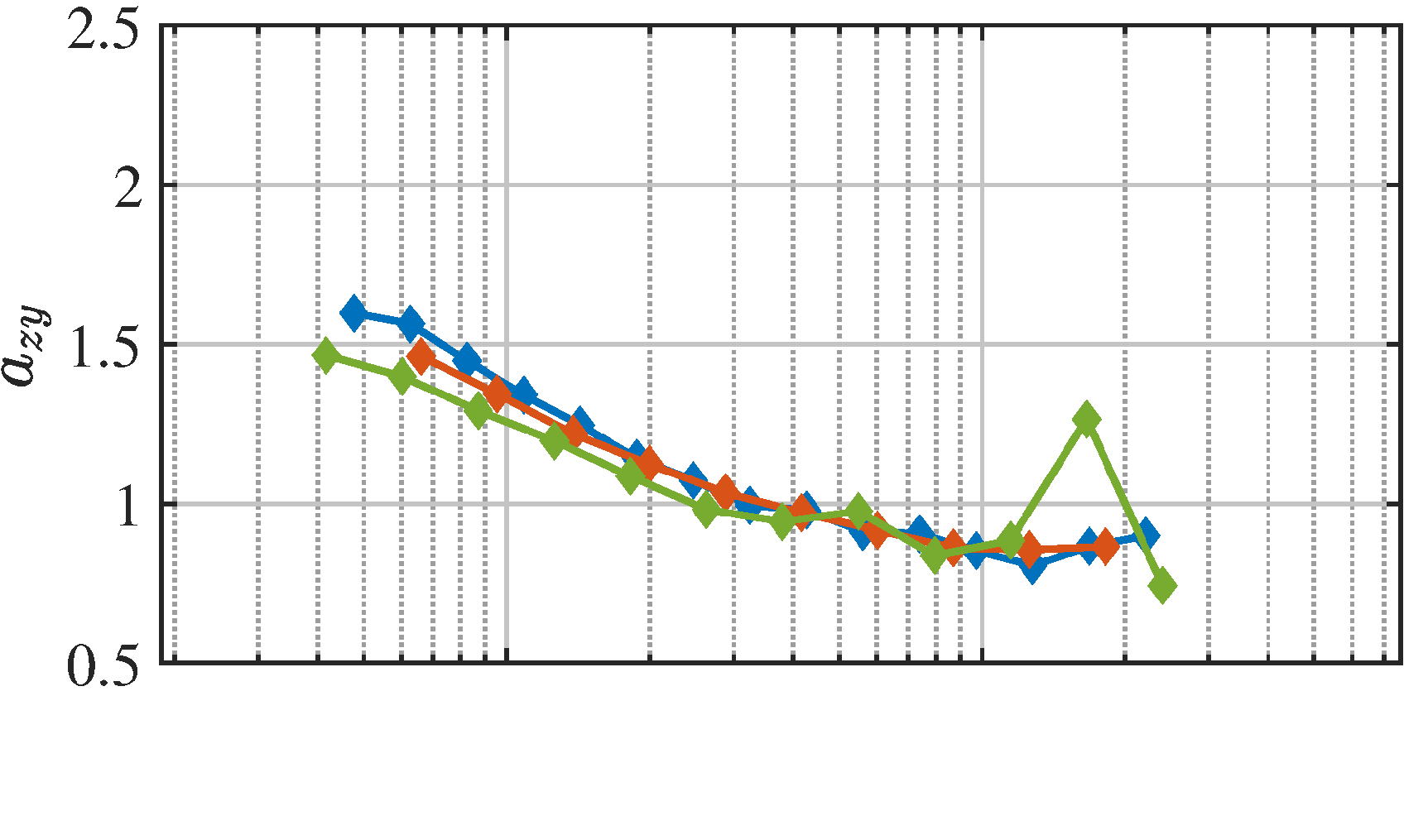}};
			\node[anchor=north west, inner sep=5pt, xshift=-0.25cm, yshift=-1pt] at (c.north west) { $(c)$};
			
			\node[anchor=south west] (d) at (\gridwidth + \hspacing, -\gridheight - \vspacing) {\includegraphics[width=\gridwidth]{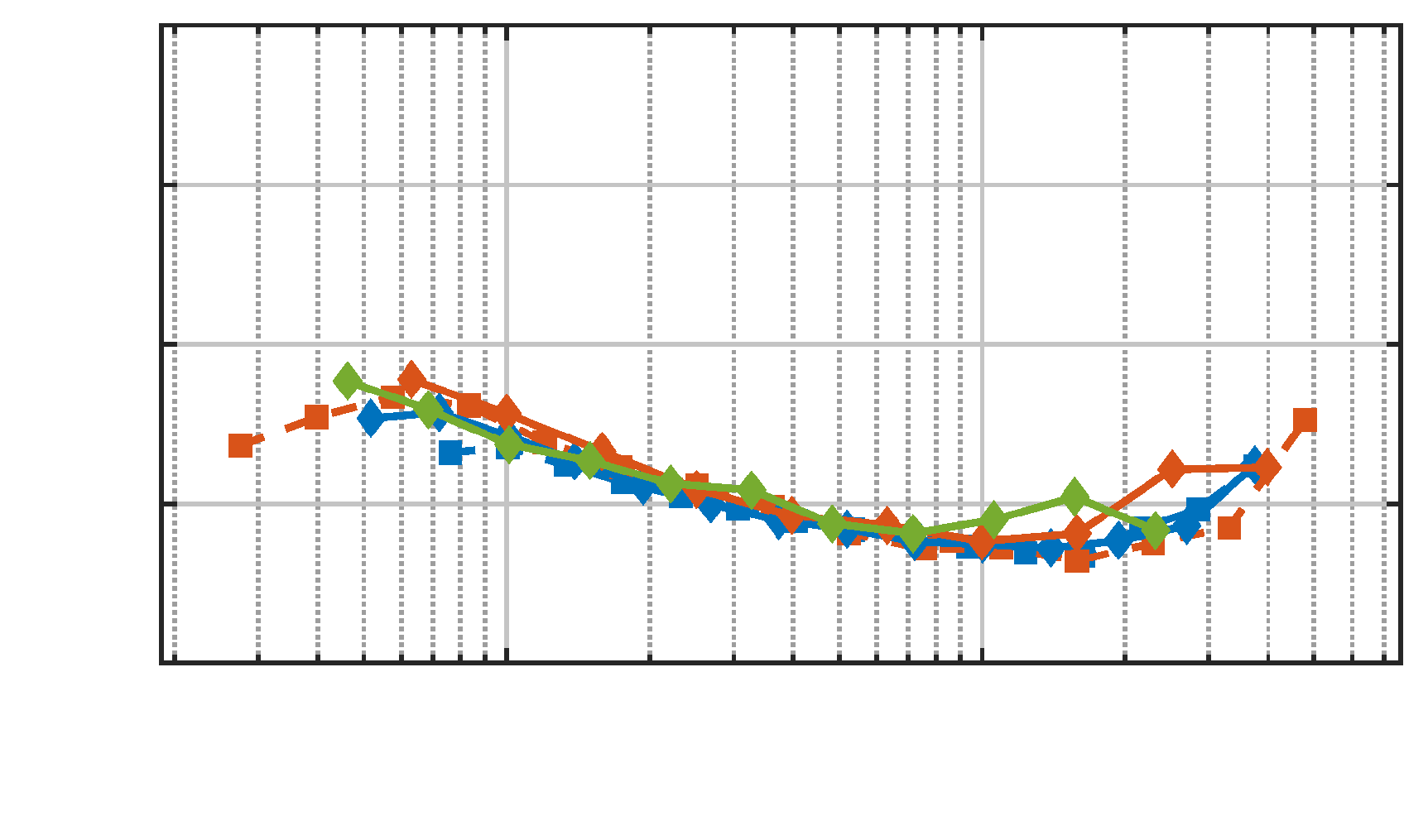}};
			\node[anchor=north west, inner sep=5pt, xshift=-0.1cm, yshift=-1pt] at (d.north west) { $(d)$}; 
			
			\node[anchor=south west] (e) at (0, -2 * \gridheight - 2 * \vspacing) {\includegraphics[width=\gridwidth]{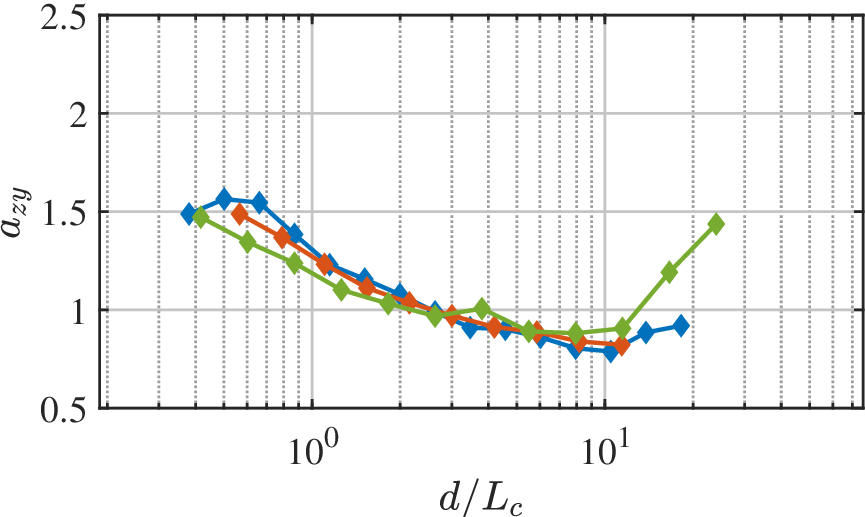}};
			\node[anchor=north west, inner sep=5pt, xshift=-0.25cm, yshift=-1pt] at (e.north west) { $(e)$};

			\node[anchor=south west] (f) at (\gridwidth + \hspacing, -2 * \gridheight - 2 * \vspacing) {\includegraphics[width=\gridwidth]{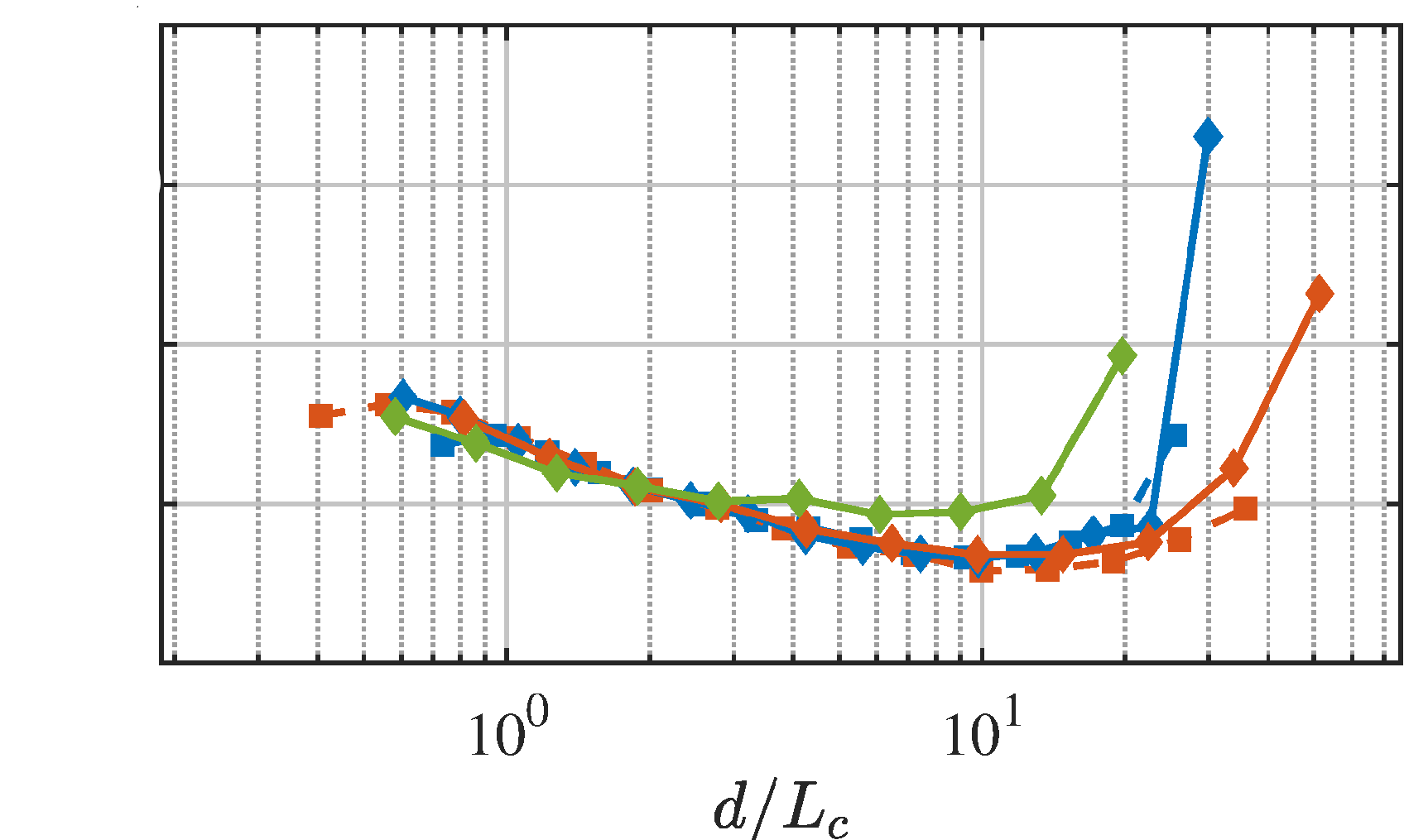}};
			\node[anchor=north west, inner sep=5pt, xshift=-0.1cm, yshift=-1pt] at (f.north west) { $(f)$}; 
		\end{pgfonlayer}
	\end{tikzpicture}
	\caption{Average aspect ratio, $a_{zy} =\Delta_z / \Delta_y$, of the circumscribing boxes for detached $Q^{-}$ structures whose center is between $y_c / \delta = 0.3$ and 0.8 as a function of the box diagonal of the structure for all $Q^{-} \mathrm{s}$ combined $(a, b), \mathrm{Q}2\mathrm{s}(c, d)$ and $\mathrm{Q}4\mathrm{s}(e, f)$. The box diagonal is normalized with $L_{c}$.}
	\label{fig::aspectratiozy}
\end{figure*}

\begin{figure*}[h!]
	\centering
	\begin{tikzpicture}

		\pgfdeclarelayer{background}
		\pgfdeclarelayer{middle2}
		\pgfdeclarelayer{middle1}
		\pgfdeclarelayer{foreground}
		\pgfdeclarelayer{title}
		\pgfsetlayers{background,middle2,middle1,foreground,title}

		\def\gridwidth{4.5cm} 
		\def\gridheight{4.5cm} 
		\def\hspacing{-.75cm} 
		\def\vspacing{-3.5cm} 

		\begin{pgfonlayer}{title} 
			\node[anchor=center] at (-0.14*\gridwidth + 0.6*\gridwidth, 4.5cm) {{Q2 - Q2}};
			\node[anchor=center] at (-0.14*\gridwidth +1.6*\gridwidth + \hspacing, 4.5cm) {{Q4 - Q4}};
			\node[anchor=center] at (-0.14*\gridwidth +2.6*\gridwidth + 2*\hspacing, 4.5cm) {{Q2 - Q4}};
			\node[anchor=center] at (-0.14*\gridwidth +3.6*\gridwidth + 3*\hspacing, 4.5cm) {{Q4 - Q2}};
		\end{pgfonlayer}

		\begin{pgfonlayer}{foreground} 
			\node[anchor=south west] at (-0.15*\gridwidth, 0) {\includegraphics[width=\gridwidth]{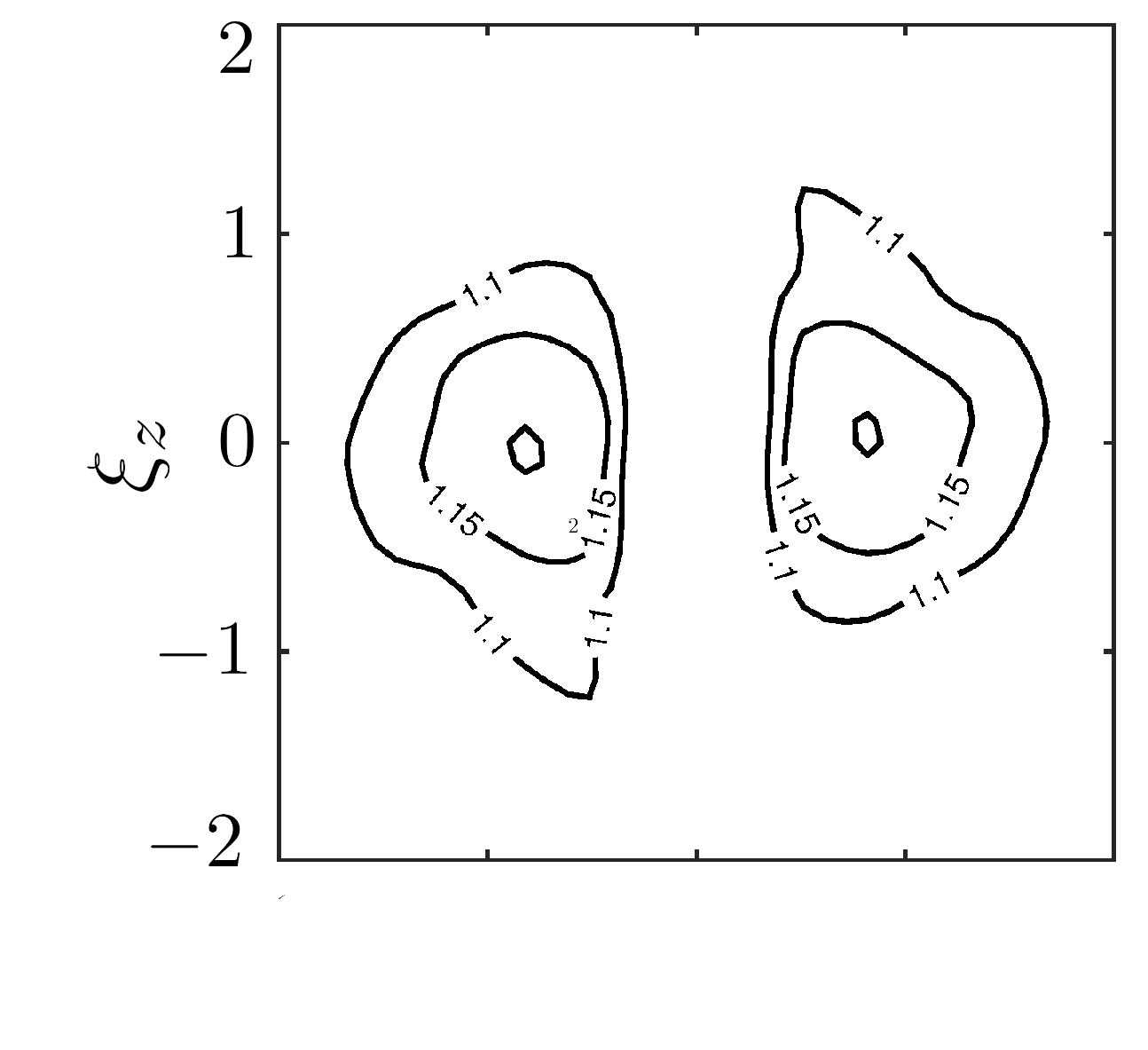}};
		\end{pgfonlayer}
		\begin{pgfonlayer}{middle1} 
			\node[anchor=south west] at (-0.15*\gridwidth + \gridwidth + \hspacing, 0) {\includegraphics[width=\gridwidth]{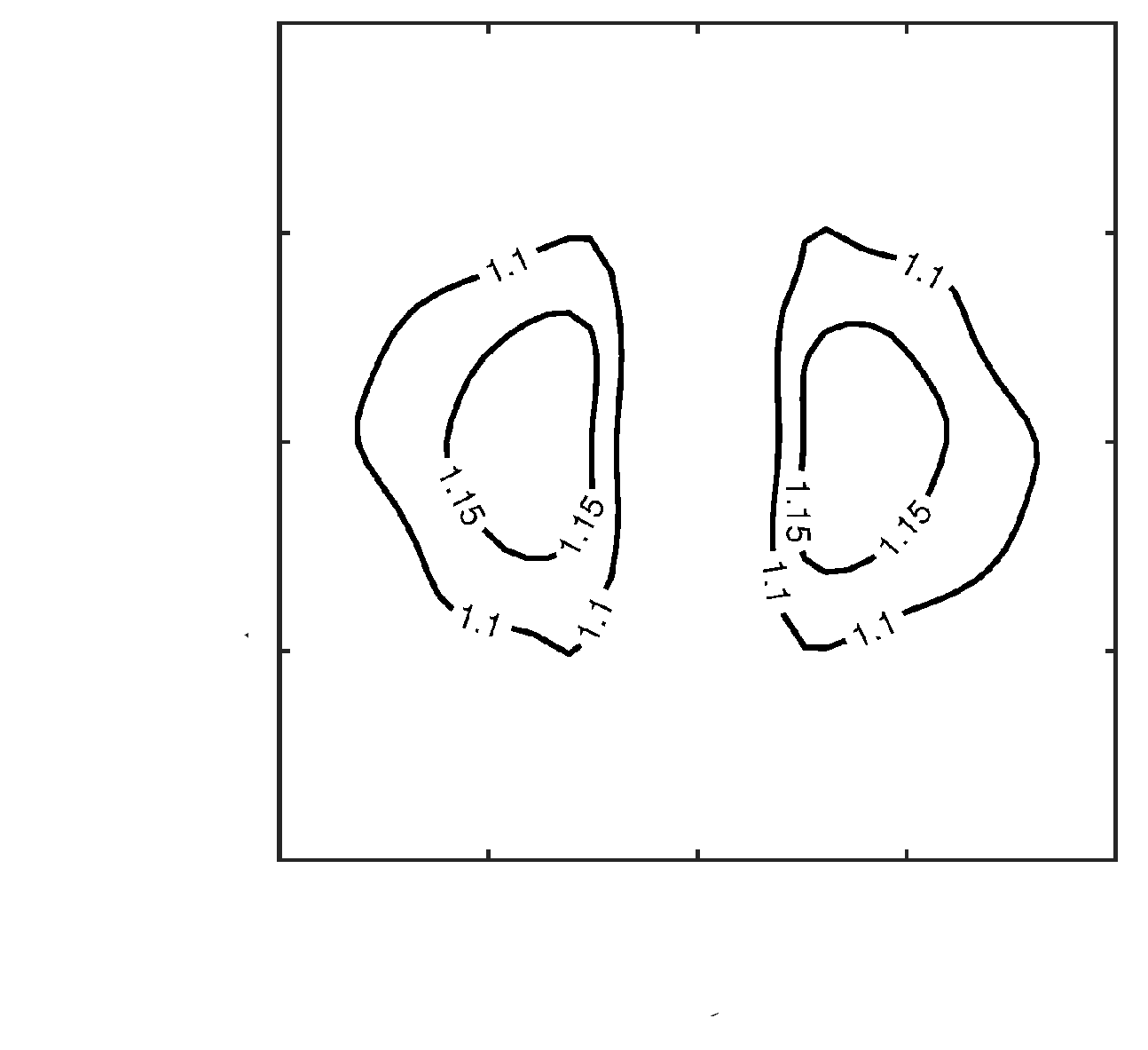}};
		\end{pgfonlayer}
		\begin{pgfonlayer}{middle2} 
			\node[anchor=south west] at (-0.15*\gridwidth + 2*\gridwidth + 2*\hspacing, 0) {\includegraphics[width=\gridwidth]{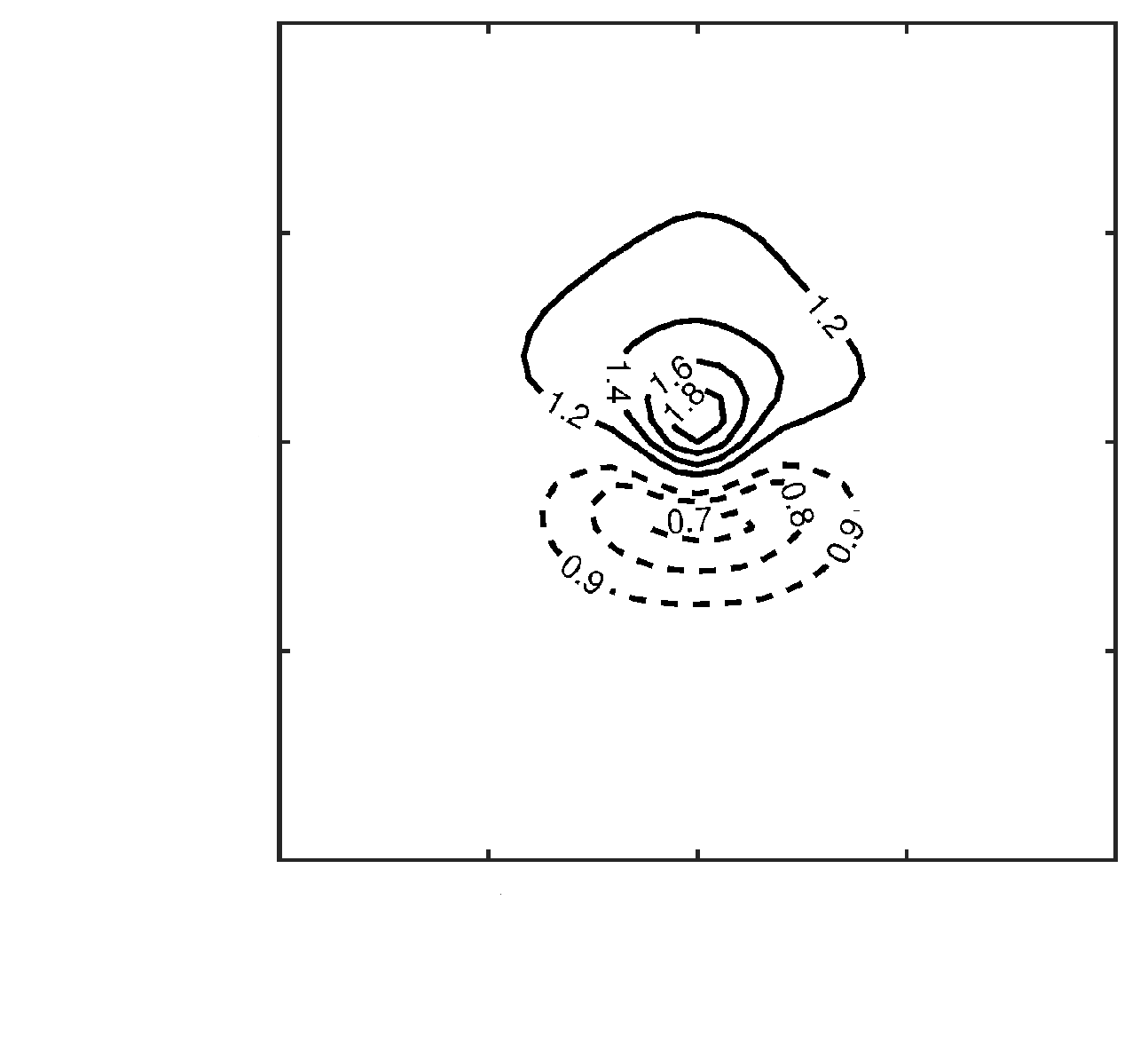}};
		\end{pgfonlayer}
		\begin{pgfonlayer}{background}
			\node[anchor=south west] at (-0.15*\gridwidth + 3*\gridwidth + 3*\hspacing, 0) {\includegraphics[width=\gridwidth]{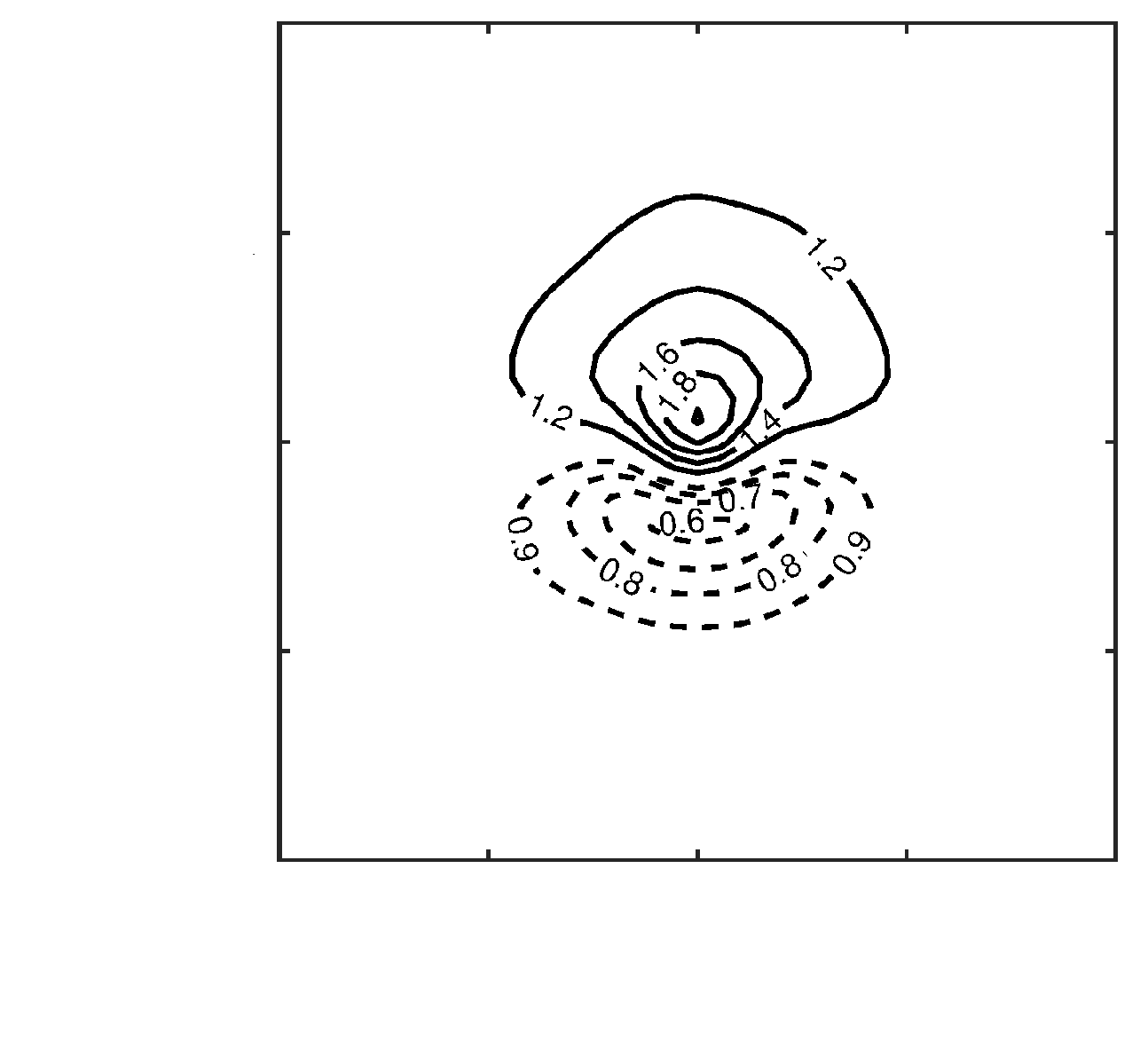}};
		\end{pgfonlayer}

		\begin{pgfonlayer}{foreground}
			\node[anchor=east] at (-0.6, \gridheight/2) {{DNS23 - APG1}};
		\end{pgfonlayer}

		\begin{pgfonlayer}{foreground}
			\node[anchor=south west] at (-0.15*\gridwidth, \vspacing) {\includegraphics[width=\gridwidth]{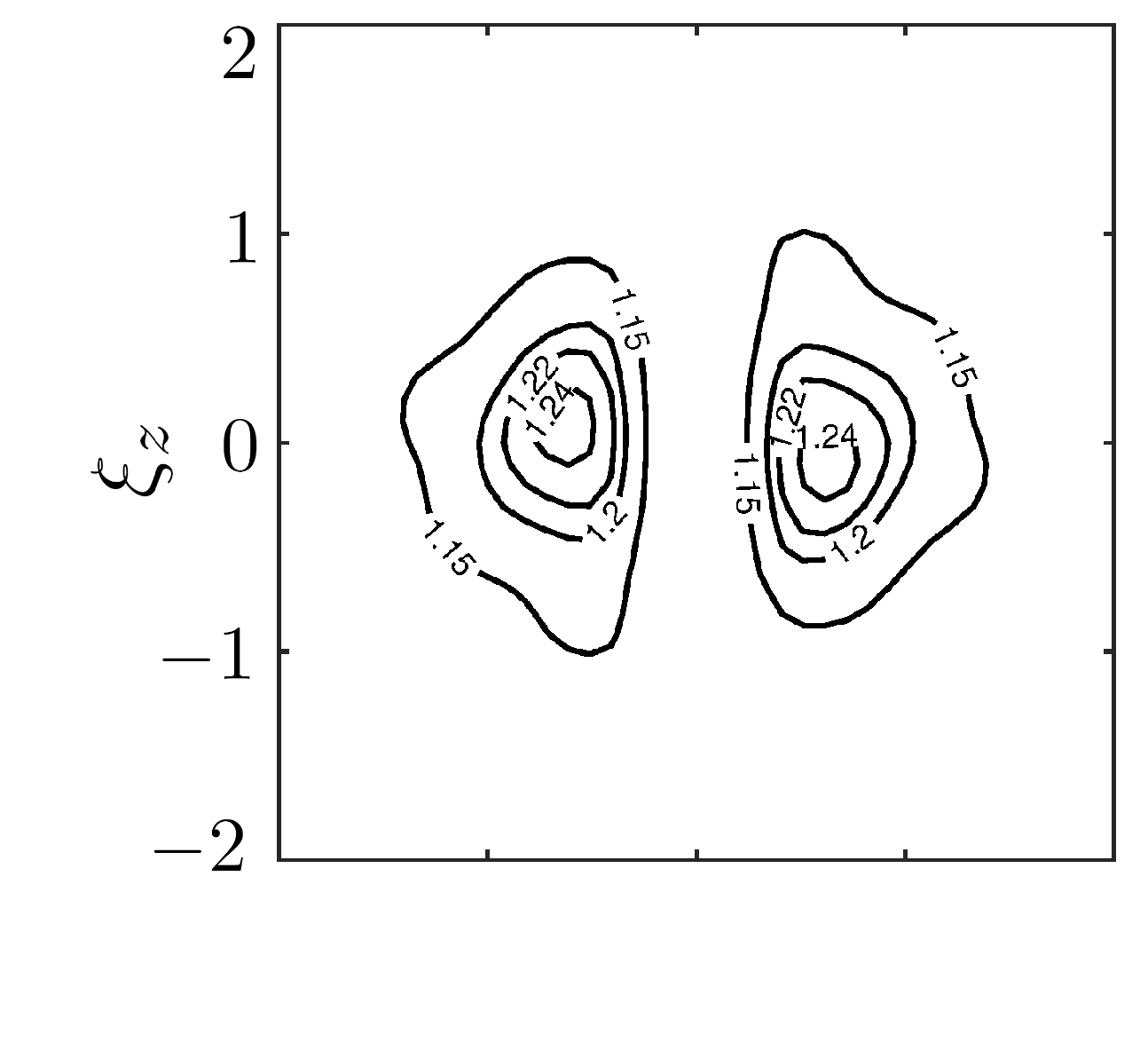}};
		\end{pgfonlayer}
		\begin{pgfonlayer}{middle1} 
			\node[anchor=south west] at (-0.15*\gridwidth + \gridwidth + \hspacing, \vspacing) {\includegraphics[width=\gridwidth]{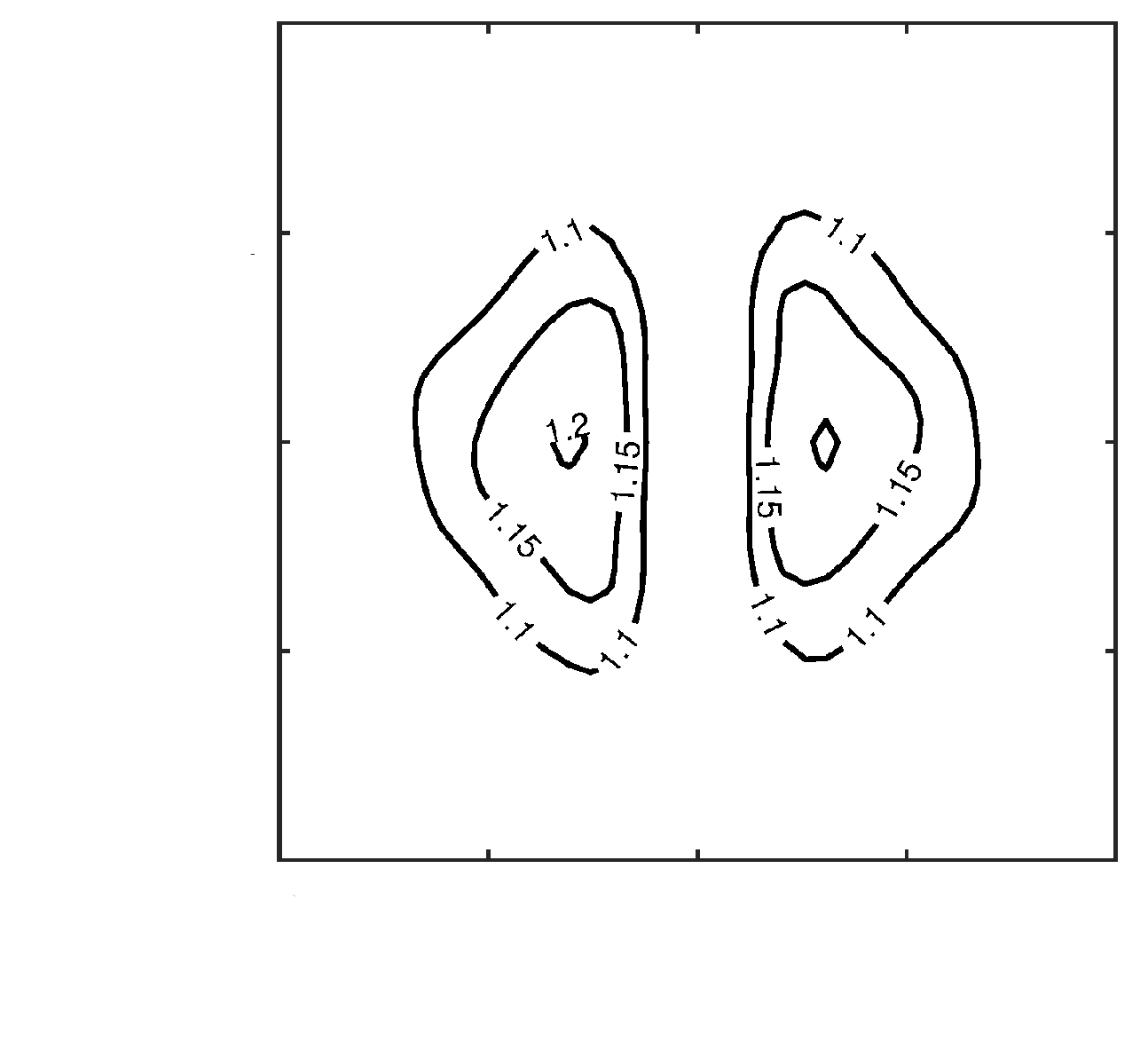}};
		\end{pgfonlayer}
		\begin{pgfonlayer}{middle2} 
			\node[anchor=south west] at (-0.15*\gridwidth + 2*\gridwidth + 2*\hspacing, \vspacing) {\includegraphics[width=\gridwidth]{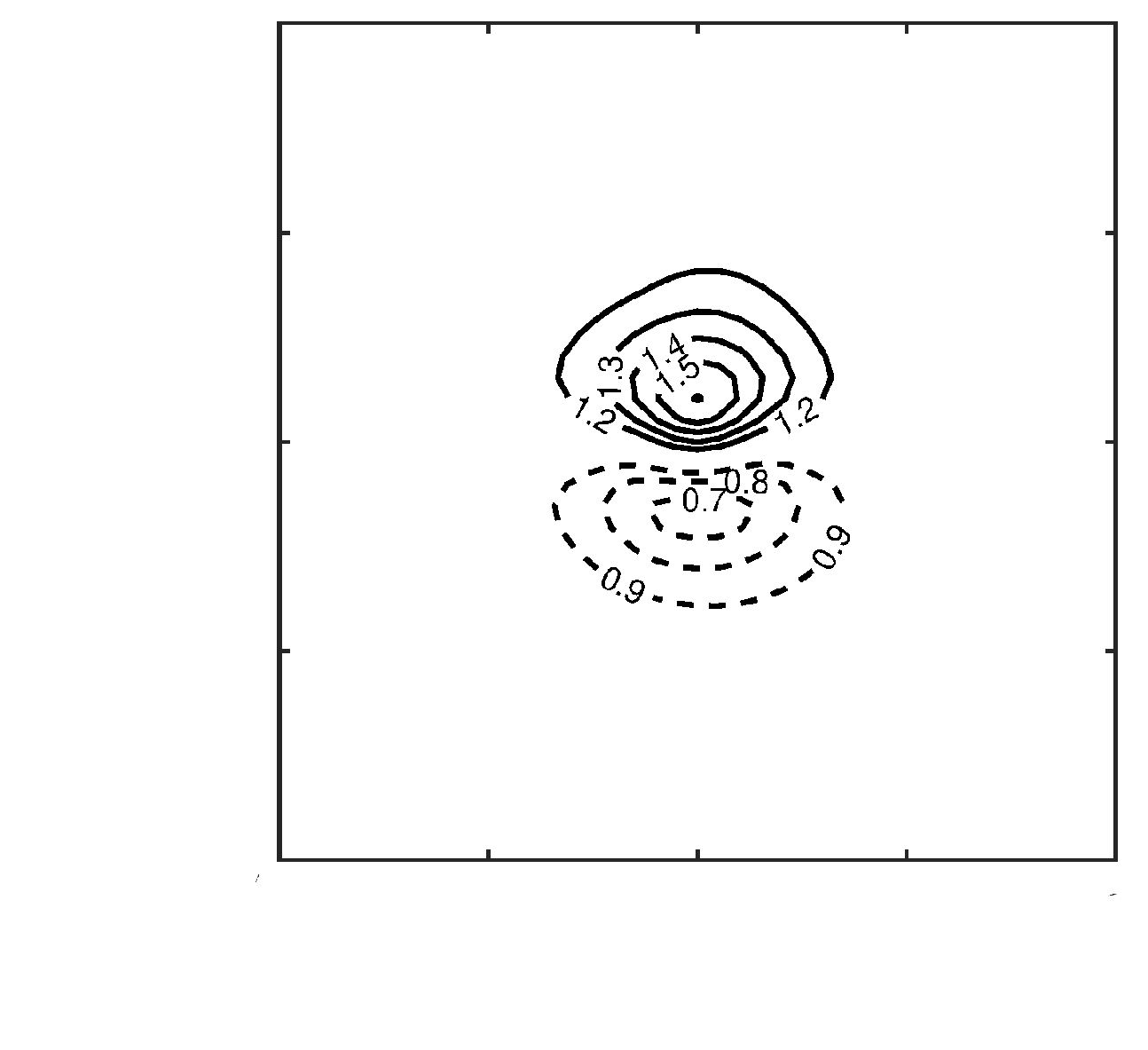}};
		\end{pgfonlayer}
		\begin{pgfonlayer}{background} 
			\node[anchor=south west] at (-0.15*\gridwidth + 3*\gridwidth + 3*\hspacing, \vspacing) {\includegraphics[width=\gridwidth]{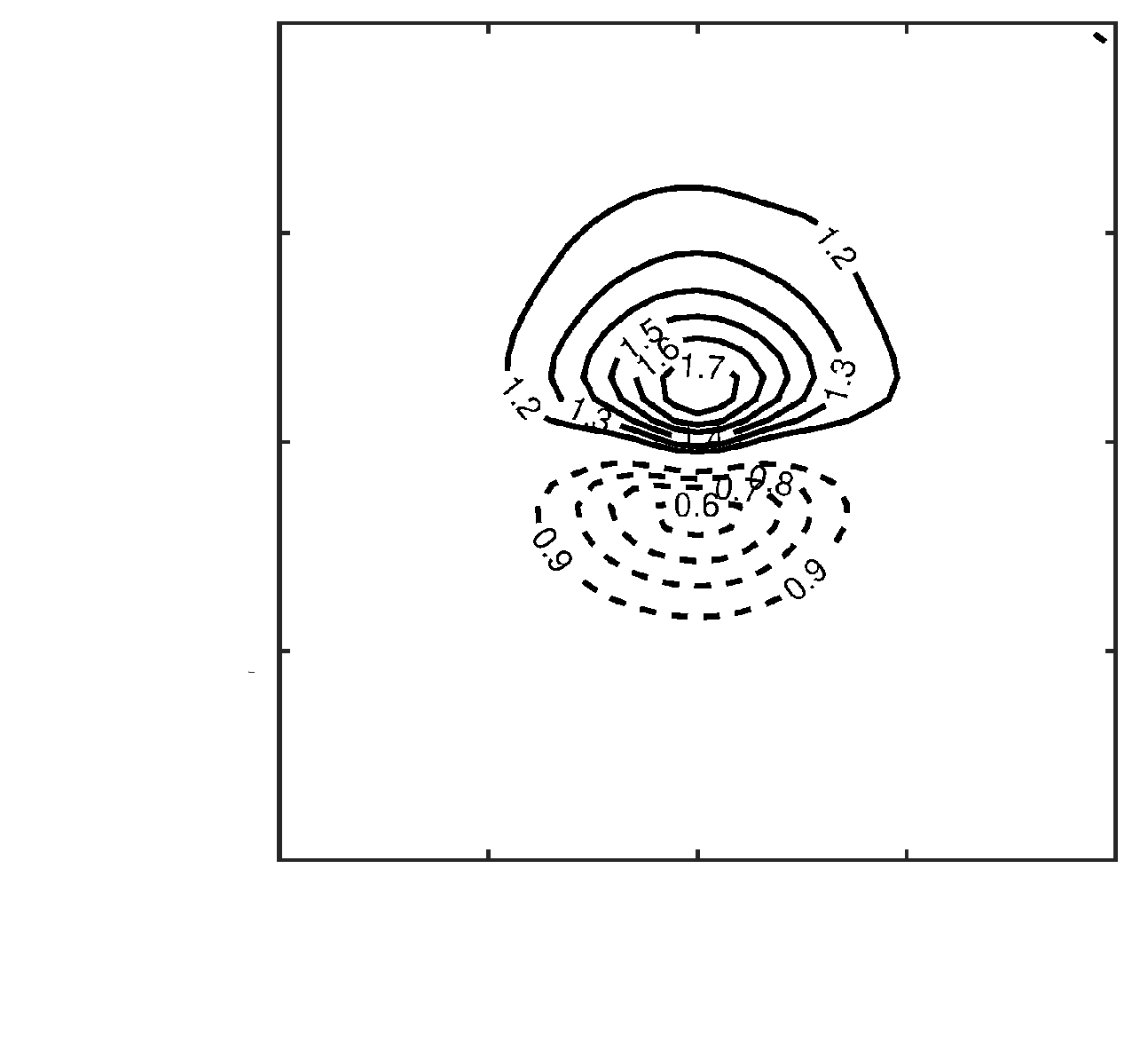}};
		\end{pgfonlayer}

		\begin{pgfonlayer}{foreground}
			\node[anchor=east] at (-0.6, \vspacing + \gridheight/2) {{DNS23 - APG2}};
		\end{pgfonlayer}

		\begin{pgfonlayer}{foreground}
			\node[anchor=south west] at (-0.15*\gridwidth, 2*\vspacing) {\includegraphics[width=\gridwidth]{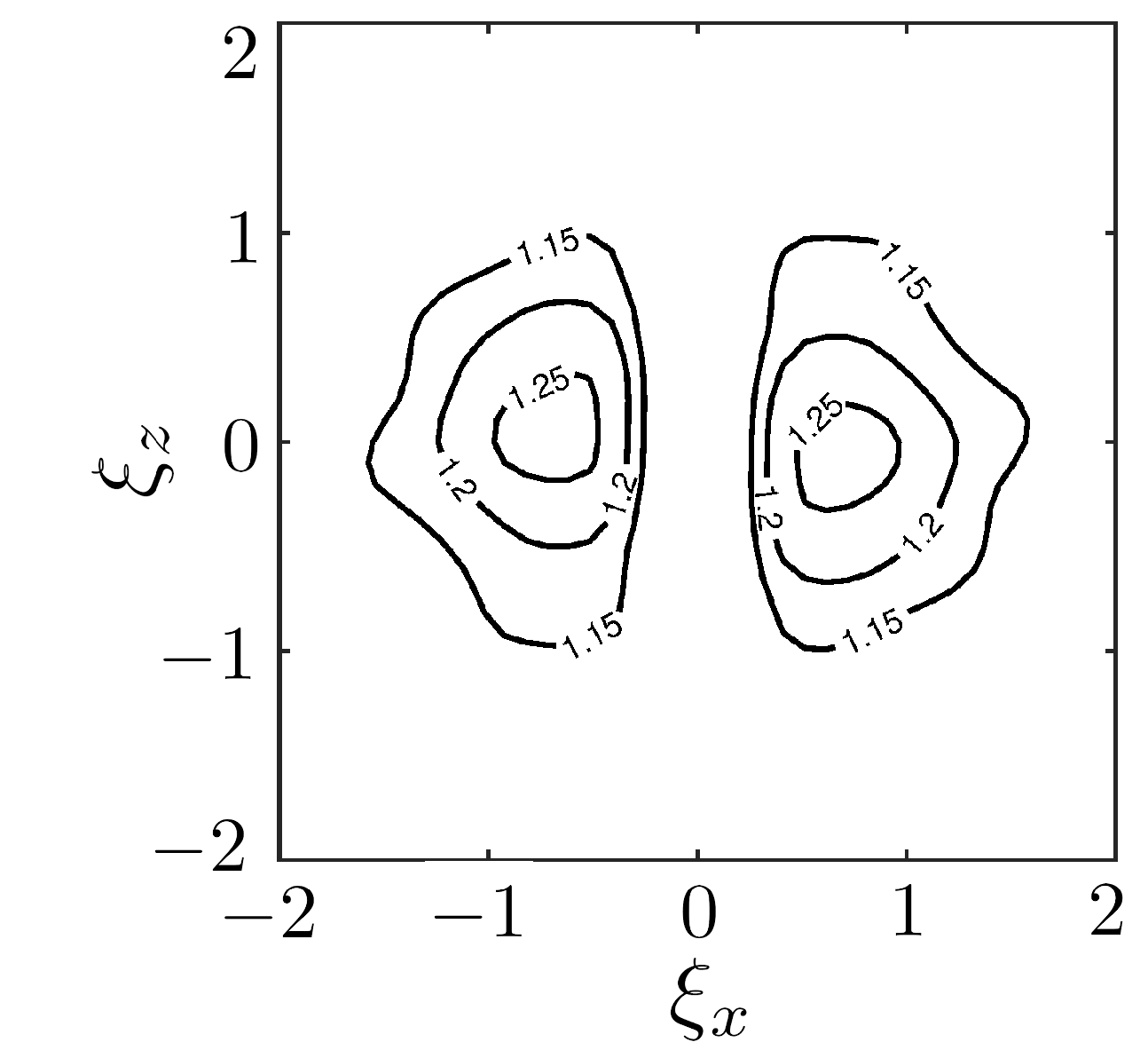}};
		\end{pgfonlayer}
		\begin{pgfonlayer}{middle1} 
			\node[anchor=south west] at (-0.15*\gridwidth + \gridwidth + \hspacing, 2*\vspacing) {\includegraphics[width=\gridwidth]{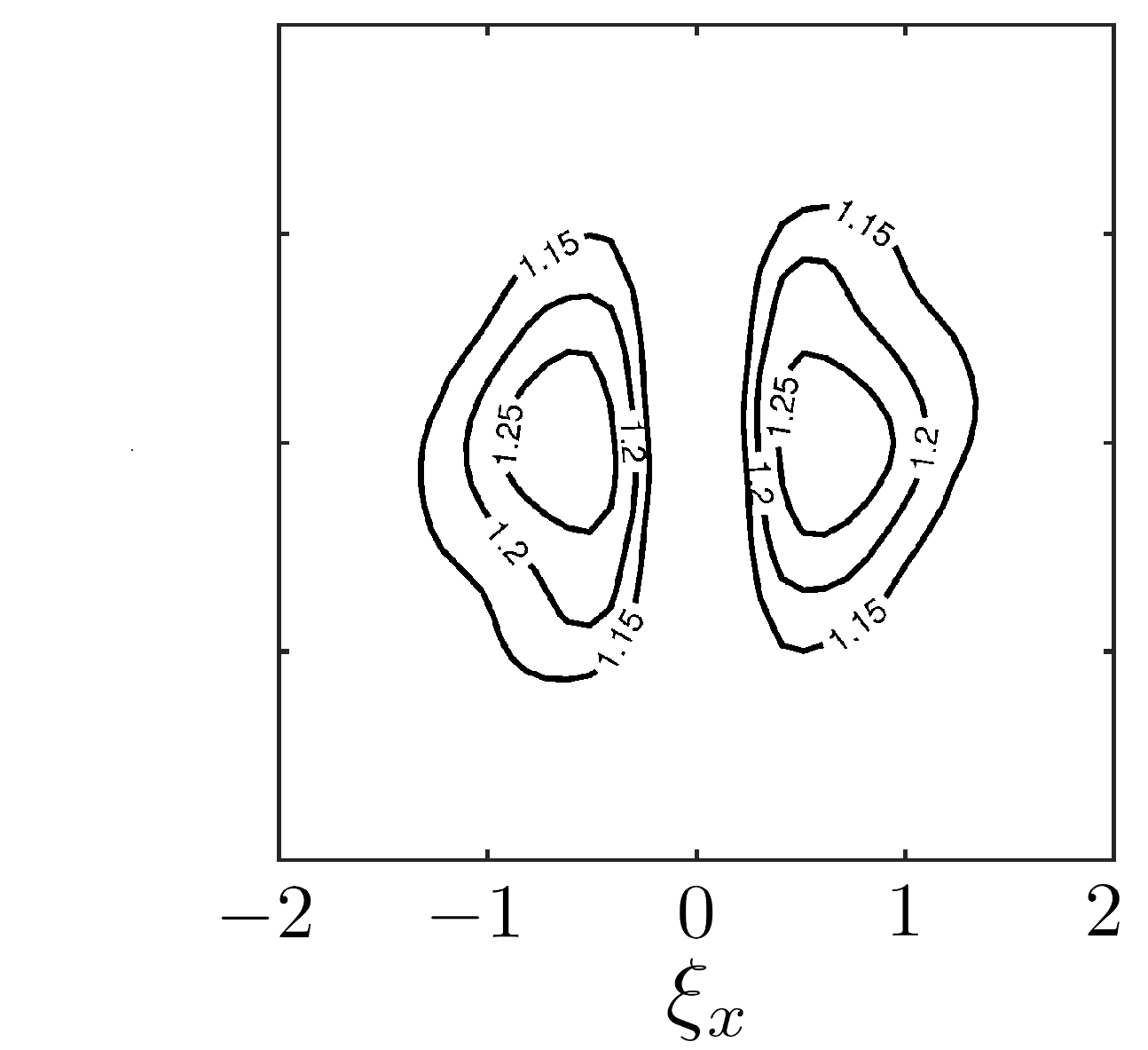}};
		\end{pgfonlayer}
		\begin{pgfonlayer}{middle2} 
			\node[anchor=south west] at (-0.15*\gridwidth + 2*\gridwidth + 2*\hspacing, 2*\vspacing) {\includegraphics[width=\gridwidth]{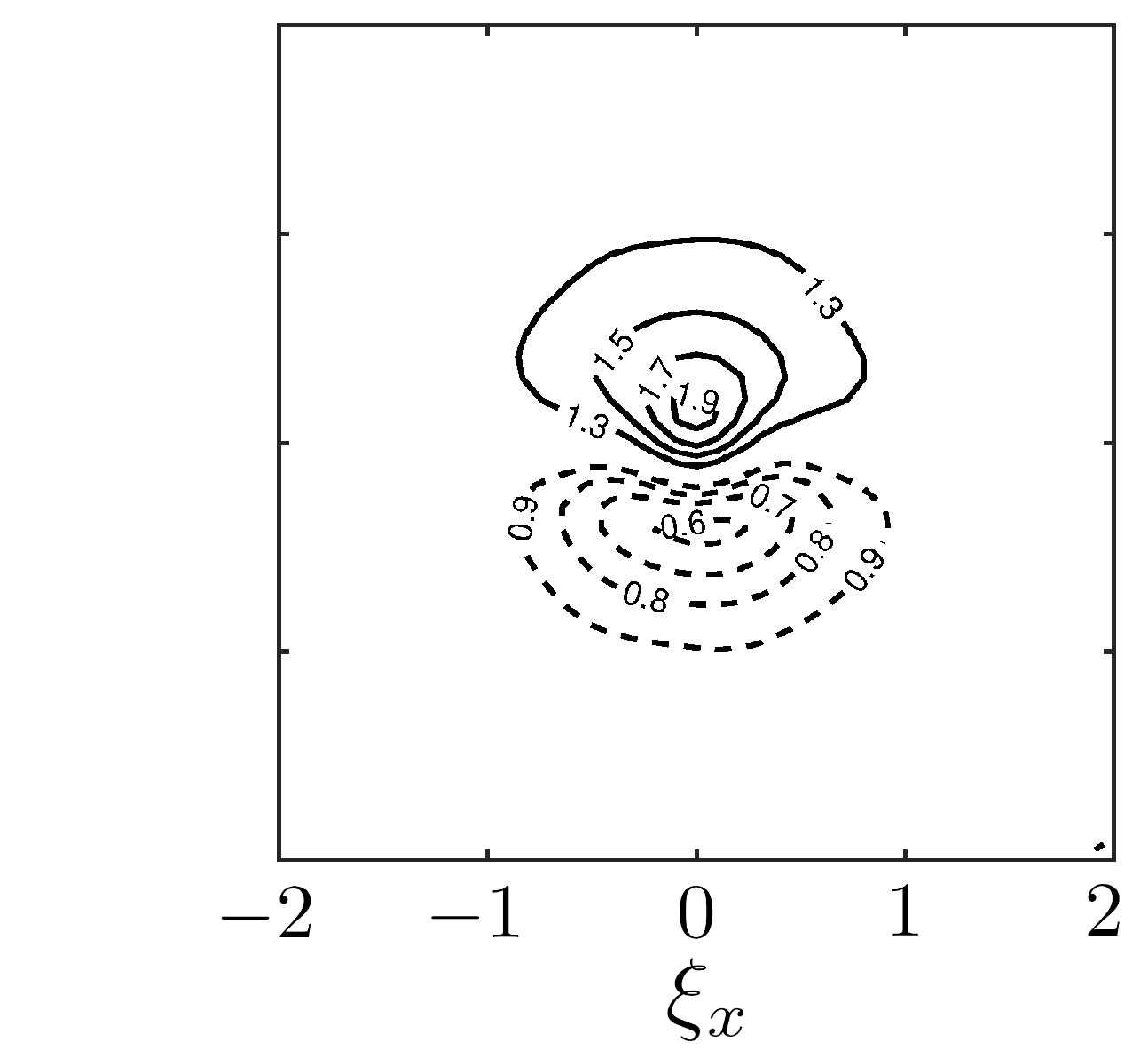}};
		\end{pgfonlayer}
		\begin{pgfonlayer}{background} 
			\node[anchor=south west] at (-0.15*\gridwidth + 3*\gridwidth + 3*\hspacing, 2*\vspacing) {\includegraphics[width=\gridwidth]{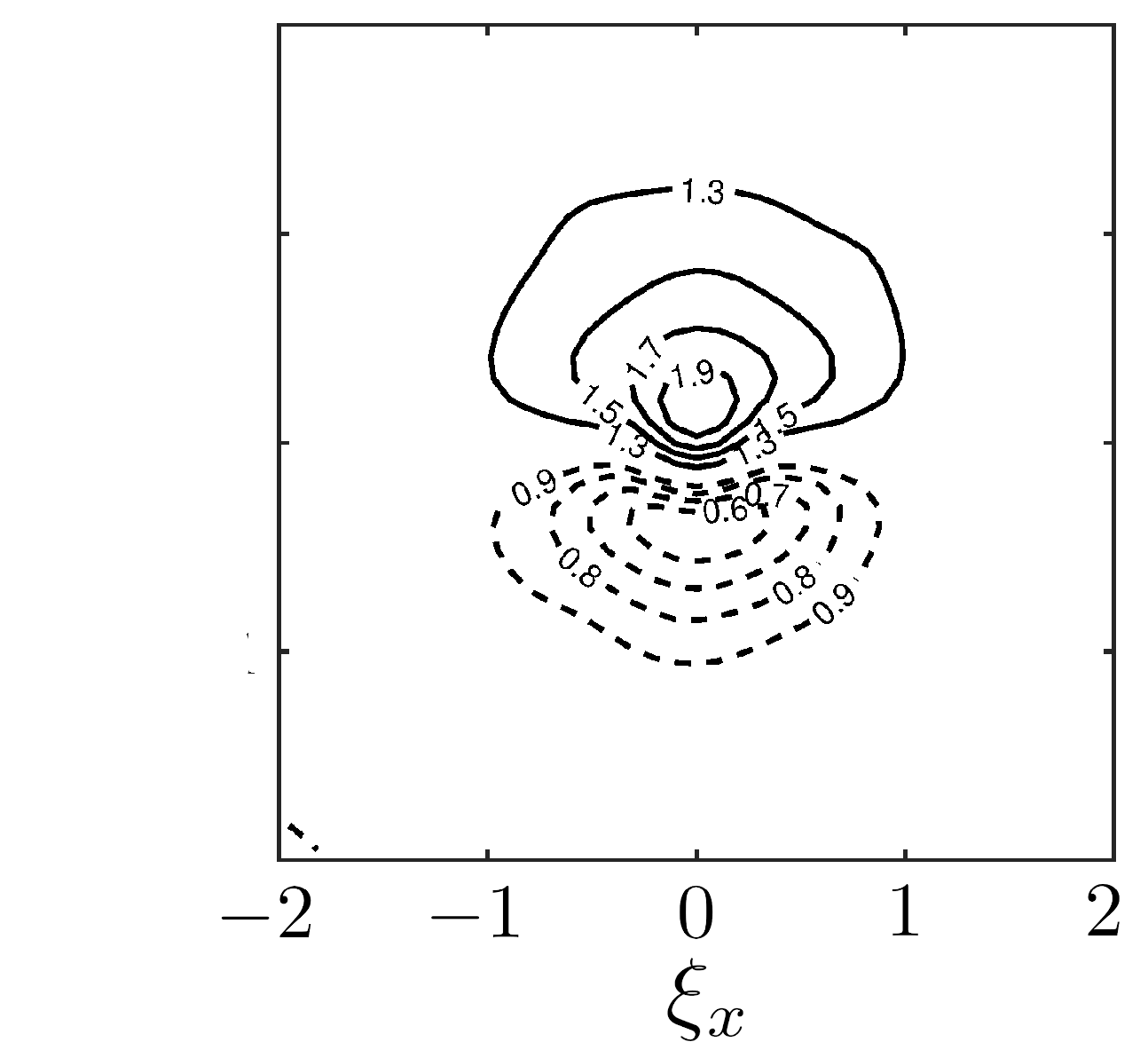}};
		\end{pgfonlayer}
		
		\begin{pgfonlayer}{foreground}
			\node[anchor=east] at (-0.8, 2*\vspacing + \gridheight/2) {{DNS23 - FPG}};
		\end{pgfonlayer}
	\end{tikzpicture}
	
	\caption{Joint PDFs of relative positions of Q2 structures with respect to Q2 structures (first column), Q4 structures with respect to Q4 structures (second column), Q2 structures with respect to Q4 structures (third column), Q4 structures with respect to Q2 structures (fourth column). Top to bottom: DNS23 APG1, APG2, FPG. Contours are normalized with the probability at long distance, $p^{ij} / p^{ij}_{far}$. Dashed lines corresponds to  $p^{ij} / p^{ij}_{far} < 1$. }
	\label{fig:relpos}
\end{figure*}

Fig.~\ref{fig:jpdfaspectratio} presents the joint PDFs of the logarithms of the aspect ratios $\left(a_{i j}=\Delta_{i} / \Delta_{j}\right)$ of the boxes circumscribing the Q2 and Q4 structures, with the first row obtained from fully-spatial data and the second row from spatio-temporal data. In low-defect cases, spatio-temporal and fully-spatial data agree with each other, showing structures that are slightly streamwise elongated. As observed in the joint PDFs of $\Delta_x$ and $\Delta_y$ in Fig.~\ref{fig:jpdfs_combined}a, the large-defect case reveals a discrepancy between the spatio-temporal and fully-spatial data. The spatio-temporal data suggests that Q2 structures tend to be approximately as wide as they are long. However, fully-spatial data shows that both Q2 and Q4 structures are slightly streamwise elongated, with similar aspect ratios.

Fig.~\ref{fig::aspectratioxy} and \ref{fig::aspectratiozy} show the average aspect ratio for all cases as a function of the diagonal size of the box circumscribing the structure $\left(d=\sqrt{\Delta_{x}^{2}+\Delta_{y}^{2}+\Delta_{z}^{2}}\right)$. The box diagonal $d$ is normalised with Corrsin length scale, $L_{c}$. In addition to the various PG TBL cases, we present the aspect ratios of Q2s and Q4s combined for the two homogeneous shear turbulence cases studied by \citet{dong}, which differ only in Reynolds number. This allows for a comparison with flows driven solely by shear, without the influence of a wall.

Fig. $\ref{fig::aspectratioxy}(a,b)$ and $\ref{fig::aspectratiozy}(a,b)$ depicts the average aspect ratio of $a_{xy}$ and $a_{zy}$ for all detached $Q^{-}$s within the region $0.3\delta < y_c < 0.8\delta$. The trend for $a_{xy}$ is very similar for all TBL cases for spatio-temporal and fully-spatial data between $2 L_c$ and $10 L_c$, with $a_{xy}$ ranging from 1.2 to 1.3. However, in fully-spatial data, structures with $d / L_c < 1$ tend to be more streamwise elongated than the spatio-temporal ones. In spatio-temporal data, the sampling frequency of the flow field limits the resolution of small-scale structures, leaving those with lifespans shorter than the sampling time unresolved. In addition, very elongated structures with $a_{xy}>2$ and sizes $d/L_c>20$ are absent in the fully-spatial data, unlike in the spatio-temporal data. This discrepancy arises because, due to the non-equilibrium nature of the flow, data collection is limited to 2$\delta$ boxes in the fully-spatial case. As a result, the size of the structures is constrained by the length of the extraction box.

The $a_{z y}$ curves collapse almost perfectly for structures whose diagonals are between $2 L_{c}$ and $10 L_{c}$ for both spatio-temporal and fully-spatial cases. The aspect ratio is slightly above 1 and then decreases to slightly below 1 with increasing structure size until approximately $d=10 L_{c}$. The diverging trends between HSTs and TBLs for large-scale structures above $d=10 L_{c}$ are attributed to the distinct boundaries of these two types of flows. In HSTs, the ratio $a_{z y}$ decreases due to the unbounded nature of structures in the $y$ direction \citep{dong}. Conversely, in TBLs, $a_{z y}$ increases as the structures are bounded in $y$. Overall, the similarity is less pronounced for $a_{x y}$ compared to $a_{z y}$. This difference may happen because the TBLs are in non-equilibrium in the streamwise direction, which is not the case for the HSTs. Comparison with equilibrium TBLs would provide further insight into this difference.

Fig. $\ref{fig::aspectratioxy}(c,d,e,f)$ and $\ref{fig::aspectratiozy}(c,d,e,f)$ separately illustrate the average aspect ratios for Q2s and Q4s. For $a_{xy}$, the fully-spatial data shows a similar trend across all TBL cases between $d/L_c = 2$ and $10$, with only a slight difference in aspect ratio values between Q2s and Q4s. In contrast, spatio-temporal data show discrepancies in the aspect ratio of Q2s and Q4s. While Q4 structures maintain nearly the same aspect ratios across all flow cases between $d/L_c = 1$ and $10$, the aspect ratios of Q2 structures decrease as the defect increases. The latter trend is absent in the fully-spatial values of $a_{xy}$ of Q2s. It could be due to the use of Taylor's hypothesis in the spatio-temporal case. A similar discrepancy was previously noted in the joint PDFs of $a_{xy}$ for Q2s, as shown in Fig.~\ref{fig:jpdfaspectratio}. The values and trend for $a_{z y}$ are almost identical for both Q2s and Q4s. The findings above show that the Q structures in TBLs and HSTs behave very similarly and almost identically in some cases when their diagonal is scaled with $L_{c}$. This clearly underscores the significant role of mean shear in shaping Q structures and, consequently, in driving turbulence production.

\subsection{Spatial organization of Q2 and Q4 structures}

The spatial organization of Q2s and Q4s holds great importance to understand the self-sustaining mechanisms of turbulence in wall-bounded flows. Fig.~\ref{fig:relpos} presents the JPDFs of relative positions of Q structures of the same and different types in the APG1, APG2, and FPG cases within a wall-parallel plane. The relative $x-$ and $z-$ positions of the Qs in quadrant $j$ with respect to those of quadrant $i$ are determined as follows:

\begin{equation}
	\zeta_x = \frac{x^j - x^i}{0.5(D^j + D^i)} \hspace{1cm} \mbox{and} \hspace{1cm} \zeta_z = \frac{z^j - z^i}{0.5(D^j + D^i)}
\end{equation}
\noindent
where \( D^i = \sqrt{\Delta^{i^2}_{x} + \Delta^{i^2}_{z}} \) represents the wall-parallel diagonal length of the structure, with \( x \) and \( z \) denoting the spatial coordinates of the structures. Spatio-temporal data are utilized to obtain these relative positions in DNS23. As before, this analysis focuses exclusively on structures within the region $0.3\delta < y_c < 0.8\delta$. To compare the JPDFs, they are normalized using the probability at a far distance, defined as the average value of the joint PDF for $\zeta_{x}^{2} + \zeta_{y}^{2} > 4$. The first and second columns of Fig.~\ref{fig:relpos} shows that Q2 and Q4 structures of the same kind tend to form an upstream - downstream configuration. Note that, in all these joint PDFs, the reference structure $i$ is located at position (0,0). In all these cases, the most probable distance between $Q^-$ structures of the same type is approximately $\left|\zeta_x\right| = 0.5 - 0.8$.

The JPDFs of different kinds shown in the last two columns are intentionally biased towards positive $\zeta_z$ in order to evaluate the spanwise symmetry of these structures. The biasing is enforced by imposing $\zeta_z > 0$ for the nearest Q4 in relation to the reference Q2, in the case of the relative position of Q4s with respect to Q2s (and vice versa for Q2s relative to Q4s). A strong secondary peak appearing at negative $\zeta_z$, alongside the primary peak at positive $\zeta_z$, indicates that the Q2 structure is surrounded by two Q4s, one on each side. Conversely, if the secondary peak is weak ($p^{ij}/p^{ij}_{far} < 1$) or absent, it suggests that Q2s and Q4s are more likely arranged side-by-side in a one-sided pair configuration. In the present flow, the most probable configuration is the single pair arrangement.

Similar analyses conducted by \cite{maciel2017a,dong,lozano2012} revealed that the streamwise alignment of Q structures of the same type, along with the presence of side-by-side sweeps and ejections, are the most probable events in ZPG, APG, channel flows, and HSTs. Additionally, similar findings from DNS22 were documented in \cite{taygun2019madrid}, which also show that near-wall turbulence activity does not influence the spatial organization of outer Q structures.

\section{Conclusion}

In this study, we analyze the Reynolds-shear-stress structures and the effect of mean shear, pressure gradient and upstream flow history on these structures in two non-equilibrium TBLs from \citet{gungor2022} and \citet{gungor2024}, as well as in the HST cases from \citet{dong}. We identify individual Reynolds-shear-stress structures in the flow and examine the spatial properties of detached sweeps and ejections. For the TBLs, the region considered is the middle of the outer layer between $0.3 \delta$ and $0.8 \delta$. The spatial characteristics of these structures seem to be governed by the mean shear as their aspect ratio behaves very similarly in all the flows, when the size of the structures is normalized with the Corrsin length scale. The relative contribution of detached sweeps and ejections to the Reynolds shear stress, however, depends on the pressure gradient context in boundary layers.

The joint probability density functions of dimensions and aspect ratio of sweeps and ejections indicate that using the volume-averaged instantaneous streamwise velocity ($U$) as the convection velocity works well in Taylor's hypothesis. Nonetheless, there are some exceptions. In the large-defect cases, this method underestimated the streamwise length of sweeps when compared to the fully-spatial data. The fully-spatial data analysis shows that the streamwise lengths of sweeps (Q2) and ejections (Q4) are comparable. Additionally, the FPG case retains the characteristics of the APG TBL, as the size of the structures in the FPG case remains similar to that in the upstream large-defect case. 

Lastly, the spatial organization of sweeps and ejections confirms that structures of different kinds tend to organize side-by-side in a one-sided pair configuration, while structures of the same kind align in an upstream-downstream configuration. This consistent pattern, observed across all flow cases, indicates that local mean shear plays a dominant role in governing structural organization. Interestingly, this behavior remains unaffected by variations in pressure gradient or flow history, aligning with findings from previous studies \citep{maciel2017a,taygun2019madrid}. Possibly related to this observation, many studies of streamwise velocity $u$ structures (large-scale motions and superstructures) in the outer region of PG TBLs have shown that these structures retain similar properties across different pressure gradient conditions, although in APG TBLs they tend to be less streamwise elongated and more inclined relative to the wall \citep{rahgozar2011, rahgozar2012, maciel2017, bross2021}.

\section{Acknowledgments}

We acknowledge the EuroHPC Joint Undertaking for awarding the project access to the EuroHPC supercomputer Leonardo DCGP at CINECA, Italy, through a EuroHPC Regular Access call, PRACE for awarding us access to Marconi100 at CINECA, Italy and Calcul Québec (www.calculquebec.ca) and the Digital Research Alliance of Canada (alliancecan.ca) for awarding us access to Niagara HPC server. TRG and YM acknowledge the support of the Natural Sciences and Engineering Research Council of Canada (NSERC), project number RGPIN-2019-04194.

\bibliographystyle{elsarticle-harv} 
\bibliography{references.bib}

\begin{thebibliography}{36}
\expandafter\ifx\csname natexlab\endcsname\relax\def\natexlab#1{#1}\fi
\providecommand{\url}[1]{\texttt{#1}}
\providecommand{\href}[2]{#2}
\providecommand{\path}[1]{#1}
\providecommand{\DOIprefix}{doi:}
\providecommand{\ArXivprefix}{arXiv:}
\providecommand{\URLprefix}{URL: }
\providecommand{\Pubmedprefix}{pmid:}
\providecommand{\doi}[1]{\href{http://dx.doi.org/#1}{\path{#1}}}
\providecommand{\Pubmed}[1]{\href{pmid:#1}{\path{#1}}}
\providecommand{\bibinfo}[2]{#2}
\ifx\xfnm\relax \def\xfnm[#1]{\unskip,\space#1}\fi
\bibitem[{Abe(2020)}]{abe2020}
\bibinfo{author}{Abe, H.}, \bibinfo{year}{2020}.
\newblock \bibinfo{title}{Direct numerical simulation of a non-equilibrium
  three-dimensional turbulent boundary layer over a flat plate}.
\newblock \bibinfo{journal}{Journal of Fluid Mechanics} \bibinfo{volume}{902},
  \bibinfo{pages}{A20}.
\bibitem[{Bobke et~al.(2017)Bobke, Vinuesa, Örlü and Schlatter}]{bobke2017}
\bibinfo{author}{Bobke, A.}, \bibinfo{author}{Vinuesa, R.},
  \bibinfo{author}{Örlü, R.}, \bibinfo{author}{Schlatter, P.},
  \bibinfo{year}{2017}.
\newblock \bibinfo{title}{History effects and near equilibrium in
  adverse-pressure-gradient turbulent boundary layers}.
\newblock \bibinfo{journal}{Journal of Fluid Mechanics} \bibinfo{volume}{820},
  \bibinfo{pages}{667--692}.
\bibitem[{Bross et~al.(2021)Bross, Eich, Schanz, Novara, Schröder and
  Kähler}]{bross2021}
\bibinfo{author}{Bross, M.}, \bibinfo{author}{Eich, F.},
  \bibinfo{author}{Schanz, D.}, \bibinfo{author}{Novara, M.},
  \bibinfo{author}{Schröder, A.}, \bibinfo{author}{Kähler, C.J.},
  \bibinfo{year}{2021}.
\newblock \bibinfo{title}{Superstructures in turbulent boundary layers with
  pressure gradients}.
\newblock \bibinfo{journal}{Proceedings in Applied Mathematics and Mechanics}
  \bibinfo{volume}{20}, \bibinfo{pages}{e202000257}.
\bibitem[{Deshpande and Vinuesa(2024)}]{deshpande2024}
\bibinfo{author}{Deshpande, R.}, \bibinfo{author}{Vinuesa, R.},
  \bibinfo{year}{2024}.
\newblock \bibinfo{title}{Streamwise energy-transfer mechanisms in zero-- and
  adverse--pressure--gradient turbulent boundary layers}.
\newblock \bibinfo{journal}{Journal of Fluid Mechanics} \bibinfo{volume}{997},
  \bibinfo{pages}{A16}.
\bibitem[{Devenport and Lowe(2022)}]{devenport2022}
\bibinfo{author}{Devenport, W.}, \bibinfo{author}{Lowe, K.T.},
  \bibinfo{year}{2022}.
\newblock \bibinfo{title}{Equilibrium and non-equilibrium turbulent boundary
  layers}.
\newblock \bibinfo{journal}{Progress in Aerospace Sciences}
  \bibinfo{volume}{131}, \bibinfo{pages}{100807}.
\bibitem[{Dong et~al.(2017)Dong, Lozano-Duran, Sekimoto and Jiménez}]{dong}
\bibinfo{author}{Dong, S.}, \bibinfo{author}{Lozano-Duran, A.},
  \bibinfo{author}{Sekimoto, A.}, \bibinfo{author}{Jiménez, J.},
  \bibinfo{year}{2017}.
\newblock \bibinfo{title}{Coherent structures in statistically stationary
  homogeneous shear turbulence}.
\newblock \bibinfo{journal}{Journal of Fluid Mechanics} \bibinfo{volume}{816},
  \bibinfo{pages}{167--208}.
\bibitem[{Gungor et~al.(2016)Gungor, Maciel, Simens and Soria}]{gungor2016}
\bibinfo{author}{Gungor, A.}, \bibinfo{author}{Maciel, Y.},
  \bibinfo{author}{Simens, M.}, \bibinfo{author}{Soria, J.},
  \bibinfo{year}{2016}.
\newblock \bibinfo{title}{Scaling and statistics of large-defect adverse
  pressure gradient turbulent boundary layers}.
\newblock \bibinfo{journal}{International Journal of Heat and Fluid Flow}
  \bibinfo{volume}{59}, \bibinfo{pages}{109--124}.
\bibitem[{Gungor et~al.(2020a)Gungor, Maciel and Gungor}]{taygun2019madrid}
\bibinfo{author}{Gungor, T.}, \bibinfo{author}{Maciel, Y.},
  \bibinfo{author}{Gungor, A.}, \bibinfo{year}{2020}a.
\newblock \bibinfo{title}{Reynolds shear-stress carrying structures in
  shear-dominated flows}.
\newblock \bibinfo{journal}{Journal of Physics: Conference Series}
  \bibinfo{volume}{1522}.
\bibitem[{Gungor et~al.(2024)Gungor, Gungor and Maciel}]{gungor2024}
\bibinfo{author}{Gungor, T.R.}, \bibinfo{author}{Gungor, A.G.},
  \bibinfo{author}{Maciel, Y.}, \bibinfo{year}{2024}.
\newblock \bibinfo{title}{Turbulent boundary layer response to uniform changes
  of the pressure force contribution}.
\newblock \bibinfo{journal}{Journal of Fluid Mechanics} \bibinfo{volume}{997},
  \bibinfo{pages}{A75}.
\bibitem[{Gungor et~al.(2020b)Gungor, Maciel and Gungor}]{gungor2020}
\bibinfo{author}{Gungor, T.R.}, \bibinfo{author}{Maciel, Y.},
  \bibinfo{author}{Gungor, A.}, \bibinfo{year}{2020}b.
\newblock \bibinfo{title}{Reynolds shear-stress carrying structures in
  shear-dominated flows}.
\newblock \bibinfo{journal}{Journal of Physics: Conference Series}
  \bibinfo{volume}{1522}, \bibinfo{pages}{012009}.
\bibitem[{Gungor et~al.(2022)Gungor, Maciel and Gungor}]{gungor2022}
\bibinfo{author}{Gungor, T.R.}, \bibinfo{author}{Maciel, Y.},
  \bibinfo{author}{Gungor, A.G.}, \bibinfo{year}{2022}.
\newblock \bibinfo{title}{Energy transfer mechanisms in adverse pressure
  gradient turbulent boundary layers: production and inter-component
  redistribution}.
\newblock \bibinfo{journal}{Journal of Fluid Mechanics} \bibinfo{volume}{948},
  \bibinfo{pages}{A5}.
\bibitem[{Harun et~al.(2013)Harun, Monty, Mathis and Marusic}]{harun2013}
\bibinfo{author}{Harun, Z.}, \bibinfo{author}{Monty, J.P.},
  \bibinfo{author}{Mathis, R.}, \bibinfo{author}{Marusic, I.},
  \bibinfo{year}{2013}.
\newblock \bibinfo{title}{Pressure gradient effects on the large-scale
  structure of turbulent boundary layers}.
\newblock \bibinfo{journal}{Journal of Fluid Mechanics} \bibinfo{volume}{715},
  \bibinfo{pages}{477--498}.
\bibitem[{Jiménez(2013)}]{jimenez_nearwall}
\bibinfo{author}{Jiménez, J.}, \bibinfo{year}{2013}.
\newblock \bibinfo{title}{{Near-wall turbulence}}.
\newblock \bibinfo{journal}{Physics of Fluids} \bibinfo{volume}{25},
  \bibinfo{pages}{101302}.
\bibitem[{Jiménez(2018)}]{jimenez2018}
\bibinfo{author}{Jiménez, J.}, \bibinfo{year}{2018}.
\newblock \bibinfo{title}{Coherent structures in wall-bounded turbulence}.
\newblock \bibinfo{journal}{Journal of Fluid Mechanics} \bibinfo{volume}{842},
  \bibinfo{pages}{P1}.
\bibitem[{Jiménez et~al.(2010)Jiménez, Hoyas, Simens and
  Mizuno}]{jimenez2010}
\bibinfo{author}{Jiménez, J.}, \bibinfo{author}{Hoyas, S.},
  \bibinfo{author}{Simens, M.}, \bibinfo{author}{Mizuno, Y.},
  \bibinfo{year}{2010}.
\newblock \bibinfo{title}{Turbulent boundary layers and channels at moderate
  {Reynolds} numbers}.
\newblock \bibinfo{journal}{Journal of Fluid Mechanics} \bibinfo{volume}{657},
  \bibinfo{pages}{335--360}.
\bibitem[{Jiménez and Kawahara(2010)}]{jimenezkawahara2010}
\bibinfo{author}{Jiménez, J.}, \bibinfo{author}{Kawahara, G.},
  \bibinfo{year}{2010}.
\newblock \bibinfo{title}{Dynamics of wall-bounded turbulence}.
\newblock \bibinfo{journal}{Ten Chapters in Turbulence} ,
  \bibinfo{pages}{221--268}.
\bibitem[{Joshi et~al.(2014)Joshi, Liu and Katz}]{joshi2014}
\bibinfo{author}{Joshi, P.}, \bibinfo{author}{Liu, X.}, \bibinfo{author}{Katz,
  J.}, \bibinfo{year}{2014}.
\newblock \bibinfo{title}{Effect of mean and fluctuating pressure gradients on
  boundary layer turbulence}.
\newblock \bibinfo{journal}{Journal of Fluid Mechanics} \bibinfo{volume}{748},
  \bibinfo{pages}{36–84}.
\bibitem[{Kim et~al.(1987)Kim, Moin and Moser}]{KimMoinMoser1987}
\bibinfo{author}{Kim, J.}, \bibinfo{author}{Moin, P.}, \bibinfo{author}{Moser,
  R.}, \bibinfo{year}{1987}.
\newblock \bibinfo{title}{Turbulence statistics in fully developed channel flow
  at low {Reynolds} number}.
\newblock \bibinfo{journal}{Journal of Fluid Mechanics} \bibinfo{volume}{177},
  \bibinfo{pages}{133--166}.
\bibitem[{Kitsios et~al.(2017)Kitsios, Sekimoto, Atkinson, Sillero, Borrell,
  Gungor, Jiménez and Soria}]{kitsios2017}
\bibinfo{author}{Kitsios, V.}, \bibinfo{author}{Sekimoto, A.},
  \bibinfo{author}{Atkinson, C.}, \bibinfo{author}{Sillero, J.A.},
  \bibinfo{author}{Borrell, G.}, \bibinfo{author}{Gungor, A.G.},
  \bibinfo{author}{Jiménez, J.}, \bibinfo{author}{Soria, J.},
  \bibinfo{year}{2017}.
\newblock \bibinfo{title}{Direct numerical simulation of a self-similar adverse
  pressure gradient turbulent boundary layer at the verge of separation}.
\newblock \bibinfo{journal}{Journal of Fluid Mechanics} \bibinfo{volume}{829},
  \bibinfo{pages}{392–419}.
\bibitem[{Lee(2017)}]{lee2017}
\bibinfo{author}{Lee, J.H.}, \bibinfo{year}{2017}.
\newblock \bibinfo{title}{Large-scale motions in turbulent boundary layers
  subjected to adverse pressure gradients}.
\newblock \bibinfo{journal}{Journal of Fluid Mechanics} \bibinfo{volume}{810},
  \bibinfo{pages}{323--–361}.
\bibitem[{Lozano-Durán et~al.(2012)Lozano-Durán, Flores and
  Jiménez}]{lozano2012}
\bibinfo{author}{Lozano-Durán, A.}, \bibinfo{author}{Flores, O.},
  \bibinfo{author}{Jiménez, J.}, \bibinfo{year}{2012}.
\newblock \bibinfo{title}{The three-dimensional structure of momentum transfer
  in turbulent channels}.
\newblock \bibinfo{journal}{Journal of Fluid Mechanics} \bibinfo{volume}{694},
  \bibinfo{pages}{100–130}.
\bibitem[{Maciel et~al.(2017a)Maciel, Gungor and Simens}]{maciel2017a}
\bibinfo{author}{Maciel, Y.}, \bibinfo{author}{Gungor, A.},
  \bibinfo{author}{Simens, M.}, \bibinfo{year}{2017a}.
\newblock \bibinfo{title}{Structural differences between small and large
  momentum-defect turbulent boundary layers}.
\newblock \bibinfo{journal}{International Journal of Heat and Fluid Flow}
  \bibinfo{volume}{67}, \bibinfo{pages}{95--110}.
\bibitem[{Maciel et~al.(2017)Maciel, Gungor and Simens}]{maciel2017}
\bibinfo{author}{Maciel, Y.}, \bibinfo{author}{Gungor, A.G.},
  \bibinfo{author}{Simens, M.}, \bibinfo{year}{2017}.
\newblock \bibinfo{title}{Structural differences between small and large
  momentum-defect turbulent boundary layers}.
\newblock \bibinfo{journal}{International Journal of Heat and Fluid Flow}
  \bibinfo{volume}{67}, \bibinfo{pages}{95--110}.
\bibitem[{Maciel et~al.(2017b)Maciel, Simens and Gungor}]{maciel2017b}
\bibinfo{author}{Maciel, Y.}, \bibinfo{author}{Simens, M.},
  \bibinfo{author}{Gungor, A.}, \bibinfo{year}{2017b}.
\newblock \bibinfo{title}{Coherent structures in a non-equilibrium
  large-velocity-defect turbulent boundary layer}.
\newblock \bibinfo{journal}{Flow, Turbulence and Combustion}
  \bibinfo{volume}{98}, \bibinfo{pages}{1--20}.
\bibitem[{Maciel et~al.(2018)Maciel, Wei, Gungor and Simens}]{maciel2018}
\bibinfo{author}{Maciel, Y.}, \bibinfo{author}{Wei, T.},
  \bibinfo{author}{Gungor, A.G.}, \bibinfo{author}{Simens, M.P.},
  \bibinfo{year}{2018}.
\newblock \bibinfo{title}{Outer scales and parameters of
  adverse-pressure-gradient turbulent boundary layers}.
\newblock \bibinfo{journal}{Journal of Fluid Mechanics} \bibinfo{volume}{844},
  \bibinfo{pages}{5–35}.
\bibitem[{Marusic and Adrian(2012)}]{Marusic_Adrian_2012}
\bibinfo{author}{Marusic, I.}, \bibinfo{author}{Adrian, R.J.},
  \bibinfo{year}{2012}.
\newblock \bibinfo{title}{The Eddies and Scales of Wall Turbulence}.
  \bibinfo{publisher}{Cambridge University Press}.
\newblock pp. \bibinfo{pages}{176--220}.
\bibitem[{Moser et~al.(1999)Moser, Kim and Mansour}]{moser1999}
\bibinfo{author}{Moser, R.D.}, \bibinfo{author}{Kim, J.},
  \bibinfo{author}{Mansour, N.N.}, \bibinfo{year}{1999}.
\newblock \bibinfo{title}{Direct numerical simulation of turbulent channel flow
  up to ${Re}_{\tau}$ = 590}.
\newblock \bibinfo{journal}{Physics of Fluids} \bibinfo{volume}{11},
  \bibinfo{pages}{943--945}.
\bibitem[{Parthasarathy and Saxton-Fox(2023)}]{parthasarathy2023}
\bibinfo{author}{Parthasarathy, A.}, \bibinfo{author}{Saxton-Fox, T.},
  \bibinfo{year}{2023}.
\newblock \bibinfo{title}{A family of adverse pressure gradient turbulent
  boundary layers with upstream favourable pressure gradients}.
\newblock \bibinfo{journal}{Journal of Fluid Mechanics} \bibinfo{volume}{966},
  \bibinfo{pages}{A11}.
\bibitem[{Piomelli and Yuan(2013)}]{piomelli2013}
\bibinfo{author}{Piomelli, U.}, \bibinfo{author}{Yuan, J.},
  \bibinfo{year}{2013}.
\newblock \bibinfo{title}{Numerical simulations of spatially developing,
  accelerating boundary layers}.
\newblock \bibinfo{journal}{Physics of Fluids} \bibinfo{volume}{25},
  \bibinfo{pages}{101304}.
\bibitem[{Rahgozar and Maciel(2011)}]{rahgozar2011}
\bibinfo{author}{Rahgozar, S.}, \bibinfo{author}{Maciel, Y.},
  \bibinfo{year}{2011}.
\newblock \bibinfo{title}{Low- and high-speed structures in the outer region of
  an adverse-pressure-gradient turbulent boundary layer}.
\newblock \bibinfo{journal}{Experimental Thermal and Fluid Science}
  \bibinfo{volume}{35}, \bibinfo{pages}{1575--1587}.
\bibitem[{Rahgozar and Maciel(2012)}]{rahgozar2012}
\bibinfo{author}{Rahgozar, S.}, \bibinfo{author}{Maciel, Y.},
  \bibinfo{year}{2012}.
\newblock \bibinfo{title}{Statistical analysis of low- and high-speed
  large-scale structures in the outer region of an adverse pressure gradient
  turbulent boundary layer}.
\newblock \bibinfo{journal}{Journal of Turbulence} \bibinfo{volume}{13},
  \bibinfo{pages}{N46}.
\bibitem[{Sillero et~al.(2013)Sillero, Jiménez and Moser}]{sillero2013}
\bibinfo{author}{Sillero, J.}, \bibinfo{author}{Jiménez, J.},
  \bibinfo{author}{Moser, R.}, \bibinfo{year}{2013}.
\newblock \bibinfo{title}{One-point statistics for turbulent wall-bounded flows
  at {R}eynolds numbers up to $\delta^+ \approx $ 2000}.
\newblock \bibinfo{journal}{Physics of Fluids} \bibinfo{volume}{25},
  \bibinfo{pages}{5102}.
\bibitem[{Skåre and Krogstad(1994)}]{skarekrogstad1994}
\bibinfo{author}{Skåre, P.E.}, \bibinfo{author}{Krogstad, P.Å.},
  \bibinfo{year}{1994}.
\newblock \bibinfo{title}{A turbulent equilibrium boundary layer near
  separation}.
\newblock \bibinfo{journal}{Journal of Fluid Mechanics} \bibinfo{volume}{272},
  \bibinfo{pages}{319--–348}.
\bibitem[{Tanarro et~al.(2020)Tanarro, Vinuesa and Schlatter}]{tanarro2020}
\bibinfo{author}{Tanarro, A.}, \bibinfo{author}{Vinuesa, R.},
  \bibinfo{author}{Schlatter, P.}, \bibinfo{year}{2020}.
\newblock \bibinfo{title}{Effect of adverse pressure gradients on turbulent
  wing boundary layers}.
\newblock \bibinfo{journal}{Journal of Fluid Mechanics} \bibinfo{volume}{883}.
\bibitem[{Vinuesa et~al.(2017)Vinuesa, Örlü, Sanmiguel~Vila, Ianiro, Discetti
  and Schlatter}]{vinuesa2017}
\bibinfo{author}{Vinuesa, R.}, \bibinfo{author}{Örlü, R.},
  \bibinfo{author}{Sanmiguel~Vila, C.}, \bibinfo{author}{Ianiro, A.},
  \bibinfo{author}{Discetti, S.}, \bibinfo{author}{Schlatter, P.},
  \bibinfo{year}{2017}.
\newblock \bibinfo{title}{Revisiting history effects in
  adverse-pressure-gradient turbulent boundary layers}.
\newblock \bibinfo{journal}{Flow, Turbulence and Combustion}
  \bibinfo{volume}{99}, \bibinfo{pages}{565--587}.
\bibitem[{Volino(2020)}]{volino2020}
\bibinfo{author}{Volino, R.J.}, \bibinfo{year}{2020}.
\newblock \bibinfo{title}{Non-equilibrium development in turbulent boundary
  layers with changing pressure gradients}.
\newblock \bibinfo{journal}{Journal of Fluid Mechanics} \bibinfo{volume}{897},
  \bibinfo{pages}{A2}.

\end{thebibliography}

\newpage

\onecolumn

\end{document}